\documentclass[11pt,english,aps,tightenlinesletterpaper,groupaddress,secnumarabic,nofootinbib]{revtex4}

\usepackage[T1]{fontenc}
\usepackage[latin9]{inputenc}
\usepackage[letterpaper]{geometry}
\usepackage{xcolor}
\usepackage{babel}
\usepackage{float}
\usepackage{amsmath}
\usepackage{amssymb}
\usepackage{graphicx}
\PassOptionsToPackage{normalem}{ulem}
\usepackage{ulem}
\usepackage[unicode=true,pdfusetitle,
 bookmarks=true,bookmarksnumbered=false,bookmarksopen=false,
 breaklinks=false,pdfborder={0 0 1},backref=false,colorlinks=false]
 {hyperref}

\makeatletter

\providecolor{lyxadded}{rgb}{0,0,1}
\providecolor{lyxdeleted}{rgb}{1,0,0}

\DeclareRobustCommand{\lyxsout}[1]{\ifx\\#1\else\sout{#1}\fi}

\@ifundefined{textcolor}{}
{%
 \definecolor{BLACK}{gray}{0}
 \definecolor{WHITE}{gray}{1}
 \definecolor{RED}{rgb}{1,0,0}
 \definecolor{GREEN}{rgb}{0,1,0}
 \definecolor{BLUE}{rgb}{0,0,1}
 \definecolor{CYAN}{cmyk}{1,0,0,0}
 \definecolor{MAGENTA}{cmyk}{0,1,0,0}
 \definecolor{YELLOW}{cmyk}{0,0,1,0}
}

\makeatother

\begin{document}
\title{Large Blue Spectral Index From a Conformal Limit of a Rotating Complex
Scalar}
\author{Daniel J. H. Chung}
\email{danielchung@wisc.edu}

\affiliation{Department of Physics, University of Wisconsin-Madison, Madison, WI
53706, USA}
\author{Sai Chaitanya Tadepalli}
\email{stadepalli@wisc.edu}

\affiliation{Department of Physics, University of Wisconsin-Madison, Madison, WI
53706, USA}
\begin{abstract}
One well known method of generating a large blue spectral index for
axionic isocurvature perturbations is through a flat direction not
having a quartic potential term for the radial partner of the axion
field. In this work, we show how one can obtain a large blue spectral
index even with a quartic potential term associated with the Peccei-Quinn
symmetry breaking radial partner. We use the fact that a large radial
direction with a quartic term can naturally induce a conformal limit
which generates an isocurvature spectral index of 3. We point out
that this conformal representation is intrinsically different from
both the ordinary equilibrium axion scenario or massless fields in
Minkowski spacetime. Another way to view this limit is as a scenario
where the angular momentum of the initial conditions slows down the
radial field or as a superfluid limit. Quantization of the non-static
system in which derivative of the radial field and the derivative
of the angular field do not commute is treated with great care to
compute the vacuum state. The parametric region consistent with axion
dark matter and isocurvature cosmology is discussed.

\tableofcontents{}
\end{abstract}
\maketitle

\section{Introduction}

Having a large blue tilt in the axionic isocurvature spectrum allows
cold dark matter (CDM) density perturbations to be enhanced on short
length scales without being in conflict with the precision cosmology
that exists for scales $k/a_{0}\lesssim1$ Mpc$^{-1}$ \citep{Chluba:2013dna,Takeuchi2014,Dent:2012ne,Chung:2015pga,Chung:2015tha,Chluba:2016bvg,Chung:2017uzc,Planck:2018jri,Chabanier:2019eai,Lee:2021bmn,Kurmus:2022guy}.
The generation of isocurvature perturbations by spectator axions,
its model-specific characteristics, and the related observational
limitations have been extensively investigated in the past (see for
example \citep{Kasuya1997,Kawasaki1995,Nakayama2015,Harigaya2015,Kadota2014,Kitajima2014,Kawasaki2014,Higaki2014,Jeong2013,Kobayashi2013,Hamann2009,Hertzberg2008,Beltran2006,Fox2004,Estevez2016,Kearney2016,Nomura2015,Kadota2015,Hikage:2012be,Langlois2003,Mollerach1990,Axenides1983,Jo2020,Iso2021,Bae2018,Visinelli2017,Takeuchi:2013hza,Bucher:2000hy,Lu:2021gso,Sakharov:2021dim,Rosa:2021gbe,Jukko:2021hql,Chen:2021wcf,Jeong:2022kdr,Cicoli:2022fzy,Koutsangelas:2022lte,Kawasaki:2023zpd}).
Although there are models of axion which naturally generate large
blue tilted spectra when there are no quartic potential terms in the
radial field \citep{Kasuya:2009up,Dreiner:2014eda}, there is no previous
discussion in the literature regarding generating a large $k$ range
of very blue spectrum followed by a plateau from a well-motivated
axion models that contain a quartic term in the radial potential \citep{Kim:2008hd,DiLuzio:2020wdo}.\footnote{For models that generate a moderately large blue tilt, although not
as large as the ones considered in this paper, see \citep{Ebadi:2023xhq}.} From a model building perspective, one can therefore ask whether
the overdamped spectrum of \citep{Kasuya:2009up} (i.e. a smooth spectrum
composed of an exponentially large $k$-range with a very blue spectral
index and followed by a zero spectral index plateau without any large
bumplike features) can be a signature of flat direction models that
are distinct from the quartic radial potential models. If the answer
is affirmative, then not only is the time-dependent mass a property
that one can infer \citep{Chung:2015tha} from measuring the spectral
shape similar to that of \citep{Kasuya:2009up}, also the existence
of flat direction would be inferrable from such a measurement.

Motivated by this question and also from the desire to find novel
well-motivated beyond the Standard Model scenarios that generate a
strongly blue tilted axionic isocurvature spectra, we consider a generic
complex scalar sector with the radial direction field $\Gamma$, the
angular field $\theta$, and a quartic coupling $\lambda$. The quartic
term usually controls the axion decay constant
\begin{equation}
\Gamma_{{\rm vac}}=\sqrt{\frac{2M^{2}}{\lambda}}
\end{equation}
(what people often denote as $f_{\mathrm{PQ}}$) where the mass parameter
$M$ controls the tachyonic mass term responsible for spontaneous
breaking of Peccei-Quinn (PQ) symmetry. For the blue isocurvature
models, we require a large rolling period of the radial field $\Gamma$
because it is the time-dependence of the background fields that map
to the nontrivial positive power (i.e.~blue tilt) of $k$ in the
dimensionless power spectrum. When $\Gamma\gg\Gamma_{\mathrm{vac}}$
and the initial kinetic energy is negligible, we might naively expect
$\Gamma$ to roll to $\Gamma_{\mathrm{vac}}$ on a time scale of $\left(V''(\Gamma)\right)^{-1/2}$.
Such a fast roll for $V''(\Gamma)\gg H^{2}$ (making a large blue
tilt) would generate the range of $k$ over which the blue spectra
is produced to be 
\begin{equation}
\frac{k_{\mathrm{break}}-k_{\mathrm{longest}}}{k_{\mathrm{longest}}}\sim\frac{H}{\sqrt{V''(\Gamma)}}\ll1
\end{equation}
where $k_{\mathrm{longest}}$ is the length scale that leaves the
horizon when $\Gamma$ begins to roll and $k_{\mathrm{break}}$ is
the scale that leaves the horizon when $\Gamma$ reaches the minimum
such that modes having $k>k_{\mathrm{break}}$ will approximately
be a flat spectrum.

However, if there is an axion background field motion in the conserved
$U(1)_{\mathrm{PQ}}$ angular direction, we know that the time scale
to reach $\Gamma_{\mathrm{vac}}$ can be infinite in the limit that
all $U(1)_{\mathrm{PQ}}$ symmetry breaking terms are turned off and
$\dot{a}\rightarrow0$. Hence, in scenarios where the angular motion
in the conserved angular momentum direction is large, we might expect
to be able to have a similar radial rolling as the flat direction
model of \citep{Kasuya:2009up}. Unlike in the scenarios of \citep{Co:2019wyp,Co:2021lkc},
we will use the initial conditions where the phenomenology generating
rotations are occurring during inflation. In such angular momentum
dependent scenarios, one naively expects the main limitations to obtaining
a large $k_{\mathrm{break}}/k_{\mathrm{longest}}$ to be the dilution
of the angular momentum due to the Hubble expansion. After substituting
the kinetic derived $\dot{\theta}^{2}$ for the ``angular momentum''
$L^{2}$, there is the well known effective potential term
\begin{equation}
V_{E}(\Gamma,a)=\frac{\lambda}{4}\Gamma^{4}+\frac{1}{2}\frac{L^{2}}{a^{6}\Gamma^{2}}
\end{equation}
which naively indicates that the effect of $L^{2}$ will decay as
$a^{-6}$, becoming irrelevant too fast to be of interest. However,
the situation is a bit more interesting.

Because $\Gamma$ is decreasing when $\Gamma>\Gamma_{\mathrm{min}}$,
the denominator $a^{6}\Gamma^{2}$ initially decreases only like $a^{-4}$:
i.e. more mildly, giving intuitively a better chance for a blue spectrum
to be generated for a larger number of efolds. Furthermore, this means
that $\Gamma^{4}$ is also decreasing as $a^{-4}$, making the relative
contribution of the angular momentum not diminish with the increasing
scale factor. Indeed, this causes the potential to scale as the inverse
mass dimension of the potential, hinting that this is a conformal
limit. As will be explained in this paper, this is a time-independent
conformal limit of a special type (different from a massless scalar
field in Minkowski space or a massless equilibrium axion in dS space),
and this will be utilized to generate a blue isocurvature spectrum
for the axion field which is approximately $\delta\chi$: i.e. $\Delta_{s}^{2}\propto k^{2}$
corresponding to a spectral index of $n_{I}=3$.

In addition to constructing a novel model of generating a blue spectrum,
we also systematically quantize the fields in this background-out-of-equilibrium
situation which can be characterized by the novel nonvanishing of
the commutator $\left[\partial_{\eta}\delta\chi,\partial_{\eta}\delta\Gamma\right]$
representing velocity correlation even in the absence of non-derivative
correlations. Although the work of \citep{Creminelli:2023kze,Hui:2023pxc}
quantizes a similar theory\footnote{We do not use any methods of their quantization because their work
appeared after we had finished that part of our paper.} and agrees with our results, we present some distinct and unique
details here regarding the axion spectrum (particularly regarding
conformal symmetry representation) and apply it to isocurvature and
dark matter phenomenology. One revelation is that the radial perturbations
about the conformal background solution mix with the angular perturbations
for any eigenstate of the Hamiltonian even at the quadratic fluctuation
level, and the different energy eigenvectors (each eigenvector representing
the mixing) are not orthogonal. Indeed, it will be shown that a massive
$\delta\Gamma$ that kinetically mixes with $\delta\chi$ has nearly
identical conformal representation as $\delta\chi$. A more important
revelation is that, despite the complicated quantum mode mixing arising
from the time-dependent background, explicit quantization allows one
to construct a time-independent Hamiltonian whose ground state well-represents
the vacuum. Because of this and angular field translational symmetry,
Goldstone theorem still applies during the conformal period, and the
dispersion relationship is approximately linear in $k$ as $k\rightarrow0$
but with a different sound speed coefficient of $1/\sqrt{3}$, similar
to a relativistic perfect fluid pressure wave. Indeed, it is well
known that a quartic complex scalar with spontaneous $U(1)$ breaking
is a simple model of a superfluid (see e.g.~\citep{Leggett:1999zz}).

As far as model parameters are concerned, there are the initial conditions
of the background fields, the quartic coupling, and the usual axion
parameters which control the dark matter abundances. The main theoretical
limitation on extending this blue spectrum over a large $k$ range
is the requirement that the axion remains a spectator, which limits
the coupling and the background field initial displacement value in
the conformal regime. We also identify a range of initial condition
deformations away from the conformal limit over which the isocurvature
spectrum is approximate $k^{2}$, beyond which parametric resonance
sets in and destroys the smooth blue spectrum. We identify the parameter
regime in which this type of model can reproduce a spectrum of blue-tilt
followed by a plateau.

The order of presentation will be as follows. In Sec.~\ref{sec:Spectator-Definition-and-conformal-limit},
we define the notation for the ``vanilla'' axion model and make
general arguments of how a time-independent conformal limit and the
spectral index $n_{I}=3$ arises with the combination of large field
displacements and angular momentum. In Sec.~\ref{sec:Explicit-quantization-in},
we quantize the theory explicitly about the large phase angular momentum
to make the vacuum choice precise and to compute the resulting normalization
for the desired correlation function. We also give a simplified discussion
of how the \emph{intermediate-time} transition away from the time-independent
conformal-era will not result in a large bump in the isocurvature
spectrum. In Sec.~\ref{sec:Deformations-away-from}, we discuss how
deformations of the \emph{initial conditions} away from the time-independent
conformal limit will modify the spectrum. This will lead to oscillatory
features in the spectrum. In Sec.~\ref{sec:Discussion}, we present
example isocurvature spectra plots and the parametric ranges over
which the QCD axion phenomenology is compatible with observations.
We then conclude with a summary. Many appendices follow that provide
details of the results presented in the main body of the work. For
example, the details of the conformal field representation will be
given in Appendix \ref{sec:Conformal-limit-for} and the details of
the quantization is presented in Appendix \ref{sec:Details-of-quantization}.

\section{\label{sec:Spectator-Definition-and-conformal-limit}Spectator Definition
and Basics of the Conformal Limit}

In this section, we introduce the Lagrangian for our spectator field
in terms of a complex scalar field $\Phi$ with an underlying global
$U(1)$ PQ symmetry and lay out the basic physics central to the computation
before delving into detailed computations in the subsequent sections.

\subsection{Basic Action\label{subsec:Basic-Action}}

Consider the following action for a spectator complex scalar field
$\Phi$ containing the axion in a 4-dimensional FLRW spacetime
\begin{equation}
S=\int dtd^{3}x\sqrt{-g}\left(-\partial_{\mu}\Phi^{*}\partial^{\mu}\Phi-V\right)
\end{equation}
where the potential is composed of the usual renormalizable terms
symmetric under a global $U(1)$
\begin{equation}
V=-2M^{2}\Phi^{*}\Phi+\lambda\left(\Phi^{*}\Phi\right)^{2}
\end{equation}
with a dimensionless self-coupling constant $\lambda$ and a dimension-one
mass parameter $M$. We will assume that the background metric $ds^{2}=g_{\mu\nu}^{(0)}dx^{\mu}dx^{\nu}=-dt^{2}+a^{2}(t)|d\vec{x}|^{2}$
is driven by an inflaton whose energy dominates over the energy of
$\Phi.$ As is well known (see e.g.~\citep{Chung:2015pga}), the
non-adiabatic quantum fluctuations of $\Phi$ are diffeomorphism gauge
invariant at the linear level and govern the spectator isocurvature
perturbations that add to the usual curvature perturbations of the
inflaton.

To make the $U(1)$ angular physics manifest, parameterize $\Phi$
as usual in terms of a radial field $\Gamma$ and an axial field $\Sigma$:
\begin{equation}
\Phi=\frac{1}{\sqrt{2}}\Gamma e^{i\frac{\Sigma}{\Gamma}}
\end{equation}
where $\Gamma$ and $\Sigma$ are real scalar fields. The potential
$V$ in terms of the real field $\Gamma$ is
\begin{equation}
V=-M^{2}\Gamma^{2}+\frac{\lambda}{4}\Gamma^{4}
\end{equation}
with the stable vacuum at 
\begin{equation}
\Gamma_{{\rm vac}}=\sqrt{\frac{2M^{2}}{\lambda}}.
\end{equation}
The kinetic terms of the Lagrangian in terms of fields $\Gamma$ and
$\Sigma$ are similarly rewritten as
\begin{equation}
-\partial_{\mu}\Phi^{*}\partial^{\mu}\Phi=-\frac{1}{2}\left(\partial_{\mu}\Gamma\partial^{\mu}\Gamma-2\frac{\Sigma}{\Gamma}\partial_{\mu}\Sigma\partial^{\mu}\Gamma+\left(\frac{\Sigma}{\Gamma}\right)^{2}\partial_{\mu}\Gamma\partial^{\mu}\Gamma+\partial_{\mu}\Sigma\partial^{\mu}\Sigma\right)\label{eq:coupling}
\end{equation}
where the $\Sigma\partial_{\mu}\Sigma\partial^{\mu}\Gamma$ coupling
will later play a nontrivial role for the perturbations. We now define
a dimensionless angular variable 
\begin{equation}
\theta\equiv\frac{\Sigma}{\Gamma}\label{eq:thetadef}
\end{equation}
such that the action in terms of $\Gamma$ and $\theta$ is 
\begin{equation}
S=\int d^{4}x\sqrt{-g^{(0)}}\left(-\frac{1}{2}g^{(0)\mu\nu}\left(\partial_{\mu}\Gamma\partial_{\nu}\Gamma+\Gamma^{2}\partial_{\mu}\theta\partial_{\nu}\theta\right)-\left(-M^{2}\Gamma^{2}+\frac{\lambda}{4}\Gamma^{4}\right)\right).\label{eq:action}
\end{equation}
where we note that the kinetic terms for the fields $\Gamma$ and
$\Gamma_{\mathrm{vac}}\theta$ appear canonically normalized. This
system has a conserved background angular momentum
\begin{equation}
L\equiv a^{2}\Gamma_{0}^{2}\partial_{\eta}\theta_{0}\label{eq:Ldef}
\end{equation}
owing to the $U(1)_{{\rm PQ}}$ symmetry where the subscript $0$
indicates homogeneous background components of the fields and $\eta$
is the conformal time variable defined as 
\begin{equation}
\eta=\frac{-1}{aH}.
\end{equation}
In principle, this large angular momentum may be generated by a CP
violating non-renormalizable term as in the usual Affleck-Dine mechanism.
We define any canonically normalized scalar field $\Upsilon$ during
inflation to be a spectator if 
\begin{equation}
\rho_{\Upsilon}\ll\text{\ensuremath{\rho_{{\rm inflaton}}}}\label{eq:spectator_cond}
\end{equation}
where $\rho_{\Upsilon}$ represents the energy density of a field
$\Upsilon$. For an initial displacement of the radial field $\Gamma$
away from its vacuum state $\Gamma_{{\rm vac}}$, Eq.~(\ref{eq:spectator_cond})
translates to

\begin{equation}
\frac{1}{2}\left(\dot{\Gamma_{0}}^{2}+\Gamma_{0}^{2}\dot{\theta_{0}}^{2}\right)+\frac{\lambda}{4}\Gamma_{0}^{4}\ll3M_{P}^{2}H^{2}\label{eq:maxradial_bound}
\end{equation}
where $H=\dot{a}(t)/a(t)$ is the Hubble expansion rate. We will refer
to this condition later when we define our spectator dynamics under
different initial conditions. This will be one of the dominant constraints
on the initial radial displacement of the system.

\subsection{\label{subsec:How-conformal-limit}How conformal limit generates
a blue spectrum}

In this section, we will explain how a conformal limit spontaneously
broken by a $U(1)$ time-translation locking can generate a blue spectral
index of $3$ for the axion. The details of this section are given
in Appendix \ref{sec:Conformal-limit-for}. One nontrivial aspect
that will be explained below is how the angular time-dependence leads
to a novel conformal phase that is distinct from the massless conformal
phase of Minkowski spacetime.

Consider the $\Phi$ action Eq.~(\ref{eq:action}) in the conformal
coordinates defined by the background metric $ds^{2}=a^{2}(\eta)\left(-d\eta^{2}+|d\vec{x}|^{2}\right)$:
\begin{align}
S & =\int d\eta d^{3}x\left(\frac{-1}{2}\eta^{\mu\nu}\left(\partial_{\mu}Y\partial_{\nu}Y+Y^{2}\partial_{\mu}\theta\partial_{\nu}\theta\right)-\left[-\frac{1}{2}\frac{a''}{a}Y^{2}-M^{2}a^{2}Y^{2}+\lambda\frac{Y^{4}}{4}\right]\right).\label{eq:action1}
\end{align}
where $\eta_{\mu\nu}$ is the Minkowski metric, $\theta$ is given
by Eq.~(\ref{eq:thetadef}), and $Y\equiv a\Gamma$. Note that only
the $M^{2}a^{2}Y^{2}$ term breaks the scaling symmetry
\begin{equation}
a\rightarrow u^{-1}a\label{eq:scaling-symmetry}
\end{equation}
where $u$ is a constant while the time-dependent term $(a''/a)Y^{2}$
term does not. On the other hand $a'/a$ is time-dependent. The action
of Eq.~(\ref{eq:action1}) therefore does not know about constant
time hypersurface proper length scales or time-translation noninvariance
when both $M^{2}a^{2}Y^{2}$ and $(a''/a)Y^{2}$ can be neglected.
If we consider only the classical homogeneous background equation
of $Y(x)\approx Y_{0}(\eta)$, as shown in the Appendix \ref{sec:Conformal-limit-for},
we can go to a classical background solution of
\begin{equation}
Y_{0}=Y_{c}=\mathrm{const}\label{eq:constantYconf}
\end{equation}
\begin{equation}
\partial_{\eta}\theta_{0}=\mathrm{const}\label{eq:conflim}
\end{equation}
in the limit
\begin{equation}
\sqrt{\lambda}Y_{0}\gg Ma,\sqrt{a''/a}.\label{eq:thirdcond}
\end{equation}
Hence, dynamically, we achieve the limit of Eq.~(\ref{eq:scaling-symmetry})
and the action written in terms of $Y$ and $\theta$ (when considering
quantum fluctuations about the classical solution) does not know about
spatial proper length scales or time translation symmetry violations.
This is intuitive since when $\sqrt{\lambda}Y_{0}\gg Ma$, the conformal
factor $a(\eta)$ scaling by a constant is a classical invariance.
Furthermore, even though $a''/a$ is time-dependent (despite it being
conformally invariant) in quasi-dS spacetime, large $\sqrt{\lambda}Y_{0}$
limit allows one to neglect this term to give a static system. A key
defining characteristic of this scenario is that $\dot{\theta}_{0}\neq0$
gives a tachyonic mass contribution $-\eta^{\mu\nu}Y^{2}\partial_{\mu}\theta\partial_{\nu}\theta/2$
which is important for achieving $\sqrt{\lambda}Y_{0}\gg Ma,\sqrt{a''/a}$
at the minimum of $Y_{0}$ effective potential. This in turn leads
to a new perturbation mixing term
\begin{equation}
-\eta^{\mu\nu}Y^{2}\partial_{\mu}\theta\partial_{\nu}\theta/2\ni Y_{0}\delta Y\partial_{\eta}\theta_{0}\partial_{0}\delta\theta\label{eq:mixing}
\end{equation}
that changes the dispersion relationship. This is the reason why rotation
is important for this scenario and leads to an interesting tree-level
conformally invariant theory which is the subject of this paper. It
is also important to note that once one expands about the background
of Eqs.~(\ref{eq:constantYconf}) and (\ref{eq:conflim}), there
is a scale $\partial_{\eta}\theta_{0}$ in the theory, but because
it arises from spontaneous symmetry breaking, it transforms under
diffeomorphism that eventually will make the conformal representation
similar to that of a massive scalar field theory with the mass parameter
behaving as a spurion (see Appendix \ref{sec:Conformal-limit-for}
for more details). Moreover, owing to the $U(1)$ symmetry, the spontaneous
conformal symmetry breaking term $\partial_{\eta}\theta_{0}$ is a
constant in the conformal time coordinates (as indicated by Eq.~(\ref{eq:conflim})).\footnote{Also, it is easy to check that there is no Q-ball formation in the
current scenario.}

Now, let's consider the axion sector with a rescaling of Eq.~(\ref{eq:thetadef})
as
\begin{equation}
\theta=\frac{a\Sigma}{Y}\equiv\frac{\mathcal{A}}{Y}.\label{eq:calAdef}
\end{equation}
 The action will be of the form 
\begin{equation}
S\ni\int d\eta d^{3}x\left(\frac{-1}{2}\eta^{\mu\nu}\partial_{\mu}\mathcal{\delta A}\partial_{\nu}\delta\mathcal{A}+U(1)\mbox{ \textbf{and} time invariant mixing of }\delta Y\mbox{ and }\mathcal{\delta A}\right)\label{eq:scaleinvariant}
\end{equation}
where $\delta\mathcal{A}$ are the scaled axion fluctuations about
the constant $\partial_{\eta}\theta_{0}$ background solution that
pairs with Eq.~(\ref{eq:constantYconf}). The mixing of $\delta Y$
and $\delta\mathcal{A}$ coming from Eq.~(\ref{eq:mixing}) is the
main difference between the Minkowski spacetime's conformal massless
field and the axion here. As will be shown in Appendix \ref{sec:Conformal-limit-for},
the scaling symmetry of Eq.~(\ref{eq:scaling-symmetry}), $SO(3)$
symmetry, spatial translation invariance, and time-translational symmetry
together with PQ $U(1)$ symmetry tells us
\begin{equation}
\langle\delta\mathcal{A}(\eta,\vec{x})\mathcal{\delta A}(\eta,0)\rangle\underset{|\vec{x}|\rightarrow\infty}{\sim}\frac{c_{A}}{|\vec{x}|^{2}}\label{eq:staticscaleinv}
\end{equation}
for a constant $c_{A}$ or equivalently
\begin{equation}
\boxed{\langle\delta\left(\Gamma\theta\right)(\eta,\vec{x})\delta\left(\Gamma\theta\right)(\eta,0)\rangle\underset{|\vec{x}|\rightarrow\infty}{\sim}\frac{c_{A}}{a^{2}(\eta)|\vec{x}|^{2}}}\label{eq:scale-invariant-theory}
\end{equation}
is the physical axion correlator.

Remarkably, despite the fact that $|\vec{x}|a\gg H^{-1},$ the fluctuations
$\delta\mathcal{A}$ do not sense the spacetime curvature. In contrast,
a generic minimally coupled massless real scalar field $\varphi_{s}$
has a kinetic term for a rescaled $\Phi_{s}\equiv a\varphi_{s}$ of
\begin{equation}
S_{\varphi}=\int d\eta d^{3}x\left(\frac{1}{2}\left[\left(\partial_{\eta}\Phi_{s}\right)^{2}-\left(\frac{a''(\eta)}{a(\eta)}\right)\Phi_{s}^{2}-\sum_{i}\left(\partial_{i}\Phi_{s}\right)^{2}\right]\right)\label{eq:contrastscala}
\end{equation}
which contains time-dependence through $\left(a''/a\right)$: i.e.
this theory is $a\rightarrow au$ invariant but it is not time-translation
invariant. In such situations, one has
\begin{equation}
\langle\Phi_{s}(\eta,\vec{x})\Phi_{s}(\eta,0)\rangle\underset{|\vec{x}|\rightarrow\infty}{\sim}F\left(\frac{a''}{a},|\vec{x}|\right)
\end{equation}
where $F$ is a functional of $(a''/a)$ and a function of $|\vec{x}|$.
Since the spatial derivatives become unimportant for Eq.~(\ref{eq:contrastscala})
for long wavelength modes, the dependence of $F$ on $\left(a''/a\right)$
becomes important in this $|\vec{x}|\rightarrow\infty$ limit.\footnote{This infinity here literally means $|\vec{x}|a\gg H^{-1}$.}
The absence of this analogous time-translation invariance breaking
term for $\mathcal{\delta A}$ in Eq.~(\ref{eq:scaleinvariant})
is partly due to the axionic nature of $\mathcal{A}$ in addition
to being in the conformal radial sector discussed previously. This
is a time-independent conformal phase of the axionic theory.

The Fourier-space isocurvature spectrum corresponding to Eq.~(\ref{eq:scale-invariant-theory})
is 
\begin{equation}
\Delta_{s}^{2}\propto k^{2}\label{eq:conformalresult}
\end{equation}
which is conventionally described as having a spectral index of $n_{I}=3$.
In other words, the large $\Gamma$ limit and a conformally compatible
boundary conditions for the background field $\Gamma_{0}$ allowed
the scaled axion field $\mathcal{A}$ to settle into a tree-level
conformal theory that does not see the expanding universe. Explicit
mode computations shown in the subsequent sections will support this
general expectation based on conformality arguments. We should also
note that Eq.~(\ref{eq:action1}) indicates that the field $\delta Y=a\delta\Gamma$
is expected to behave as a massive field in the long wavelength limit
owing to the mass scale provided by $Y^{2}\left(\partial_{\eta}\theta_{0}\right)^{2}$
with a large $\partial_{\eta}\theta_{0}$ supporting a large Eq.~(\ref{eq:constantYconf}).
This implies that $\delta Y$ two-point function in the long wavelength
limit will behave as the massive correlator in flat space giving $\left\langle \delta Y\delta Y\right\rangle \propto k^{3}$
which implies 
\begin{equation}
\left\langle \delta\Gamma\delta\Gamma\right\rangle =C_{Y}k^{3}/a^{2}/(\partial_{\eta}\theta_{0})\label{eq:delgamcorr}
\end{equation}
(where $C_{Y}$ is a constant) during the time-independent conformal
era when $\partial_{\eta}\theta_{0}=\sqrt{\lambda}\Gamma_{0}(\eta_{i})a(\eta_{i})$
(see Appendix~\ref{sec:Conformal-limit-for} that explains the appearance
of $\partial_{\eta}\theta_{0}$ from a conformal representation perspective).
As explained in the Appendix~\ref{sec:Conformal-limit-for}, we cannot
read off $k^{2}$ behavior of Eq.~(\ref{eq:conformalresult}) from
conformal invariance alone because of the spontaneous breaking scale
$\partial_{\eta}\theta_{0}$: it is a result of knowing \emph{time-independent}
conformal invariance and masslessness of the $\mathcal{\delta A}$
field (the latter coming from the Goldstone property of the spontaneously
broken $U(1)_{PQ}$ symmetry).

The mixing of $\delta Y$ with $\delta\mathcal{A}$ through the $\eta^{\mu\nu}Y^{2}\partial_{\mu}\theta\partial_{\nu}\theta$
term after the spontaneous symmetry breaking term $\partial_{\eta}\theta_{0}$
is turned on in Eq.~(\ref{eq:action1}) leads to an interesting dispersion
relationship. Instead of the dispersion relationship of a free massless
theory, it will be that of a relativistic perfect fluid acoustic wave:
i.e.
\begin{equation}
\frac{d\omega}{dk}=\frac{1}{\sqrt{3}}\label{eq:newsoundspeed}
\end{equation}
where $\omega$ is the frequency associated with the lighter eigenmode.
This is an indication that the axion here is a perturbation about
a nontrivially interacting background medium. One obvious consequence
of this is that the perturbations freeze out a bit earlier when $k\approx a(t_{k})\sqrt{3}H$
during inflation in contrast with the situation when $\sqrt{3}\rightarrow1$.
The fact that the dispersion relationship here is linear in $k$ is
just as for a Nambu-Goldstone boson: the shift symmetry is still intact
even though there is a nontrivial mixing. In field theoretic situations
where the system acts approximately as an isotropic, adiabatic fluid,
one expects the trace of the energy momentum tensor to vanish if the
system is conformal $P\approx\rho/3$ which implies that the sound
speed is as given by Eq.~(\ref{eq:newsoundspeed}).\footnote{This follows from the conservation of dilatation current $j_{\mu}=T_{\mu\nu}x^{\nu}$
if the dilatation is assumed to arise from a recoordinatization. More
about the relationship between the diffeomorphism representation and
the spurion representation is explained in Appendix.~\ref{sec:Conformal-limit-for}.}

Before moving on to the details, we should also remark about what
the usual axion conformal phase is after the initial time-independent
conformal phase ends and the $\Gamma$ has settled to its minimum
leading to the ordinary axion quantum fluctuation physics. The theory
in that case is that of a spacetime-curvature induced massive scalar
field
\begin{equation}
S\approx\int d^{4}x\frac{1}{2}\left\{ -\eta^{\mu\nu}\partial_{\mu}\left(a\Sigma\right)\partial_{\nu}(a\Sigma)+\left(\frac{a''}{a}\right)\left(a\Sigma\right)^{2}\right\} \label{eq:axion}
\end{equation}
which does have manifest conformal invariance of Eq.~(\ref{eq:scaling-symmetry}),
but not time translation invariance nor masslessness since during
inflation $\left(a''/a\right)=2/\eta^{2}$ which leads to a well-known
tachyonic mass for $\mathcal{A}=a\Sigma$. Hence, we see that the
theory which we are analyzing in detail in this paper is a theory
that goes from a time-independent spontaneously broken conformal phase
to a time-dependent conformal phase, latter of which is the usual
axionic isocurvature quantum fluctuation theory during inflation.

\section{\label{sec:Explicit-quantization-in}Explicit quantization in the
conformal limit}

Although we have given a conformal limit argument in Sec.~\ref{subsec:How-conformal-limit}
for the spectral index $n_{I}=3$, we have not justified the selection
of the vacuum state in the situation in which the background field
$\partial_{\eta}\theta_{0}$ is large. Also, given the fast rotation
which kinetically mixes the radial mode with the angular mode, we
expect the dispersion relationship to change from the standard one
leading to order unity changes in the power spectrum. To address these
issues, we quantize the theory in the conformal limit explicitly.\footnote{After the quantization part of this work was completed, the work \citep{Hui:2023pxc,Creminelli:2023kze}
appeared which quantizes a similar theory and agrees with the results
here. One main difference is that we present more details here regarding
the axion spectrum and apply it to isocurvature and dark matter phenomenology.}

\subsection{Conformal limit power quantization and power spectra}

As shown in the Appendix \ref{sec:Details-of-quantization}, we can
quantize the two real scalar degrees of freedom 
\begin{equation}
\delta\psi^{n}=\left(\delta Y,\delta X\right)^{n}\equiv\left(a\delta\Gamma,a\delta\chi\right)^{n}\equiv\left(a\delta\Gamma,a\Gamma_{0}\delta\theta\right)^{n}
\end{equation}
governed by the quadratically expanded action
\begin{align}
S_{2} & =\int d\eta d^{3}x\left\{ -\frac{1}{2}\eta_{\mu\nu}\partial^{\mu}\delta Y\partial^{\nu}\delta Y-\frac{1}{2}\eta_{\mu\nu}\partial^{\mu}\delta X\partial^{\nu}\delta X\right.\nonumber \\
 & -2\delta Y\eta_{\mu\nu}\partial^{\mu}\delta X\partial^{\nu}\theta_{0}+\frac{2\delta X\delta Y}{Y_{0}}\eta_{\mu\nu}\partial^{\mu}Y_{0}\partial^{\nu}\theta_{0}+\frac{\delta X}{Y_{0}}\eta_{\mu\nu}\partial^{\mu}\delta X\partial^{\nu}Y_{0}+\frac{\delta Y\eta_{\mu\nu}\partial^{\mu}\delta Y\partial^{\nu}a}{a}\nonumber \\
 & +\frac{1}{2}\left(\delta X\right)^{2}\eta_{\mu\nu}\left(\frac{\partial^{\mu}Y_{0}\partial^{\nu}Y_{0}}{Y_{0}^{2}}+2\frac{\partial^{\mu}a\partial^{\nu}a}{a^{2}}-2\frac{\partial^{\mu}a\partial^{\nu}Y_{0}}{aY_{0}}\right)\nonumber \\
 & \left.-\frac{1}{2}\left(\delta Y\right)^{2}\eta_{\mu\nu}\left(\partial^{\mu}\theta_{0}\partial^{\nu}\theta_{0}+\frac{\partial^{\mu}a\partial^{\nu}a}{a^{2}}\right)-\left(-\frac{2M^{2}a^{2}}{2}+\frac{3\lambda}{2}Y_{0}^{2}\right)\left(\delta Y\right)^{2}\right\} .\label{eq:Sreduced}
\end{align}
in the coordinates $ds^{2}=a^{2}(\eta)\left(-d\eta^{2}+|d\vec{x}|^{2}\right)$
using
\begin{align}
\left[\delta\psi^{n}(\eta,\vec{x}),\delta\psi^{m}(\eta,\vec{x})\right] & =0,\\
\left[\pi^{n}(\eta,\vec{x}),\pi^{m}(\eta,\vec{x})\right] & =0,\label{eq:picommutator}\\
\left[\delta\psi^{n}(\eta,\vec{x}),\pi^{m}(\eta,\vec{x})\right] & =i\delta^{nm}\delta^{(3)}(\vec{x}-\vec{y})
\end{align}
as usual. What is special in the scenario considered in this paper
is that the coefficients involving $\{X_{0},Y_{0},a\}$ are generally
time-dependent, but in the conformal limit described by Eqs.~(\ref{eq:constantYconf}),
(\ref{eq:conflim}), and (\ref{eq:thirdcond}), the coefficients become
time-independent: e.g.
\begin{equation}
Y_{0}\equiv a\Gamma_{0}\approx Y_{c}=\frac{L^{1/3}}{\lambda^{1/6}}=\mathrm{constant}\label{eq:backgroundsol}
\end{equation}
which follows from the conditions given in Eqs.~(\ref{eq:Ldef})
and (\ref{eq:constant}). Here, $Y_{c}$ represents a constant conformal
background radial solution.

Since we are going to compute the quantum correlator to $0$th order
in $\lambda$ while Eq.~(\ref{eq:backgroundsol}) does not allow
us to set $\lambda=0$, an explanation of the expansion is in order.
Note that this conformal limit background is a solution to the classical
equation of motion
\begin{equation}
\frac{\delta S}{\delta\Phi^{*}}=0
\end{equation}
which corresponds to leading $\hbar\rightarrow0$ field path. Keeping
the nonlinear interactions for the classical equation means we are
treating $\lambda\left|\Phi\right|^{2}\Phi$ to be on equal footing
as $M^{2}\Phi$. On the other hand we are computing the quantum dynamics
with $\lambda\rightarrow0$ in considering the quadratic quantum fluctuations
for the quantum correlator. Hence, we are taking the limit 
\begin{equation}
O(\lambda|\Phi|^{2}a^{2})\sim O(\lambda Y_{0}^{2})\gtrsim O(M^{2}a^{2})
\end{equation}
in the quadratic computation. Eq.~(\ref{eq:backgroundsol}) then
says we are in the parametric region in which
\begin{equation}
\lambda^{2/3}L^{2/3}\gtrsim O(M^{2}a^{2})
\end{equation}
which will break down when $a^{-2}$ has sufficiently diluted $L^{2/3}$.

In this regime when $Y_{0}=Y_{c}$ and $\partial_{\eta}\theta_{0}$
are constants, the Hamiltonian density simplifies to 
\begin{equation}
\mathcal{H}=\frac{1}{2}\left(\partial_{\eta}\delta Y\right)^{2}+\frac{1}{2}\left(\partial_{\eta}\delta X\right)^{2}+\frac{1}{2}\left(\partial_{i}\delta\Gamma\right)^{2}+\frac{1}{2}\left(\partial_{i}\delta\chi\right)^{2}-\frac{1}{2}\left(\delta Y\right)^{2}\left(\partial_{\eta}\theta_{0}\right)^{2}+\left(\frac{3\lambda}{2}Y_{c}^{2}\right)\left(\delta Y\right)^{2}.
\end{equation}
The Fock state diagonalizing the Hamiltonian can be constructed using
the ladder operators as
\begin{equation}
\delta\psi^{n}=\int\frac{d^{3}p}{(2\pi)^{3/2}}\left[a_{\vec{p}}^{++}c_{++}V_{++}^{n}e^{-i\omega_{++}\eta}+a_{\vec{p}}^{+-}c_{+-}V_{+-}^{n}e^{-i\omega_{+-}\eta}+h.c.\right]e^{i\vec{p}\cdot\vec{x}}
\end{equation}
where
\begin{equation}
V_{++}^{n}=\left(\begin{array}{c}
1\\
\mathcal{R}_{++}
\end{array}\right),\,\,V_{+-}^{n}=\left(\begin{array}{c}
1\\
\mathcal{R}_{+-}
\end{array}\right),
\end{equation}
\begin{align}
\mathcal{R}_{++} & \equiv i\left(\frac{-2\left(\frac{L}{Y_{c}^{2}}\right)\omega_{++}}{\frac{1}{2}\left(\omega_{++}^{2}-\omega_{+-}^{2}\right)+\left(\lambda Y_{c}^{2}\right)+2\left(\frac{L}{Y_{c}^{2}}\right)^{2}}\right),\label{eq:A++}\\
\mathcal{R}_{+-} & \equiv i\left(\frac{2\left(\frac{L}{Y_{c}^{2}}\right)\omega_{+-}}{\frac{1}{2}\left(\omega_{++}^{2}-\omega_{+-}^{2}\right)-\left(\lambda Y_{c}^{2}\right)-2\left(\frac{L}{Y_{c}^{2}}\right)^{2}}\right),\label{eq:A+-}
\end{align}
\begin{equation}
\omega_{s_{1}s_{2}}\equiv s_{1}\sqrt{k^{2}+3\lambda Y_{c}^{2}+s_{2}Y_{c}\sqrt{\lambda\left(4k^{2}+9\lambda Y_{c}^{2}\right)}},\label{eq:freq}
\end{equation}
\begin{align}
c_{++}c_{++}^{*} & =-\frac{\left(1-\mathcal{R}_{+-}^{2}\right)\omega_{+-}-2i\partial_{\eta}\theta_{0}\mathcal{R}_{+-}}{2\left(\mathcal{R}_{+-}\omega_{+-}-\mathcal{R}_{++}\omega_{++}\right)\left(\mathcal{R}_{+-}\omega_{++}-\mathcal{R}_{++}\omega_{+-}\right)},\label{eq:cpp and cpm}\\
c_{+-}c_{+-}^{*} & =-\frac{\left(1-\mathcal{R}_{++}^{2}\right)\omega_{++}-2i\partial_{\eta}\theta_{0}\mathcal{R}_{++}}{2\left(\mathcal{R}_{+-}\omega_{+-}-\mathcal{R}_{++}\omega_{++}\right)\left(\mathcal{R}_{+-}\omega_{++}-\mathcal{R}_{++}\omega_{+-}\right)}.
\end{align}
Note that $V_{++}$ and $V_{+-}$ are not orthogonal. In the IR limit
$1\ll k^{2}\ll\lambda Y_{c}^{2}$, the two distinct frequency-squared
values are
\begin{equation}
\omega_{\pm-}^{2}\approx\frac{k^{2}}{3}+O\left(\frac{k^{4}}{\lambda Y_{c}^{2}}\right),
\end{equation}
and
\begin{equation}
\omega_{\pm+}^{2}\approx6\lambda Y_{c}^{2}+\frac{5k^{2}}{3}+O\left(\frac{k^{4}}{\lambda Y_{c}^{2}}\right)
\end{equation}
corresponding to low and high frequency solutions and are separated
by a large $O\left(\lambda Y_{c}^{2}/k^{2}\right)$ hierarchy. In
the UV limit,
\begin{equation}
\lim_{k\gg\lambda Y_{c}^{2}}\omega_{\pm\pm}^{2}\rightarrow k^{2}
\end{equation}
and the two frequency solutions become degenerate. When excited with
the lighter normal frequency $\omega_{\pm-}$ the fluctuations $\delta\Gamma,\delta\chi$
have a group velocity
\begin{equation}
\lim_{k\ll\sqrt{\lambda}Y_{c}}\frac{d\omega_{\pm-}}{dk}\approx\frac{1}{\sqrt{3}}\label{eq:group_vel}
\end{equation}
corresponding to a radiation fluid with sound speed squared $c_{s}^{2}\approx1/3$.
This is what we naively expect from the conformal limit discussed
in Sec.~\ref{subsec:How-conformal-limit} if the conformal limit
of this interacting system is behaving like a relativistic perfect
fluid which has an equation of state $P=\rho/3$ owing to the conformal
symmetry current conservation. As is well known, such a fluid has
a sound speed
\begin{equation}
c_{s}^{2}=\frac{\partial P}{\partial\rho}=\frac{1}{3}
\end{equation}
which implies that acoustic waves travel with speed $1/\sqrt{3}$
matching the group velocity Eq.~(\ref{eq:group_vel}). Another interesting
analogy comes from the tightly coupled cosmic microwave background
radiation to the electrons just before recombination. In that situation,
in the limit that the baryon loading vanishes, the speed of sound
is $1/\sqrt{3}$. Physically, the fast scattering of the electrons
is inducing photon pressure on the nonrelativistic electrons, setting
up an acoustic wave, similar to how a $\partial_{\eta}\theta_{0}$-induced
mixing is generating an axion pressure-supported acoustic wave in
the mixture of axions and heavy radial fields.

To choose the vacuum, we define it as usual as
\begin{equation}
a_{\vec{p}}^{+\pm}|0\rangle=0
\end{equation}
since the ladder operators diagonalize the Hamiltonian. The nonadiabaticity
at the end of the conformal period may lead to particle production
since the WKB vacuum after the conformal period will be different
from this vacuum. Any particles produced during such time periods
will be inflated away.\footnote{Any effects on nongaussianities from this will be left for a future
work.}

An interesting consequence of this quantized system is that the $\pi^{n}$
commutator equation of Eq.~(\ref{eq:picommutator}) which usually
is not very constraining induces a special constraint of 
\begin{equation}
\left[\partial_{\eta}\delta X,\partial_{\eta}\delta Y\right]=-2i\partial_{\eta}\theta_{0}\delta^{(3)}(\vec{x}-\vec{y})
\end{equation}
which makes the kinetic contributions to the isocurvature \textbf{cross}
correlators of radial and angular mode temporarily nonzero as long
as $\partial_{\eta}\theta_{0}$ is not negligible. This leads to $\left\langle \partial_{\eta}\delta\Gamma\partial_{\eta}\delta\chi\right\rangle \neq0$
even though $\left\langle \delta\Gamma\delta\chi\right\rangle =0$
during this time-independent conformal era when $\partial_{\eta}\theta_{0}$
is constant. Eventually, the time-independent conformal era ends when
the background radial modes reaches the minimum of the effective potential
making $\partial_{\eta}\theta_{0}\rightarrow0$ and in turn causing
this kinetic cross correlation to disappear as the system leaves the
time-independent conformal phase to enter the usual time-dependent
conformal phase of the stabilized axions.

\begin{figure}
\begin{centering}
\includegraphics[scale=0.6]{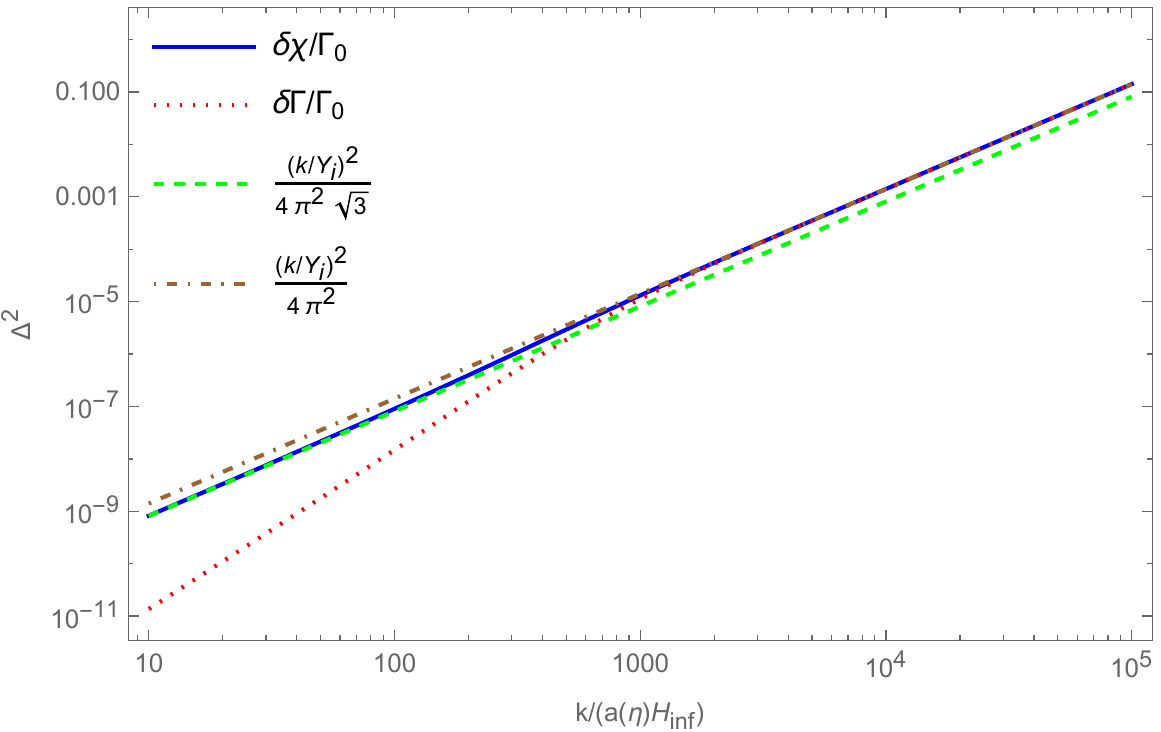}
\par\end{centering}
\caption{\label{fig:spectral-dependence}Plot illustrating the spectral dependence
of the correlation functions given in Eqs.~(\ref{eq:dtheta-spectrum-during CR})
and (\ref{eq:dgamma-spectrum-during CR}) corresponding respectively
to the angular (solid blue) and radial (dotted red) directions in
the time-independent conformal era. For modes $k\ll\partial_{\eta}\theta_{0}$,
the radial and angular isocurvature fluctuations exhibit spectral
dependencies of $k^{3}$and $k^{2}$ respectively. The green dashed
curve represents our analytic approximation taken from Eq.~(\ref{eq:dchi-dchi-spectrum}),
where $Y_{i}=\Gamma_{0}(\eta_{i})a(\eta_{i})$. Note the presence
of an additional normalizing factor of $1/\sqrt{3}$, resulting from
the angular modes behaving like a radiation fluid with a sound speed
$c_{s}^{2}=1/3$. Modes with $k\gg\partial_{\eta}\theta_{0}$ do not
see the effective potential and thus resemble massless modes. These
modes must be normalized with the usual Bunch-Davies (BD) vacuum state.
The brown dotdashed curve provides an analytic approximation for pure
massless modes normalized with the BD vacuum solution. The plot highlights
smooth transition from the vacuum state for the strongly coupled axion,
obtained from the minimization of the Hamiltonian density during the
time-independent conformal era, to the usual BD solution. Notably,
the spectral dependence of angular fluctuations $\propto k^{2}$ highlights
that in the rotating axion model, angular fluctuations maintain conformality
across all scales during $t<t_{{\rm tr}}$.}
\end{figure}

The correlation functions of radial and angular directions in the
time-independent conformal region (before the transition of the radial
field at time $t_{{\rm tr}}$ to $f_{{\rm PQ}}$) is
\begin{align}
\Delta_{\frac{\delta\Gamma}{\Gamma_{0}}\frac{\delta\Gamma}{\Gamma_{0}}}^{2}(\eta<\eta_{{\rm tr}}) & =\frac{1}{\Gamma_{0}^{2}(\eta)a^{2}(\eta)}\frac{k^{3}}{2\pi^{2}}\left(\left|c_{++}\right|^{2}+\left|c_{+-}\right|^{2}\right)\label{eq:dgamma-spectrum-during CR}
\end{align}
\begin{align}
\Delta_{\frac{\delta\chi}{\Gamma_{0}}\frac{\delta\chi}{\Gamma_{0}}}^{2}(\eta<\eta_{{\rm tr}}) & =\frac{1}{\Gamma_{0}^{2}(\eta)a^{2}(\eta)}\frac{k^{3}}{2\pi^{2}}\left(\left|c_{++}\mathcal{R}_{++}\right|^{2}+\left|c_{+-}\mathcal{R}_{+-}\right|^{2}\right)\label{eq:dtheta-spectrum-during CR}
\end{align}
where $c_{+\pm}$ and $\mathcal{R}_{+\pm}$ depend on $k$. In Fig.~\ref{fig:spectral-dependence},
we plot the correlation functions given in Eqs.~(\ref{eq:dgamma-spectrum-during CR})
and (\ref{eq:dchi-dchi-spectrum}) and compare them with the analytic
approximations. We illustrate that for modes $k\ll\partial_{\eta}\theta_{0}$,
the radial and angular isocurvature fluctuations during $t<t_{{\rm tr}}$
exhibit spectral dependencies of $k^{3}$and $k^{2}$ respectively.
 The complicated $k$-dependence simplifies to
\begin{align}
\lim_{k\ll\partial_{\eta}\theta_{0}}\Delta_{\frac{\delta\Gamma}{\Gamma_{0}}\frac{\delta\Gamma}{\Gamma_{0}}}^{2}(\eta<\eta_{{\rm tr}}) & \approx\left(\frac{1}{3^{1/2}2^{3/2}}\right)\frac{1}{\Gamma_{0}^{2}(\eta)a^{2}(\eta)}\frac{k^{3}}{2\pi^{2}\partial_{\eta}\theta_{0}},
\end{align}

\begin{align}
\lim_{k\ll\partial_{\eta}\theta_{0}}\Delta_{\frac{\delta\chi}{\Gamma_{0}}\frac{\delta\chi}{\Gamma_{0}}}^{2}(\eta<\eta_{{\rm tr}}) & \approx\frac{1}{3^{1/2}}\left(\frac{H/(2\pi)}{\Gamma_{0}(\eta_{i})}\right)^{2}\left(\frac{k}{a(\eta_{i})H}\right)^{2}\label{eq:dchi-dchi-spectrum}
\end{align}
on large length scales. Although the spectral indices here can be
inferred from the symmetry representations and minimal dynamical considerations
of Appendix \ref{sec:Conformal-limit-for}, the details of $1/\sqrt{3}$
and normalization factors appearing here are difficult to predict
without explicit quantization. Although one may naively think $\partial_{\eta}\theta_{0}$
here acts as a scale similar to the horizon scale in ordinary curvature
perturbations, making the spectral amplitude freeze out, the spectral
amplitude is actually always approximately frozen during the time-independent
conformal period. Since $\partial_{\eta}\theta_{0}\gg a(\eta_{i})H$
is typical for the parametric region of phenomenological interest,
we can have a frozen subhorizon $k^{2}$ spectrum for a massless field.
The $\delta\Gamma$ spectral index minus one is $3$ while the $\delta\chi$
spectral index is characterized by $n_{I}-1=2$ as anticipated. This
says that the $\delta\chi$ correlator dominates over the $\delta\Gamma$
correlator by a factor of $\partial_{\eta}\theta_{0}/k$. The fact
that $1/\sqrt{3}$ appears even for the massive $\delta\Gamma$ correlator
is indicative of the sound speed changing due to the presence of $\partial_{\eta}\theta$
induced mixing.

At approximately the time $t_{{\rm tr}}$, the time-independent conformal
regime in this strongly mixed model comes to an end, and the radial
field settles to the minimum of the potential at $f_{{\rm PQ}}$.
From the EoM provided in Eq.~(\ref{eq:axial}) for the axial perturbations
$\delta\chi$, we infer that around this time, the axial perturbations
transition to a massless axion state entering a time-dependent conformal
era. Because $\Gamma_{0}(\eta)$ tends to follow $\delta\chi$ mode
on superhorizon scales (see Appendix \ref{sec:Relationship-between-radial})
and because the radial kinetic energy is too small to generate nontrivial
resonances, there is no evolution of Eq.~(\ref{eq:dchi-dchi-spectrum})
after the transition to the vacuum at time $t=t_{\mathrm{tr}}$. Therefore,
the dimensionless power spectrum Eq.~(\ref{eq:dchi-dchi-spectrum})
can be used for $\eta>\eta_{\mathrm{tr}}$ as well. For modes that
exit the horizon a long time after the radial field has settled to
the minimum at $f_{{\rm PQ}}$, the spectrum is scale invariant. For
these modes the initial amplitude of the axial field fluctuations
is normalized with the usual BD vacuum state as $\sim1/\sqrt{2k}$.
Hence, we approximate the spectrum as 
\begin{equation}
\lim_{k\gg\partial_{\eta}\theta_{0}}\Delta_{\frac{\delta\chi}{\Gamma_{0}}\frac{\delta\chi}{\Gamma_{0}}}^{2}(\eta\rightarrow0)\approx\left(\frac{H/(2\pi)}{f_{PQ}}\right)^{2}\label{eq:massless_plateau}
\end{equation}
which is the same as the usual equilibrium spectrum. In matching Eqs.~(\ref{eq:dchi-dchi-spectrum})
and (\ref{eq:massless_plateau}), there is a sound-speed related factor
shift in where the blue tilt region will match the plateau, and this
is the hallmark of our current model flowing from one rotating phase
conformal field theory to the usual Goldstone case which from the
perspective Eq.~(\ref{eq:axion}) corresponds to a time-dependent
scenario.

Note that Bunch-Davies boundary condition is in the limit $k/(aH)\rightarrow\infty$.
If $\infty$ is interpreted modestly as $\partial_{\eta}\theta_{0}/k\gtrsim1$,
we see that the $\delta\Gamma$ correlation function dominates in
the UV. This indicates that the kinetic term of $\delta\Gamma$ is
important in the Bunch-Davies limit and $\delta\Gamma$ cannot be
integrated out in this limit. Indeed, one can explicitly compute
\begin{equation}
\Delta_{\partial_{\eta}\delta\Gamma\partial_{\eta}\delta\chi}^{2}(k,\eta<\eta_{{\rm tr}})=\frac{1}{a^{2}(\eta)}\frac{k^{3}}{2\pi^{2}}\left(i\partial_{\eta}\theta_{0}\right)
\end{equation}
which says that there is a strong mixing between $\delta\chi$ and
$\delta\Gamma$ in the modest kinematic range reasonable for standard
Bunch-Davies boundary conditions.\footnote{Because $\left\langle \partial_{\eta}\Gamma(\vec{x})\partial_{\eta}\chi(\vec{y})\right\rangle $
is not a Hermitian correlator, it does not have a direct measurability:
the measurable correlation $\left\langle \partial_{\eta}\Gamma(\vec{x})\partial_{\eta}\chi(\vec{y})+\partial_{\eta}\chi(\vec{y})\partial_{\eta}\Gamma(\vec{x})\right\rangle $
vanishes at least at this order in perturbation theory. Nonetheless,
the kinetic correlations will appear in quantum dynamics including
interactions. We will leave this topic to a future investigation.} Even more impressively, we know that \textbf{even} in the \emph{small}
$k$ limit, there is essentially no distinction between the $\delta\Gamma$
kinetic correlator and the $\delta\chi$ kinetic correlator
\begin{align}
\lim_{k\ll\partial_{\eta}\theta_{0}}\Delta_{\partial_{\eta}\delta\Gamma\partial_{\eta}\delta\Gamma}^{2}(k,\eta<\eta_{{\rm tr}}) & \approx\frac{1}{a^{2}(\eta)}\sqrt{\frac{3}{2}}\frac{k^{3}}{2\pi^{2}}\left(\partial_{\eta}\theta_{0}\right),\\
\lim_{k\ll\partial_{\eta}\theta_{0}}\Delta_{\partial_{\eta}\delta\chi\partial_{\eta}\delta\chi}^{2}(k,\eta<\eta_{{\rm tr}}) & \approx\frac{1}{a^{2}(\eta)}\sqrt{\frac{2}{3}}\frac{k^{3}}{2\pi^{2}}\left(\partial_{\eta}\theta_{0}\right),
\end{align}
in the time-independent conformal phase. This says you cannot really
integrate out $\delta\Gamma$ for any $k$ if you care about the kinetic
term.\footnote{Of course kinetic terms become more important for larger $k$ values.}
This implies that the only way to justify completely integrating out
the $\delta\Gamma$ mode in this cosmological context is not to use
the standard Bunch-Davies conditions, but require a non-standard restriction
of modes that satisfy
\begin{equation}
\frac{k}{\partial_{\eta}\theta_{0}}\ll1\label{eq:justifyintegratingout}
\end{equation}
and neglect kinetic aspects of inhomogeneity correlator physics.

\subsection{Post-time-independent-conformal-era time evolution}

After the time-independent conformal era ends, what happens to these
spectra? In this section, we will consider the time evolution of our
coupled system during inflation and determine the dimensionless power
spectrum for the axial and radial fields as $\eta\rightarrow0$. We
consider the equation of motion for the background field $\Gamma_{0}$
derived from Eq.~(\ref{eq:action}) and substitute $\partial_{\eta}\theta_{0}$
with the conserved angular momentum $L$ defined by Eq.~(\ref{eq:Ldef}):
\begin{equation}
\partial_{\eta}^{2}\Gamma_{0}+2\frac{\partial_{\eta}a}{a}\partial_{\eta}\Gamma_{0}+\left(-2M^{2}a^{2}+\lambda\Gamma_{0}^{2}a^{2}-\left(\frac{L}{a^{2}\Gamma_{0}^{2}}\right)^{2}\right)\Gamma_{0}=0.\label{eq:Gamma0-EoM}
\end{equation}
Note the in the limit $\eta\rightarrow0$, the radial field settles
to the minimum
\begin{equation}
\Gamma_{0,{\rm min}}=M\sqrt{\frac{2}{\lambda}}\equiv f_{{\rm PQ}}
\end{equation}
and the angular velocity decays as $H^{3}/a^{2}$. Using the full
solution for the background field $\Gamma_{0}$, we can obtain the
evolution of the linear perturbations $\delta\phi_{k}^{n}=\left(\delta\Gamma_{k},\delta\chi_{k}\right)^{n}$
for each mode $k$ by solving the mode function $h_{k}^{jm}(\eta)$
using the Eq.~(\ref{eq:orig_eom}) derived in Appendix \ref{sec:Details-of-quantization}:
\begin{equation}
\left[\delta^{nj}\partial_{\eta}^{2}+\kappa^{nj}\partial_{\eta}+\left(\mathcal{W}^{2}\right)^{nj}\right]h_{k}^{jm}(\eta)=0
\end{equation}
where $n$ is the flavor index and $m$ runs over distinct frequencies.
We set the initial conditions at $\eta_{i}$ for each of the two frequency
solutions as follows:
\begin{align}
\left(\begin{array}{c}
h_{k}^{1}(\eta)\\
h_{k}^{2}(\eta)
\end{array}\right)_{\eta=\eta_{i}}^{++} & =c_{++}\left(\begin{array}{c}
V_{++}^{1}(\eta_{i})\\
V_{++}^{2}(\eta_{i})
\end{array}\right)e^{-i\omega_{++}\eta_{i}},\qquad\left(\begin{array}{c}
\partial_{\eta}h_{k}^{1}(\eta)\\
\partial_{\eta}h_{k}^{2}(\eta)
\end{array}\right)_{\eta=\eta_{i}}^{++}=-i\omega_{++}\left(\begin{array}{c}
h_{k}^{1}(\eta)\\
h_{k}^{2}(\eta)
\end{array}\right)_{\eta=\eta_{i}}^{++}\label{eq:IC-1}
\end{align}
and 
\begin{align}
\left(\begin{array}{c}
h_{k}^{1}(\eta)\\
h_{k}^{2}(\eta)
\end{array}\right)_{\eta=\eta_{i}}^{+-} & =c_{+-}\left(\begin{array}{c}
V_{+-}^{1}(\eta_{i})\\
V_{+-}^{2}(\eta_{i})
\end{array}\right)e^{-i\omega_{+-}\eta_{i}},\qquad\left(\begin{array}{c}
\partial_{\eta}h_{k}^{1}(\eta)\\
\partial_{\eta}h_{k}^{2}(\eta)
\end{array}\right)_{\eta=\eta_{i}}^{+-}=-i\omega_{+-}\left(\begin{array}{c}
h_{k}^{1}(\eta)\\
h_{k}^{2}(\eta)
\end{array}\right)_{\eta=\eta_{i}}^{+-}.\label{eq:IC-2}
\end{align}
For modes $k\ll\partial_{\eta}\theta_{0}$, the above initial conditions
correspond to the mode amplitudes 
\begin{equation}
\left(\begin{array}{c}
h_{k}^{1}(\eta)\\
h_{k}^{2}(\eta)
\end{array}\right)_{\eta=\eta_{i}}^{++}\approx\frac{1}{6^{3/4}\sqrt{\partial_{\eta}\theta_{0}}}\left(\begin{array}{c}
\sqrt{3}\\
-i\sqrt{2}
\end{array}\right)
\end{equation}
and 
\begin{equation}
\left(\begin{array}{c}
h_{k}^{1}(\eta)\\
h_{k}^{2}(\eta)
\end{array}\right)_{\eta=\eta_{i}}^{+-}\approx\frac{1}{\sqrt{2}3^{3/4}}\left(\begin{array}{c}
\sqrt{k}/\partial_{\eta}\theta_{0}\\
i\sqrt{3}/\sqrt{k}
\end{array}\right)
\end{equation}
such that the canonical field amplitudes have the ratios
\begin{equation}
\left|\frac{\delta\Gamma_{k}(\eta_{i})}{\delta\chi_{k}(\eta_{i})}\right|^{++}\approx\sqrt{\frac{3}{2}},
\end{equation}
and 
\begin{equation}
\left|\frac{\delta\Gamma_{k}(\eta_{i})}{\delta\chi_{k}(\eta_{i})}\right|^{+-}\approx\frac{k}{\sqrt{3}\partial_{\eta}\theta_{0}}
\end{equation}
for the $++$ and $+-$ frequency solutions respectively. The fact
that the radial mode amplitude vanishes in the $k\rightarrow0$ limit
indicates that the smaller frequency mode is primarily made of the
angular mode at the initial time. Using these initial conditions for
the mode functions $h_{k}^{nr}$, we evolve the coupled system from
$\eta_{i}$ to a late time $\eta_{f}$ when the background radial
field $\Gamma_{0}$ is settled at $f_{{\rm PQ}}$ and all modes of
interest $k$ are super-horizon. The above results also imply that
the amplitude of the axial fluctuations for modes with $k\ll\partial_{\eta}\theta_{0}$
is dominated by the lighter frequency $\left(\omega_{+-}\right)$
solutions since
\begin{equation}
\lim_{k\ll\partial_{\eta}\theta_{0}}\frac{\left|\delta\chi_{k}(\eta_{i})\right|^{++}}{\left|\delta\chi_{k}(\eta_{i})\right|^{+-}}\approx O\left(\sqrt{\frac{k}{\partial_{\eta}\theta_{0}}}\right).
\end{equation}

\subsubsection{Adiabatic time-evolution example\label{subsec:Adiabatic-time-evolution-example}}

Let us consider an example where we initialize the background radial
field $\Gamma_{0}$ at $\eta_{i}$ with the time-independent conformal
solution 
\begin{equation}
a(\eta_{i})\Gamma_{0}(\eta_{i})=Y_{c}\equiv\frac{L^{1/3}}{\lambda^{1/6}}.
\end{equation}
Furthermore, in this example, we set $H=H_{\inf}$, $\lambda=1$ and
$f_{{\rm PQ}}=10H_{\inf}$ such that $M=f_{{\rm PQ}}/\sqrt{2}$. We
choose the conserved angular momentum $L=\sqrt{\lambda}10^{9}H_{\inf}^{3}a^{3}(\eta_{i})$,
hence $\Gamma_{0}(\eta_{i})=1000H_{\inf}$. Note that even though
the radial field has a large displacement away from the minimum along
a quartic potential, the effective radial mass is only order $H_{{\rm inf}}$
due to the effects of the angular momentum. This cancellation is nothing
more than the statement that stable orbits not passing through the
origin exist with angular momentum conservation, and if the space
does not expand, this orbit can persist indefinitely.

\begin{figure}
\begin{centering}
\includegraphics[scale=0.65]{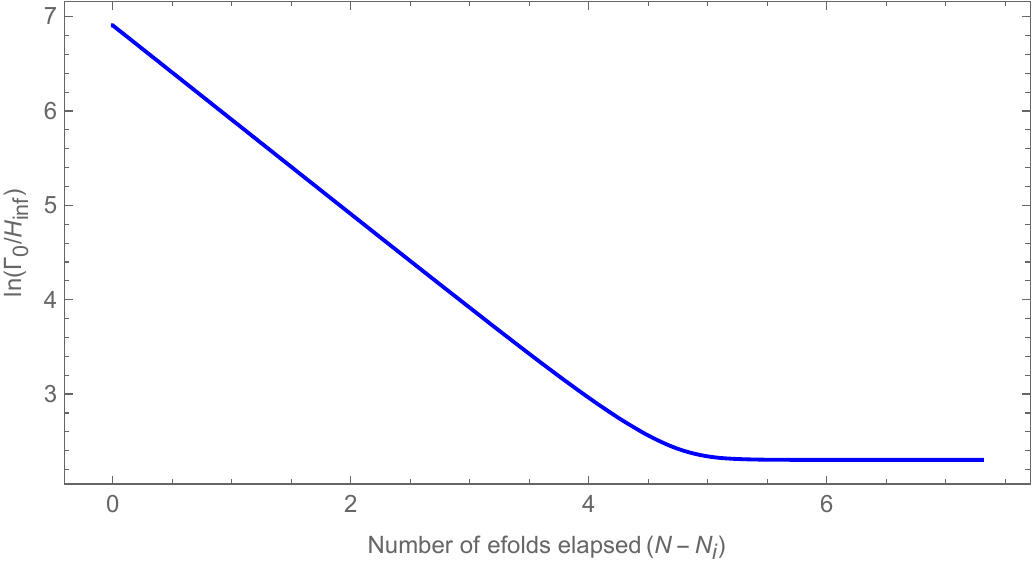}
\par\end{centering}
\caption{\label{fig:radial_time_evolution_ex1}Plot showing the time evolution
of the background radial field $\Gamma_{0}(t)$ during the quasi-dS
phase of inflation. Starting from $\Gamma_{0}(\eta_{i})=1000H_{\inf}$,
the field takes approximately $5$ efolds to settle to the minimum
$M\sqrt{2/\lambda}=10H_{\inf}$. The initial evolution of the field
$a(\eta)\Gamma(\eta)={\rm constant}.$}
\end{figure}
\begin{figure}
\begin{centering}
\includegraphics[scale=0.7]{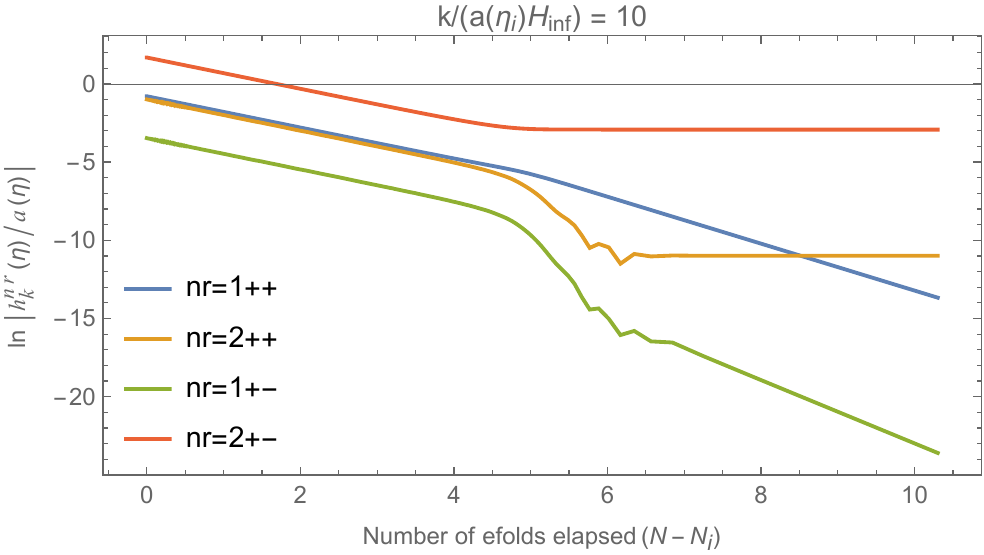}
\par\end{centering}
\caption{\label{fig:time_evolution_hknr}Plot showing the time evolution of
the mode functions $h_{k}^{nr}/a$ during the quasi de-Sitter phase
of inflation for a fiducial mode $k/a(\eta_{i})=10H_{{\rm inf}}$
where the background radial field moves along a trajectory as expected
in the time-independent conformal period as shown in Fig.~\ref{fig:radial_time_evolution_ex1}.
The mode amplitudes $h_{k}^{2+\pm}/a$ corresponding to the axial
field freeze out when the background radial field settles to the minimum
while the radial perturbations corresponding to $h_{k}^{1+\pm}$ persist
in their massive state, undergoing continued decay.}
\end{figure}

In Fig.~\ref{fig:radial_time_evolution_ex1}, we show the time evolution
of the background radial field. We observe that for our choice of
fiducial values, the radial field takes approximately 5 efolds to
settle to the minimum. This is close to the analytic estimate
\begin{equation}
\Delta N_{{\rm settle}}\approx\ln\left(\frac{Y_{c}(\eta_{i})/a(\eta_{i})}{f_{{\rm PQ}}\sqrt{1+H^{2}/M^{2}}}\right).\label{eq:N-settle}
\end{equation}
In Fig.~\ref{fig:time_evolution_hknr}, we show the time evolution
of the corresponding radial and axial mode functions for the two frequency
solutions $\omega_{+\pm}$ for a fiducial wavenumber $k/a(\eta_{i})=10$.
The mode amplitudes $h_{k}^{2r}$ corresponding to the axial field
$(n=2)$ freeze out when the background radial field settles to the
minimum at time $t_{tr}$. In contrast, the radial perturbations corresponding
to $h_{k}^{1r}$ persist in their massive state, undergoing continued
decay. For the modes $k<\partial_{\eta}\theta_{0}\equiv\sqrt{\lambda}Y_{c}$,
the mode function $h_{k}^{2r}$ corresponding to the lower frequency
solution $\omega_{r}=\omega_{+-}$ has the larger amplitude. This
is mainly due to the overall normalization factor of $1/\sqrt{\omega_{r}}$.
Hence, a lower frequency yields a comparatively larger mode amplitude.

\section{Deformations away from time-independent conformal limit\label{sec:Deformations-away-from}}

The previous section described a special initial condition leading
to a time-independent $\Gamma_{0}(\eta)a(\eta)$, leading to a time-independent
conformal theory. Such cases are the analogs of circular orbits in
mechanics. In this section, we describe deformations away from this
time-independent conformal limit such that $\Gamma_{0}(\eta)a(\eta)$
will have oscillations. These will imprint oscillations into the power
spectrum as we will show.

\subsection{Equations of motion\label{subsec:Equation-of-motions}}

Eq.~(\ref{eq:action}) implies the background equations of motion
(EoM) for the radial and angular degrees of freedom of
\begin{align}
\partial_{\eta}^{2}\Gamma_{0}+2\frac{\partial_{\eta}a}{a}\partial_{\eta}\Gamma_{0}+\left(-2M^{2}a^{2}+\lambda\Gamma_{0}^{2}a^{2}-\left(\partial_{\eta}\theta_{0}\right)^{2}\right)\Gamma_{0} & =0\label{eq:radial_eom}\\
\partial_{\eta}^{2}\theta+2\frac{\partial_{\eta}a}{a}\partial_{\eta}\theta_{0}+2\frac{\partial_{\eta}\Gamma_{0}}{\Gamma_{0}}\partial_{\eta}\theta_{0} & =0\label{eq:angular_eom}
\end{align}
where we have as usual assumed the background field to be spatially
independent like the background metric. We distinguish the spatially
inhomogeneous fluctuations of quantities $Q$ with $\delta Q.$ Eq.~(\ref{eq:angular_eom})
leads to the conserved angular momentum $L$ as defined in Eq.~(\ref{eq:Ldef}).
This can be interpreted as there being a comoving homogeneous $U(1)$
charge density $\Gamma_{0}^{2}\partial_{t}\theta_{0}$ that dilutes
as $a^{-3}$. We also note from Eq.~(\ref{eq:radial_eom}) that the
background radial field has a force from $\left(\partial_{\eta}\theta_{0}\right)^{2}\Gamma_{0}$
that can cancel a part of $\left(\lambda\Gamma_{0}^{3}-2M^{2}\Gamma_{0}\right)a^{2}$
depending on the size of the angular momentum $L$. This is the key
cancellation that allows the radial roll to be slow similar to the
flat direction situation of \citep{Kasuya:2009up} which is only lifted
by $O\left(H^{2}\right)$ mass terms. The linear order perturbation
variables $\delta\Gamma$ and $\delta\chi\equiv\Gamma_{0}\delta\theta$
in Fourier space satisfy the mode equations given in Eqs.~(\ref{eq:radial})
and (\ref{eq:axial}). Here, we note that one can easily identify
most of the axial fluctuation $\delta\Sigma$ with $\delta\chi$ since
\begin{equation}
\delta\Sigma=\Gamma\delta\theta+\theta\delta\Gamma
\end{equation}
and in the limit $\delta\Gamma/\Gamma\ll\delta\theta/\theta$, the
axial fluctuations are given as $\delta\Sigma\approx\delta\chi$.
Eqs.~(\ref{eq:radial}) and (\ref{eq:axial}) show that the fluctuations
in the radial and angular directions are coupled at linear order via
the dominant quadratic interaction term $\mathcal{L}_{{\rm int}}\supset-2\delta\Gamma a^{2}\eta_{\mu\nu}\partial^{\mu}\delta\chi\partial^{\nu}\theta_{0}$.
This spontaneous conformal symmetry breaking induced coupling is a
novel feature of field fluctuations about a rotating background.

In Sec.~\ref{sec:Explicit-quantization-in} we highlighted that the
coupled $\delta\Gamma-\delta\chi$ system can be diagonalized with
two sets of normal frequencies denoted as $\omega_{++}$ and $\omega_{+-}$.
In the IR limit corresponding to modes with $k^{2}\ll\lambda Y_{c}^{2}$,
the lowest frequency $\omega_{+-}$ has a dispersion relationship
that is linear in $k$ and the associated mode function resembles
a Goldstone mode. In this limiting case corresponding to a Goldstone
mode, it is possible to integrate out the radial mode $\delta\Gamma$
and obtain a decoupled EoM for the axial field fluctuations $\delta\chi$.
To this end, we rewrite the EoM for the scaled radial fluctuation
$\delta Y$ from Eq.~(\ref{eq:EoM-dYdX}) and neglect the kinetic
term $\partial_{\eta}^{2}\delta Y$: 
\begin{equation}
-\partial_{i}^{2}\delta Y-2\partial_{\eta}\theta_{0}\partial_{\eta}\delta X+\left(-2M^{2}a^{2}+3\lambda Y_{0}^{2}-\left(\partial_{\eta}\theta_{0}\right)^{2}-\frac{\partial_{\eta}^{2}a}{a}\right)\delta Y+2\partial_{\eta}\theta_{0}\left(\frac{\partial_{\eta}Y_{0}}{Y_{0}}\right)\delta X=0.
\end{equation}
As noted in the discussion around Eq.~(\ref{eq:justifyintegratingout}),
we can justify this step using the fact that we are not concerned
with kinetic correlators. Going to the Fourier space and evaluating
within the conformal regime, we obtain the expression
\begin{equation}
\left(k^{2}-2M^{2}a^{2}+3\lambda Y_{0}^{2}-\left(\partial_{\eta}\theta_{0}\right)^{2}-\frac{\partial_{\eta}^{2}a}{a}\right)\delta Y_{k}=2\partial_{\eta}\theta_{0}\partial_{\eta}\delta X_{k}.
\end{equation}
The above expression allows us to replace $\delta Y_{k}$ in the EoM
for the axial field. Thus, we arrive at the following decoupled EoM
for the scaled axial field $\delta X$: 
\begin{align}
\partial_{\eta}^{2}\delta X_{k}+2\partial_{\eta}\theta_{0}\,\partial_{\eta}\left(\frac{2\partial_{\eta}\theta_{0}\partial_{\eta}\delta X_{k}}{\left(k^{2}-2M^{2}a^{2}+3\lambda Y_{0}^{2}-\left(\partial_{\eta}\theta_{0}\right)^{2}-\frac{\partial_{\eta}^{2}a}{a}\right)}\right)\qquad\qquad\nonumber \\
+\left(k^{2}-2M^{2}a^{2}+\lambda Y_{0}^{2}-\left(\partial_{\eta}\theta_{0}\right)^{2}-\frac{\partial_{\eta}^{2}a}{a}\right)\delta X_{k} & =0
\end{align}
which can be rewritten as
\begin{align}
\partial_{\eta}^{2}\delta X_{k}\left(1+\frac{4\left(\partial_{\eta}\theta_{0}\right)^{2}}{\left(k^{2}-2M^{2}a^{2}+3\lambda Y_{0}^{2}-\left(\partial_{\eta}\theta_{0}\right)^{2}-\frac{\partial_{\eta}^{2}a}{a}\right)}\right) \qquad\qquad\nonumber \\
+\left(k^{2}-2M^{2}a^{2}+\lambda Y_{0}^{2}-\left(\partial_{\eta}\theta_{0}\right)^{2}-\frac{\partial_{\eta}^{2}a}{a}\right)\delta X_{k}\approx0\label{eq:decoupled-dX}
\end{align}
where the factor associated with the kinetic term $\partial_{\eta}^{2}\delta X_{k}$
evaluates to
\begin{equation}
\left(1+\frac{4\left(\partial_{\eta}\theta_{0}\right)^{2}}{\left(k^{2}-2M^{2}a^{2}+3\lambda Y_{0}^{2}-\left(\partial_{\eta}\theta_{0}\right)^{2}-\frac{\partial_{\eta}^{2}a}{a}\right)}\right)\approx3
\end{equation}
in the IR limit $\left(k\ll\sqrt{\lambda}Y_{c}\right)$ of the time-independent
conformal solution $\left(\sqrt{\lambda}Y_{c}=\partial_{\eta}\theta_{0}\right)$.
Therefore, in this limiting scenario, the strong coupling with the
radial mode only changes the overall normalization for the kinetic
term of $\delta\chi$.

In time coordinate $t$, the mass-squared term for the perturbations
$\delta\chi$ can be identified from the decoupled EoM as
\begin{equation}
m_{\delta\chi}^{2}=-\frac{\partial_{t}^{2}\Gamma_{0}}{\Gamma_{0}}-3\frac{\partial_{t}\Gamma_{0}}{\Gamma_{0}}=-2M^{2}+\lambda\Gamma_{0}^{2}-\frac{L^{2}}{a^{6}\Gamma^{4}}\label{eq:masssqform}
\end{equation}
which goes to zero when the radial field settles to its vacuum state,
say at time $t_{{\rm tr}},$ and the angular momentum term is negligible.
In this limit when $\delta\chi$ becomes massless, the quantum fluctuation
mode $\delta\chi$ does not decay any further, whereas modes decay
when $m_{\delta\chi}^{2}$ is non-negligible (even if the modes are
superhorizon). The isocurvature spectrum has a $k$ dependence that
is usually parameterized by the isocurvature spectral index $n_{I}$:
\begin{equation}
\Delta_{s}^{2}(k)\propto k^{n_{I}-1}.
\end{equation}
For a slowly varying mass of the linear spectator fluctuation $\delta\chi$,
the decoupled EoM in Eq.~(\ref{eq:decoupled-dX}) suggests that the
spectral index $n_{I}$ can be evaluated as 
\begin{equation}
n_{I}(k)-1=3-3\sqrt{1-\frac{4}{9}m_{\delta\chi}^{2}(k)}\label{eq:spectral_index}
\end{equation}
where $m_{\delta\chi}^{2}(k)$ is the effective mass-squared function
from Eq.~(\ref{eq:masssqform}) evaluated at a time $t_{k}$ when
$k/a(t_{k})\approx1$. Hence, $m_{\delta\chi}^{2}(k)$ must be at
least $O\left(H^{2}\right)$ for a blue isocurvature power spectrum,
which can be achieved early in the evolution of $\Gamma(t)$ due to
the cancellation between $\lambda\Gamma_{0}^{2}$ and $\dot{\theta}^{2}$.
For the time-independent conformal solution where $Y_{0}\equiv a\Gamma_{0}={\rm constant}$
until $t<t_{tr}$, we have 
\begin{equation}
-2M^{2}+\lambda\Gamma_{0}^{2}-\frac{L^{2}}{a^{6}\Gamma^{4}}=\frac{\partial_{\eta}^{2}a}{a^{3}}=2H^{2}\quad\forall t<t_{tr}
\end{equation}
which yields
\begin{equation}
m_{\delta\chi}^{2}(k<k_{tr})=2H^{2}
\end{equation}
and a spectral index
\begin{equation}
n_{I}(k<k_{tr})-1=2
\end{equation}
where the scale associated with the transition $t_{tr}$ is given
as
\begin{equation}
k_{tr}=\frac{a(\eta_{tr})}{a(\eta_{i})}k_{i}.
\end{equation}

For $k$-modes such that $t_{k}\gtrsim t_{tr}$ and $\dot{\theta}(t_{k})$
is negligible, the $\delta\chi(t>t_{k})$ spectator is massless, the
$\delta\chi$ power spectrum flattens out and becomes scale invariant.
This region is recognized as a massless plateau characterized by the
familiar $[H^{2}/(2\pi f_{{\rm PQ}})]^{2}$ isocurvature amplitude.
In contrast, the fluctuation $\delta\Gamma$ in the radial field has
an effective mass-squared term $m_{\delta\Gamma}^{2}=m_{\delta\chi}^{2}+2\lambda\Gamma_{0}^{2}$.
In the limit $\Gamma\rightarrow f_{{\rm PQ}}$ and negligible angular
velocity, the superhorizon fluctuations in $\delta\Gamma$ can continue
to decay if $2\lambda f_{{\rm PQ}}^{2}>9/4H^{2}$ and will not contribute
significantly to the power spectrum due to the decay. Preventing the
decay will require $2\lambda f_{{\rm PQ}}^{2}<O\left(H^{2}\right)$
such that radial fluctuations can also contribute to the overall power
spectrum. However, these cases do not give rise to a time-independent
conformal background solution as explained in Sec.~\ref{sec:Spectator-Definition-and-conformal-limit}
and thus do not result in an extremely blue spectral index (e.g.~$n_{I}>2.4$
\citep{Chung:2015tha}).

In Sec.~\ref{subsec:Rotating-scenario}, we will study how deviations
away from the time-independent conformal solution impact the effective
mass-squared parameter $m_{\delta\chi}^{2}$ in Eq.~(\ref{eq:masssqform}),
and thus determine a particular parametric window of initial conditions
within which one can obtain large blue isocurvature power spectrum
for a rotating spectator field $\Phi$.

\subsection{\label{subsec:Non-rotating-scenario}Non-rotating scenario}

Before we present the rotating case, we will briefly comment on the
non-rotating complex scalar dynamics during inflation in the context
of a quartic potential. In such cases, the angular velocity is taken
to be zero and hence the net angular momentum is negligible. During
inflation if the radial field $\Gamma$ is frozen at some large displacement
$\Gamma_{i}$, at some initial time $t_{i}$, away from the stable
vacuum $\Gamma_{\mathrm{vac}}$, the isocurvature fluctuations in
the angular direction are scale-invariant and can be suppressed due
to largeness of $\Gamma_{i}$. After $\Gamma$ starts to roll towards
$\Gamma_{\mathrm{vac}}$ due to the Hubble expansion rate dropping
below the large mass of order $\sqrt{\lambda}\Gamma$, based on arguments
similar to that presented around Eq.~(\ref{eq:action1}), one might
naively conclude that there is a scale invariant isocurvature during
the roll towards the minimum. However, this would be incorrect since
Eq.~(\ref{eq:constant}) shows that $Y_{0}=0$ with $\dot{\theta}=0$
(i.e. rotations turned off), which contradicts the time-independent
conformality requirement $Y_{0}\gg\sqrt{a''/a}$.

Furthermore, after reaching the minimum, large amplitude oscillations
of the background radial field can lead to parametric resonant enhancement
of the angular fluctuations $\delta\chi$. Alternatively, if we consider
large radial displacements such that $\lambda\Gamma^{2}/H^{2}\gg1$
during inflation, where the maximum radial displacement is bounded
by the spectator condition given in Eq.~(\ref{eq:maxradial_bound}),
then the radial field is not frozen and oscillations of the radial
field during inflation will give rise to similar parametric resonance
(PR) effects for the isocurvature fluctuations.

More explicitly, the solution to the non-rotating background radial
EoM in Eq.~(\ref{eq:radial_eom}) for a quartic potential is given
by elliptic functions. When the amplitude is large such that $\lambda\Gamma_{0}^{2}/H^{2}\gg1$,
the elliptic solution can be approximated as (\citep{Greene:1997fu,Kawasaki:2013iha})
\begin{equation}
\Gamma_{0}(t)\approx\Gamma_{i}e^{-H\left(t-t_{i}\right)}\cos\left(c\sqrt{\lambda}\Gamma_{i}/H\left(1-e^{-H(t-t_{i})}\right)\right)\label{eq:radialsol_nonrotating}
\end{equation}
where $c\approx0.847$, $\Gamma_{i}$ is the radial displacement at
an initial time $t_{i}$, and we have considered a quasi de-Sitter
scale factor $a(t)/a(t_{i})=\exp(H(t-t_{i}))$ during inflation for
an approximately constant inflationary Hubble parameter. As noted
in the introduction, the field rolls down to the minimum in a Hubble
time. Due to the oscillating background radial field, the linearized
EoM for the axial fluctuations $\delta\chi$ in Eq.~(\ref{eq:axial})
now has a large amplitude oscillating mass-squared term
\begin{equation}
m_{\delta\chi}^{2}\approx\lambda\Gamma_{i}^{2}e^{-2H\left(t-t_{i}\right)}\,\cos^{2}\left(c\sqrt{\lambda}\Gamma_{i}/H\left(1-e^{-H(t-t_{i})}\right)\right).\label{eq:mass_squared_nonrotating}
\end{equation}
In the absence of angular rotations, the EoM in Eq.~(\ref{eq:EoM-dYdX})
for the scaled linear fluctuations $X=a\delta\chi$ takes the form
\begin{equation}
\partial_{\eta}^{2}X+\left(k^{2}+\left(-2M^{2}a^{2}+\lambda Y_{i}^{2}\cos^{2}\left(c\sqrt{\lambda}Y_{i}\left(\eta-\eta_{i}\right)\right)-\frac{\partial_{\eta}^{2}a}{a}\right)\right)X(\eta)=0
\end{equation}
where we have substituted $Y_{0}=a\Gamma_{0}$ from the expression
in Eq.~(\ref{eq:radialsol_nonrotating}). The above expression can
be reframed in the form of a general Mathieu differential equation\textbf{:
\begin{equation}
\partial_{z}^{2}X+\left(A_{\chi}+2q_{\chi}\cos\left(2z\right)\right)X=0
\end{equation}
}for
\begin{equation}
z=c\sqrt{\lambda}Y_{i}\left(\eta-\eta_{i}\right),
\end{equation}
\begin{align}
A_{\chi} & =\frac{k^{2}-2M^{2}a^{2}-\frac{\partial_{\eta}^{2}a}{a}}{c^{2}\lambda Y_{i}^{2}}+2q_{\chi},\\
q_{\chi} & =\frac{1}{4c^{2}}.
\end{align}

We note that the Mathieu parameter $q_{\chi}$ appears constant only
as long as $\lambda\Gamma^{2}/H^{2}\gg1$. As $\lambda\Gamma^{2}/H^{2}\rightarrow1$,
the background radial field cannot be approximated through elliptic
functions and hence the angular fluctuations do not satisfy Mathieu
equation anymore. Substituting for the value of $c$ from above, we
infer that $q_{\chi}\approx0.35$ and $A_{\chi}\approx2q_{\chi}$
for modes with $k^{2}\ll\lambda Y_{i}^{2}$. For these modes, PR occurs
in the first instability band. This results in a large exponential
amplification of the fluctuations and may result in the formation
of axion strings until back-reaction ceases PR. However, the inflation
eventually dilutes these away and any disastrous cosmological effects
from them are generically avoided.\footnote{In principle, these may produce gravity waves. We will defer the investigation
of this issue to a future work.} Modes that lie barely outside the instability band do not undergo
PR and extend over a short $\Delta k$-range of approximately $\Delta k/k_{i}\sim O(10)$.
In Sec.~\ref{sec:Discussion}, we will revisit the concept of PR
resulting from minor deviations from the time-independent conformal
solution and provide a thorough discussion on the underlying Mathieu
system.

\subsection{\label{subsec:Rotating-scenario}Rotating scenario}

As discussed around Eq.(\ref{eq:spectral_index}), to achieve a large
blue isocurvature power spectrum we require that the background radial
field $\Gamma$ or the angular fluctuations $\delta\chi$ have an
effective mass of $O(H)$ for a suitable $N_{{\rm blue}}$ number
of e-folds during inflation. The blue-tilted part of the isocurvature
spectrum has an approximate $\Delta k/k_{i}$-range equal to $\exp\left(N_{{\rm blue}}\right)$
and hence $N_{{\rm blue}}$ provides a parametric cutoff for the transition
of the isocurvature power spectrum from a blue region to a massless
plateau. The above requirements can be easily fulfilled for a tuned
rotating complex scalar field $\Phi$ during inflation. We will now
discuss this parametric range and dynamics in detail, computing an
analytic estimate of the isocurvature spectrum as well as an expression
for the parametric tuning. The perturbations away from the $n_{I}=3$
limit can be viewed as perturbing the boundary conditions away from
the time-independent conformal limit case presented in Appendix~
\ref{sec:Conformal-limit-for}.

We begin with the EoM for the background field $Y_{0}=a\Gamma_{0}$:
\begin{equation}
\partial_{\eta}^{2}Y_{0}+\left(-2M^{2}a^{2}-\frac{\partial_{\eta}^{2}a}{a}+\lambda Y_{0}^{2}-\left(\frac{L}{Y_{0}^{2}}\right)^{2}\right)Y_{0}=0\label{eq:Y0_EoM}
\end{equation}
The above EoM implies that the time-independent conformal radial field
$Y_{0}$ has an effective potential
\begin{align}
V_{Y_{0}}(\eta) & =\frac{1}{2}\left(-\frac{2M^{2}+2H^{2}}{H^{2}\eta^{2}}+\frac{1}{2}\lambda Y_{0}^{2}+\frac{L^{2}}{Y_{0}^{4}}\right)Y_{0}^{2}+{\rm constant}.\label{eq:Veff_for_Y0}
\end{align}
For a constant background solution, the potential is driven by the
quartic (self-interaction) term with an appropriately large angular
velocity and a comparatively negligible Hubble friction and mass terms\textbf{:
\begin{equation}
\lambda Y_{0}^{2}\gg\frac{2M^{2}+2H^{2}}{H^{2}\eta^{2}}\Longrightarrow\lambda\Gamma_{0}^{2}\left(\eta\right)\gg2M^{2}+2H^{2}.
\end{equation}
}Thus, for a large angular velocity, the effective potential has a
time-dependent local minimum at
\begin{equation}
Y_{0}(\eta)=\left(1+\delta(\eta)\right)Y_{c}
\end{equation}
where $|\delta(\eta)|\ll1$.

To study the quantum inhomogeneities about a background solution where
$\delta(\eta)$ is nontrivial, we will below introduce two parameters
$\{\kappa,\epsilon_{L}\}$ that control the deformations away from
the time-independent conformal limit. Consider an initial displacement
of the scaled radial field
\begin{equation}
Y_{0}(\eta_{i})=Y_{i}\gg f_{{\rm PQ}}a(\eta_{i}).
\end{equation}
and parameterize initial radial velocity as
\begin{equation}
\left.\partial_{\eta}Y_{0}\right|_{\eta_{i}}=\kappa\sqrt{6\lambda}Y_{i}^{2}\label{eq:epsRkinetic}
\end{equation}
at some initial time $\eta_{i}$ and where $\kappa$ is a dimensionless
number. For a rotating ($L\neq0$) complex scalar field, we parameterize
the angular velocity as
\begin{equation}
\left.\partial_{\eta}\theta_{0}\right|_{\eta_{i}}=\left(1-\epsilon_{L}\right)\sqrt{\lambda}Y_{i}.
\end{equation}
The corresponding value of the conserved angular momentum $L$ is
given as
\begin{equation}
L=\left(1-\epsilon_{L}\right)\sqrt{\lambda}Y_{i}^{3}\label{eq:L_parameterization}
\end{equation}
With this parameterization, a value of $\kappa=\epsilon_{L}=0$ refers
to the situation where the angular kinetic gradient approximately
cancels with the radial potential gradient term at $t_{i}$ with the
residual $2M^{2}\ll\lambda\Gamma_{i}^{2}$. This is the time-independent
conformal limit boundary condition presented earlier. As we will show
now, the approximate cancellation with $|\epsilon_{L}|\ll1$ results
in a pseudo-flat direction in radial dynamics that is only lifted
by an $O(H^{2})$ mass-squared term after integrating out residual
UV degree of freedoms which arise as a result $\epsilon_{L}\neq0$.
Hence, there exists a parametric window for $\epsilon_{L}$ within
which the leading approximation of a time-independent conformal background
solution is stable against UV oscillations.\footnote{A similar parameterization was given in \citep{Co:2020dya} where
the authors found numerically that the PR doesn't occur if $\left|\epsilon_{L}\right|\lesssim0.2$.
Therein, the authors show that post-inflation if $\left|\epsilon_{L}\right|\lesssim0.2$,
the rotating PQ field can lead to kinetic misalignment mechanism for
axion production. However, in this paper, we are interested in rotations
that occur during inflation and decay before the end of inflation.}

With the above parameterization, the new time-independent conformal
background solution is
\begin{align}
Y_{c} & =\left(1-\epsilon_{L}\right)^{1/3}Y_{i}.\label{eq:Yc_with_epsilon}
\end{align}
Next we write the complete solution of the background radial field
as
\begin{equation}
Y_{0}(\eta)=Y_{c}+\Delta Y_{0}(\eta)
\end{equation}
and substitute into Eq.~(\ref{eq:Y0_EoM}) to obtain an EoM for $\Delta Y_{0}$:
\begin{equation}
\partial_{\eta}^{2}\left(Y_{c}+\Delta Y_{0}\right)+\left(-2M^{2}a^{2}-\frac{\partial_{\eta}^{2}a}{a}+\lambda\left(Y_{c}^{2}+\Delta Y_{0}^{2}+2Y_{c}\Delta Y_{0}\right)-\frac{L^{2}}{\left(Y_{c}+\Delta Y_{0}\right)^{4}}\right)\left(Y_{c}+\Delta Y_{0}\right)=0.\label{eq:DeltaY+Y0_EoM}
\end{equation}
Considering initial displacements and velocities not significantly
deviating from a conformal solution $Y_{c}$, and thus parameterized
by small values of $\epsilon_{L}$ and $\kappa$, we can examine small-amplitude
oscillations $\Delta Y_{0}\ll Y_{c}$. Hence, we linearize the EoM
for $\Delta Y_{0}$ as
\begin{equation}
\partial_{\eta}^{2}\left(\Delta Y_{0}\right)+\left(6\lambda Y_{c}^{2}\right)\Delta Y_{0}\approx\left(2M^{2}a^{2}+\frac{\partial_{\eta}^{2}a}{a}\right)Y_{c}
\end{equation}
where the $O\left(H^{2}\right)$ terms on the RHS induce supplementary
small amplitude deviations away from a constant background solution
even when $\epsilon_{L}=\kappa=0$. Using the initial condition $\Delta Y_{0}(\eta_{i})=Y_{i}-Y_{c}$
and $\left.\partial_{\eta}\Delta Y_{0}\right|_{\eta_{i}}=\kappa\sqrt{6\lambda}Y_{i}^{2}$
and by defining a new frequency parameter
\begin{equation}
f=\sqrt{6\lambda}Y_{c}\equiv\sqrt{6\lambda}Y_{i}\left(1-\epsilon_{L}\right)^{1/3}\label{eq:define_f}
\end{equation}
we obtain the approximate solution
\begin{align}
\Delta Y_{0}(\eta) & \approx\left(Y_{i}-Y_{c}\right)\cos\left(f\left(\eta-\eta_{i}\right)\right)+\frac{\sqrt{6}Y_{c}\left(2M^{2}/H^{2}+2\right)+6\kappa\sqrt{\lambda}Y_{i}^{2}\eta_{i}}{\sqrt{6}f\eta_{i}}\sin\left(f\left(\eta-\eta_{i}\right)\right)\nonumber \\
 & +Y_{c}\left(2M^{2}/H^{2}+2\right)\left(\cos\left(f\eta\right)\left(Ci\left(f\eta_{i}\right)-Ci\left(f\eta\right)\right)+\sin\left(f\eta\right)\left(Si\left(f\eta_{i}\right)-Si\left(f\eta\right)\right)\right)
\end{align}
where we have taken $\frac{\partial_{\eta}^{2}a}{a}=2H^{2}$ and $Ci,Si$
are cosine- and sine-integral functions respectively defined as
\begin{equation}
Ci(z)\equiv-\int_{z}^{\infty}\frac{dt}{t}\cos t
\end{equation}
\begin{equation}
Si(z)\equiv\int_{0}^{z}\frac{dt}{t}\cos t\,.
\end{equation}
 The oscillations have a constant frequency $f$ in conformal time
coordinate. During the conformal regime when $f\eta\gg1$, we can
reduce the $Ci,Si$ functions in the above solution to obtain an approximate
result:
\begin{align}
\Delta Y_{0}(\eta) & \approx\left(Y_{i}-Y_{c}\right)\cos\left(f\left(\eta-\eta_{i}\right)\right)+\frac{\kappa Y_{i}^{2}}{Y_{c}}\sin\left(f\left(\eta-\eta_{i}\right)\right)\nonumber \\
 & +Y_{c}\left(\frac{\left(2M^{2}+2H^{2}\right)}{f^{2}\eta^{2}H^{2}}-\frac{\left(2M^{2}+2H^{2}\right)}{f^{2}\eta_{i}^{2}H^{2}}\cos\left(f\left(\eta-\eta_{i}\right)\right)\right).\label{eq:delta_Y0_solution}
\end{align}
 Therefore, an approximate analytic solution for the background radial
solution is
\begin{align}
Y_{0}(\eta) & \approx Y_{c}\left(1+\left(\frac{1-\left(1-\epsilon_{L}\right)^{1/3}}{\left(1-\epsilon_{L}\right)^{1/3}}\right)\cos\left(f\left(\eta-\eta_{i}\right)\right)+\frac{\kappa}{\left(1-\epsilon_{L}\right)^{2/3}}\sin\left(f\left(\eta-\eta_{i}\right)\right)\right)\nonumber \\
 & +Y_{c}\left(\frac{\left(2M^{2}/H^{2}+2\right)}{f^{2}\eta^{2}}-\frac{\left(2M^{2}/H^{2}+2\right)}{f^{2}\eta_{i}^{2}}\cos\left(f\left(\eta-\eta_{i}\right)\right)\right).\label{eq:full_radial_solution_Y0}
\end{align}
By comparing our analytic solution with the numerical results, we
find that modifying $f$ from Eq.~(\ref{eq:define_f}) to
\begin{equation}
f=\sqrt{6\lambda}Y_{c}\left(1+\delta\right)\equiv\sqrt{6\lambda}Y_{i}\left(1-\epsilon_{L}\right)^{1/3}\left(1+\delta\right)
\end{equation}
with $\delta=0.1137\epsilon_{L}^{2.178}$ leads to a sub-percent level
accuracy for $\eta<\eta_{{\rm tr}}$. For instance, $\delta\approx0.006$
for $\epsilon_{L}=0.25$. The additional empirical factor $\left(1+\delta\right)$
accounts for minor correction to the frequency due to the residual
nonlinear effects of our original nonlinear differential system in
Eq.~(\ref{eq:DeltaY+Y0_EoM}). We note that the kinetic energy induced
oscillatory terms vanish in the limit $\{\kappa\rightarrow0,\epsilon_{L}\rightarrow0\}$
corresponding to the time-independent conformal boundary conditions.
When the set $\{\kappa,\epsilon_{L}\}$ is nontrivial, then the effective
action Eq.~(\ref{eq:actioneffective}) obtains a time-dependent conformal
representation: i.e. even in the $a''/a$ neglected approximation
\begin{align}
S_{2} & \approx\int d\eta d^{3}x\left\{ -\frac{1}{2}\eta_{\mu\nu}\partial^{\mu}\delta Y\partial^{\nu}\delta Y-\frac{1}{2}\eta_{\mu\nu}\partial^{\mu}\delta X\partial^{\nu}\delta X-\frac{\delta X}{Y_{0}}\partial_{0}\delta XY_{0}'(\eta)\right.\nonumber \\
 & \left.+\frac{2L}{Y_{0}^{2}(\eta)}\left[\partial_{0}\delta X\delta Y-\frac{Y_{0}'(\eta)}{Y_{0}(\eta)}\delta X\delta Y\right]-\frac{1}{2}\frac{\left(Y_{0}'(\eta)\right)^{2}}{Y_{0}^{2}(\eta)}\left(\delta X\right)^{2}+\frac{1}{2}\left[\left(\frac{L}{Y_{0}^{2}(\eta)}\right)^{2}-3\lambda Y_{0}^{2}(\eta)\right]\left(\delta Y\right)^{2}\right\} \label{eq:Sreduced-1}
\end{align}
the $Y_{0}$ dependent terms of this equation are violating time-translation
invariance.\footnote{Recall we had a simpler $a''/a$ time-dependent conformal representation
in Eq.~(\ref{eq:axion}).} According to Eq.~(\ref{eq:full_radial_solution_Y0}), for $|\kappa|,|\epsilon_{L}|\ll1$,
the radial field oscillates around the mean $\Gamma_{i}\left(a_{i}/a\right)\left(1-\epsilon_{L}\right)^{1/3}$
with an amplitude
\begin{align}
\Gamma_{\mathrm{amp}} & =\Gamma_{i}\left(a_{i}/a\right)\sqrt{\left(1-\left(1-\epsilon_{L}\right)^{1/3}\right)^{2}+\frac{\kappa^{2}}{\left(1-\epsilon_{L}\right)^{2/3}}}\\
 & =\Gamma_{i}\left(a_{i}/a\right)\sqrt{\left(\epsilon_{L}/3\right)^{2}+\kappa^{2}}+O\left(\left\{ \epsilon_{L}^{2},\kappa^{2},\epsilon_{L}\kappa\right\} \right)\label{eq:small_amp_oscillations}
\end{align}
 and a large frequency $O\left(f\right)\gg H$. These oscillations
are small if
\begin{align}
|\epsilon_{L}| & \ll1\label{eq:epsLsmall}\\
|\kappa| & \ll1.\label{eq:subdominant_conds}
\end{align}
Although there is an asymmetry of how fast $\epsilon_{L}$ rises for
$\epsilon_{L}>0$ versus $\epsilon_{L}<0$ due to the fact that $\frac{d\Gamma_{\mathrm{amp}}}{d\epsilon_{L}}$
diverges at $\epsilon_{L}=1$, the asymmetry magnitude is typically
small. Since the parameters $\text{\ensuremath{\epsilon_{L}}}$ and
$\kappa$ induce similar deviations, we can remove this degeneracy
by taking $\kappa\rightarrow0$.

The boundary of $\Delta Y_{0}\lesssim0.1Y_{c}$  for small oscillations
corresponds to $\left|\epsilon_{L}\right|\lesssim0.3$ for $\kappa=0$.
As $\Delta Y_{0}$ increases due to an increasing $|\epsilon_{L}|$,
the oscillating mass-squared term at the linear order can lead to
PRs. The onset of PR for the radial and axial fluctuations of rotating
complex spectator will be discussed in Sec.~\ref{sec:Discussion}.
There we will show that the mass-squared term for the radial modes
$\delta Y_{k}$ leads to PR in the blue-tilted region of the spectrum
if $\epsilon_{L}\gtrsim0.25$ or $\epsilon_{L}\lesssim-0.37$. Including
effects due to the strong coupling with the axial field, the PR can
be avoided for $|\epsilon_{L}|\lesssim0.1$.

Next we evaluate the mass-squared quantity $m_{\delta\chi}^{2}$.
Using the analytic solution for the background radial field given
in Eq.~(\ref{eq:full_radial_solution_Y0}) and substituting $Y_{c}$
from Eq.~(\ref{eq:backgroundsol}), we evaluate $m_{\delta\chi}^{2}$
in Eq.~(\ref{eq:masssqform}) up to linear order in $\Delta Y_{0}$
as
\begin{align}
m_{\delta\chi}^{2} & =\left(-2M^{2}a^{2}+\lambda Y_{0}^{2}-\frac{L^{2}}{Y_{0}^{4}}\right)a^{-2}\\
 & =-2M^{2}+6\lambda Y_{c}\Delta Y_{0}a^{-2}+O\left(9\lambda\Delta Y_{0}^{2}a^{-2}\right).
\end{align}
Substituting the solution for $\Delta Y_{0}$ from Eq.~(\ref{eq:delta_Y0_solution}),
we obtain the expression for the mass-squared $m_{\delta\chi}^{2}$
as
\begin{align}
m_{\delta\chi}^{2} & \approx-2M^{2}+a^{-2}\left(6\lambda Y_{c}Y_{i}\left(1-Y_{c}/Y_{i}\right)\cos\left(f\left(\eta-\eta_{i}\right)\right)+\kappa6\lambda Y_{i}^{2}\sin\left(f\left(\eta-\eta_{i}\right)\right)\right)\nonumber \\
 & +\left(\frac{\left(2M^{2}/H^{2}+2\right)}{\eta^{2}}a^{-2}-a^{-2}\frac{\left(2M^{2}/H^{2}+2\right)}{\eta_{i}^{2}}\cos\left(f\left(\eta-\eta_{i}\right)\right)\right),\nonumber \\
m_{\delta\chi}^{2} & \approx2H^{2}+a^{-2}\left(f^{2}\left(\left(1-\epsilon_{L}\right)^{-1/3}-1\right)-\frac{\left(2M^{2}/H^{2}+2\right)}{\eta_{i}^{2}}\right)\cos\left(f\left(\eta-\eta_{i}\right)\right)\nonumber \\
 & +f^{2}a^{-2}\kappa\left(1-\epsilon_{L}\right)^{-2/3}\sin\left(f\left(\eta-\eta_{i}\right)\right).\label{eq:mr2_eqn1}
\end{align}
The last two terms in the above expression are fast oscillating large
amplitude (since $6\lambda Y_{i}^{2}\gg H^{2}$) contributions to
the effective mass-squared function $m_{\delta\chi}^{2}$ . As shown
in Appendix \ref{sec:WKB-approximation-for} and also discussed in
\citep{Chung:2021lfg}, if the ratio of the amplitude to frequency-squared
of the oscillatory terms are much less than $1$, then the effective
mass-squared quantity is dominated by the slow-varying terms. Hence,
from Eq.~(\ref{eq:mr2_eqn1}), we have the ratios of amplitude to
frequency-squared for the two oscillating terms as approximately $O\left(\kappa\right)$
and $O\left(\max\left[\epsilon_{L}/3,\left(f_{{\rm PQ}}/\Gamma_{i}\right)^{2}\right]\right)$
respectively. Since small radial oscillations require bounds given
in Eqs.~(\ref{eq:epsLsmall}-\ref{eq:subdominant_conds}), averaging
over (which we will refer to as integrating out) the UV fluctuations
to obtain an effectively slowly varying equation as discussed in Appendix
\ref{sec:WKB-approximation-for} is justified.

Finally after integrating out the UV oscillations, the effective $O(H^{2})$
mass-squared term up to zeroth order in $\epsilon_{L},\kappa$ is
\begin{equation}
m_{\delta\chi}^{2}\approx2H^{2},\label{eq:mr2_eqn2}
\end{equation}
and using the definition given in Eq.~(\ref{eq:spectral_index})
the isocurvature power spectrum for the rotating complex scalar has
a blue spectral index of
\begin{equation}
n_{I}\approx3.\label{eq:resultbrute}
\end{equation}
In view of the conformal limit discussion of Sec.~\ref{subsec:How-conformal-limit},
this is simply a stability statement indicating that the UV oscillations
do not change the leading approximation of conformal behavior when
Eqs.~(\ref{eq:epsLsmall}) and (\ref{eq:subdominant_conds}) are
satisfied.

In addition to the conformal arguments given in Sec.~\ref{subsec:How-conformal-limit}
and Eq.~(\ref{eq:resultbrute}), here is yet another way to view
the power spectrum from a horizon exit perspective. If we approximate
the quantum fluctuations $\delta\theta$ in the angular modes as $H/\left(2\pi\Gamma_{k}\right)$
where $\Gamma_{k}$ is the radial amplitude when the relevant mode
exits the horizon at some time $t_{k}$, the isocurvature power spectrum
is approximately
\begin{equation}
\Delta_{s}^{2}(k)\sim\left(\frac{H}{2\pi\Gamma_{k}\theta_{i}}\right)^{2}\label{eq:delta-s}
\end{equation}
where $\theta_{i}$ is the final misalignment angle when the radial
field settles to its stable vacuum\footnote{Consistent with its use in the literature, we denote the final misalignment
angle as $\theta_{i}$. It is important not to confuse this with the
initial value of $\theta$ at time $t_{i}$.}. Using the leading Eq.~(\ref{eq:mr2_eqn1}) (or equivalently the
conformal solution Eq.~(\ref{eq:constantYconf})), we find
\begin{align}
\Delta_{s}^{2}(k) & \sim\left(\frac{H}{2\pi\theta_{i}\Gamma_{i}\left(1-\epsilon_{L}\right)^{1/3}}\right)^{2}\left(\frac{a(t_{k})}{a_{i}}\right)^{2}\nonumber \\
 & \sim\left(\frac{H}{2\pi\theta_{i}\Gamma_{i}\left(1-\epsilon_{L}\right)^{1/3}}\right)^{2}\left(\frac{k}{k_{i}}\right)^{2}\label{eq:power_spectrum_1}
\end{align}
where $k_{i}$ corresponds to the mode exiting the horizon at $t_{i}$
or conformal time $\eta_{i}$.

To determine $\theta_{i}$, we integrate Eq.~(\ref{eq:angular_eom})
to obtain
\begin{align}
\theta(t) & =\theta(t_{i})+\int_{t_{i}}^{t}dt\frac{L}{a^{3}\Gamma^{2}}.
\end{align}
Since $L$ is a constant and the radial field decays exponentially
as $\Gamma\approx\Gamma_{i}\left(1-\epsilon_{L}\right)^{1/3}\left(a_{i}/a\right)$
until $\sqrt{\lambda}\Gamma(t)\rightarrow O\left(M,H\right)$, the
integral in the above expression is dominated at early times $\left(t<t_{{\rm tr}}\right)$
and saturates as $\Gamma\rightarrow f_{{\rm PQ}}$. Substituting Eq.~(\ref{eq:full_radial_solution_Y0})
as the analytic solution to the radial field and neglecting any $O(\epsilon_{L})$
oscillations, we can approximate
\begin{align}
\theta(t) & \approx\theta(t_{i})+\frac{L}{\Gamma_{i}^{2}\left(1-\epsilon_{L}\right)^{2/3}a_{i}^{2}}\int_{t_{i}}^{t}\frac{dt}{a}\\
 & \approx\theta(t_{i})+\frac{\sqrt{\lambda}\Gamma_{i}}{\left(1-\epsilon_{L}\right)^{1/3}}\left(\frac{1-e^{-H\left(t-t_{i}\right)}}{H}\right)
\end{align}
where we took the scale factor as $a(t)\approx\exp\left(Ht\right)$.
Hence for $t\gg t_{i}$,
\begin{equation}
\theta_{i}=\lim_{t\gg t_{i}}\theta(t)\approx\theta(t_{i})+\frac{\sqrt{\lambda}\Gamma_{i}/H}{\left(1-\epsilon_{L}\right)^{1/3}}.
\end{equation}
 Choosing to express $\theta_{i}$ in the interval $[-\pi,\pi]$ as
is sometimes customarily done, we write
\begin{equation}
\theta_{i}+\pi\approx\left(\theta(t_{i})+\frac{\sqrt{\lambda}\Gamma_{i}/H}{\left(1-\epsilon_{L}\right)^{1/3}}\right)\mod2\pi.\label{eq:thetavac}
\end{equation}
 The $\sqrt{\lambda}\Gamma_{i}/H/\left(1-\epsilon_{L}\right)^{1/3}$
merely adds to the usual uncertainty in the vacuum $\theta$ angle.

\subsubsection{Quasi-adiabatic time-evolution example\label{subsec:Quasi-adiabatic-time-evolution-e}}

Let us consider an example of a rotating complex scalar with deviations
away from the conformal solution. Similar to the example presented
in Sec.~\ref{subsec:Adiabatic-time-evolution-example}, we set $\lambda=1$
and $f_{{\rm PQ}}=10H_{\inf}$ . Further, we initialize the background
radial field $\Gamma_{0}$ at $\eta_{i}$ with the same value as in
Sec.~\ref{subsec:Adiabatic-time-evolution-example}:
\begin{equation}
\Gamma_{0}(\eta_{i})=1000H_{{\rm inf}}.
\end{equation}
To parameterize the deviations away from an adiabatic time-evolution
for a perfect conformal solution, we set $\epsilon_{L}=0.1$ and $\kappa=0$
such that the conserved angular momentum from Eq.~(\ref{eq:L_parameterization})
is given as
\begin{equation}
L=0.9\sqrt{\lambda}10^{9}H_{\inf}^{3}a^{3}(\eta_{i}).
\end{equation}
Note that with the above parameterization, Eq.~(\ref{eq:Yc_with_epsilon})
implies that the new conformal background value is $Y_{c}\approx965.49H_{{\rm inf}}a(\eta_{i})$.
In Fig.~\ref{fig:radial_time_evolution_ex1} we plot the time evolution
of $\Gamma_{0}(t)$ from an initial amplitude $\Gamma_{0}(\eta_{i})$
to $f_{{\rm PQ}}$. In the same plot (see inset) we show a comparison
of our analytic solution with the numerical result.

\begin{figure}
\begin{centering}
\includegraphics[scale=0.62]{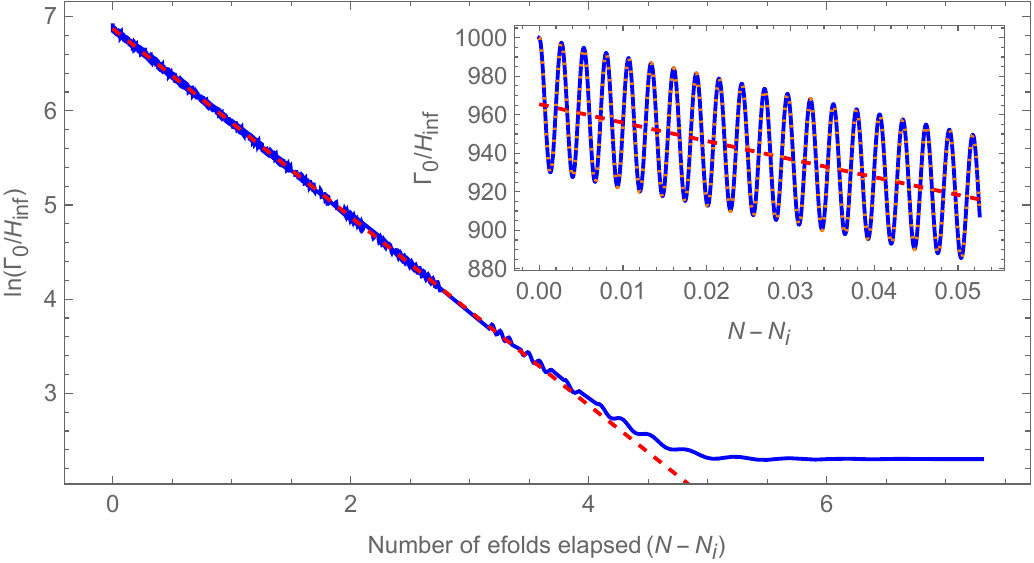}
\par\end{centering}
\caption{\label{fig:radial_time_evolution_ex2}Plot showing the time evolution
of the background radial field $\Gamma_{0}(t)$ (solid blue curve)
during the quasi de-Sitter phase of inflation for deviations from
conformal conditions, parameterized by $\text{\ensuremath{\epsilon_{L}=0.1}}$
and $\kappa=0$. Starting from $\Gamma_{0}(\eta_{i})=1000H_{\inf}$,
the radial field quickly evolves along the conformal solution $\Gamma_{c}(\eta)=\frac{L^{1/3}}{\lambda^{1/6}a(\eta)}$
(red dashed line) while undergoing small amplitude $\sim O\left(\epsilon_{L}\Gamma_{c}(\eta_{i})\right)$
oscillations along this trajectory. The oscillations have a frequency
$\approx\sqrt{6\lambda}Y_{c}$. The orange dotted curve shown in the
inset represents our analytic approximation for the small amplitude
oscillations as given in Eq.~(\ref{eq:full_radial_solution_Y0}).}
\end{figure}
\begin{figure}
\begin{centering}
\includegraphics[scale=0.7]{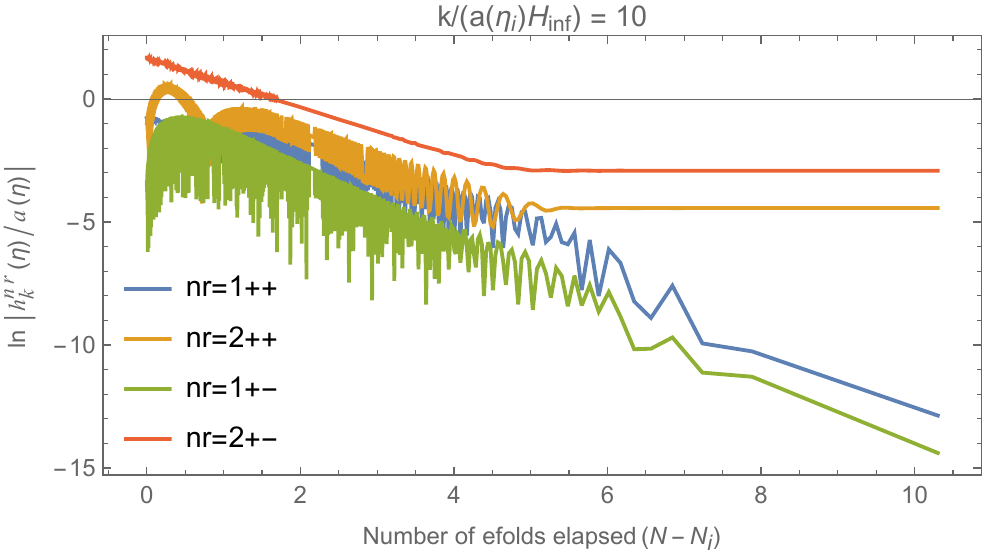}
\par\end{centering}
\caption{\label{fig:time_evolution_hknr_2}Plot showing the time evolution
of the mode functions $h_{k}^{nr}$ during the quasi de-Sitter phase
of inflation for a fiducial mode $k/a(\eta_{i})=10H_{{\rm inf}}$
in the context of quasi-adiabatic example where $\epsilon_{L}=0.1$
and $\kappa=0$. Compared to Fig.~\ref{fig:time_evolution_hknr},
we observe that the mode amplitude show small amplitude oscillations
similar to background radial field. Apart from these oscillations,
the general evolution of the mode functions is similar to the case
presented in Sec.~\ref{subsec:Adiabatic-time-evolution-example}.}
\end{figure}
To study the evolution of the linear perturbations, we note that the
time-scale of oscillations of the background radial field is much
smaller than the evolution of the mean conformal solution, i.e
\begin{equation}
T_{{\rm osc}}\sim O\left(\frac{1}{\sqrt{6\lambda}\Gamma_{0}(\eta_{i})}\right)\ll O\left(\frac{1}{H}\right).
\end{equation}
Hence, we will time-average over these rapid oscillations, and assume
an approximately conformal evolution of the background radial field.
This assumption allows us to quantize this system similar to the analysis
presented in Sec.~\ref{sec:Explicit-quantization-in}. Therefore,
we employ the same set of initial conditions for the two frequency
solutions that we presented in Eqs.~(\ref{eq:IC-1}) and (\ref{eq:IC-2})
to solve for the mode functions with a non-zero $\epsilon_{L}$. In
Fig.~\ref{fig:time_evolution_hknr_2}, we show the time evolution
of the radial and axial mode functions for a fiducial wavenumber $k/a(\eta_{i})=10$
in the context of this quasi-adiabatic system with small deviations
away from a conformal solution.

\section{Plots and Discussion\label{sec:Discussion}}

In Sec.~\ref{sec:Explicit-quantization-in} we derived analytic expressions
for the effective mass-squared term $m_{\delta\chi}^{2}$ for the
axial fluctuations at the linear order. Subsequently we showed that
for a particular conformal choice of background field boundary conditions,
the isocurvature power spectrum $\Delta_{s}^{2}$ has a blue index
$n_{I}\approx3$ and hence increases as $k^{2}$ before transitioning
to a massless plateau in agreement with the general considerations
of Sec.~\ref{subsec:How-conformal-limit}. Next, we considered deformations
away from the conformal boundary condition, which generically induces
radial background field oscillations which in turn nontrivially alter
the perturbation dynamics. In this section, we give plots of the isocurvature
power spectrum and briefly discuss the parameteric dependences. The
dimensionless superhorizon isocurvature power spectrum of the axial
field fluctuations is given as:
\begin{equation}
\Delta_{s}^{2}(k)=4\omega_{a}^{2}\frac{\Delta_{\delta\chi\delta\chi}^{2}(k,\eta_{f})}{\left(f_{{\rm PQ}}\theta_{i}\right)^{2}}\label{eq:isospec}
\end{equation}
where $\eta_{f}\rightarrow0$ and $\Delta_{\delta\chi\delta\chi}^{2}$
is given in Eq.~(\ref{eq:DELTAsq_phi_nm}) and $\omega_{a}\equiv\Omega_{\mathrm{axion}}/\Omega_{\mathrm{cdm}}$
assumes that the axions make up the CDM. Such axionic CDM would contribute
to an approximately uncorrelated photon-dark matter isocurvature inhomogeneities
in the post-inflationary cosmological evolution before the horizon
reentry. Following the discussion under Eq.~(\ref{eq:dchi-dchi-spectrum}),
we approximate the amplitude of the blue-tilted region of the spectrum
as
\begin{equation}
\Delta_{s}^{2}(k<k_{{\rm tr}})=\frac{\omega_{a}^{2}}{\pi^{2}\sqrt{3}}\left(\frac{H}{\Gamma_{0}(\eta_{i})\theta_{i}}\right)^{2}\left(\frac{k}{a(\eta_{i})H}\right)^{2}.\label{eq:analytic-blue}
\end{equation}
From Eq.~(\ref{eq:massless_plateau}), we infer that the scale invariant
part of the isocurvature spectrum is
\begin{align}
\Delta_{s}^{2}(k\gg k_{{\rm tr}}) & =\frac{\omega_{a}^{2}}{\pi^{2}}\left(\frac{H}{f_{{\rm PQ}}\theta_{i}}\right)^{2}.\label{eq:analytic-massless}
\end{align}
In the plots presented in this section, we normalize the isocurvature
power spectra $\Delta_{s}^{2}(k)$ with respect to the quantity $\left(f_{{\rm PQ}}\theta_{i}\right)^{2}/\left(\omega_{a}H\right)^{2}$.
Hence, we plot $4H^{-2}\Delta_{\delta\chi\delta\chi}^{2}(k,\eta_{f})$
on the y-axis and the analytic approximation for the normalized spectrum
$\overline{\Delta_{s}^{2}}(k)$ can be expressed as
\begin{equation}
\overline{\Delta_{s}^{2}}(k<k_{{\rm tr}})=\frac{\left(f_{{\rm PQ}}\theta_{i}\right)^{2}}{\omega_{a}^{2}H^{2}}\Delta_{s}^{2}(k<k_{{\rm tr}})=\begin{cases}
\frac{1}{\pi^{2}\sqrt{3}}\left(\frac{f_{{\rm PQ}}}{\Gamma_{0}(\eta_{i})}\right)^{2}\left(\frac{k}{a(\eta_{i})H}\right)^{2} & k<k_{{\rm tr}}\\
\frac{1}{\pi^{2}} & k\gg k_{{\rm tr}}
\end{cases}.
\end{equation}
\begin{figure}[H]
\begin{centering}
\includegraphics[scale=0.55]{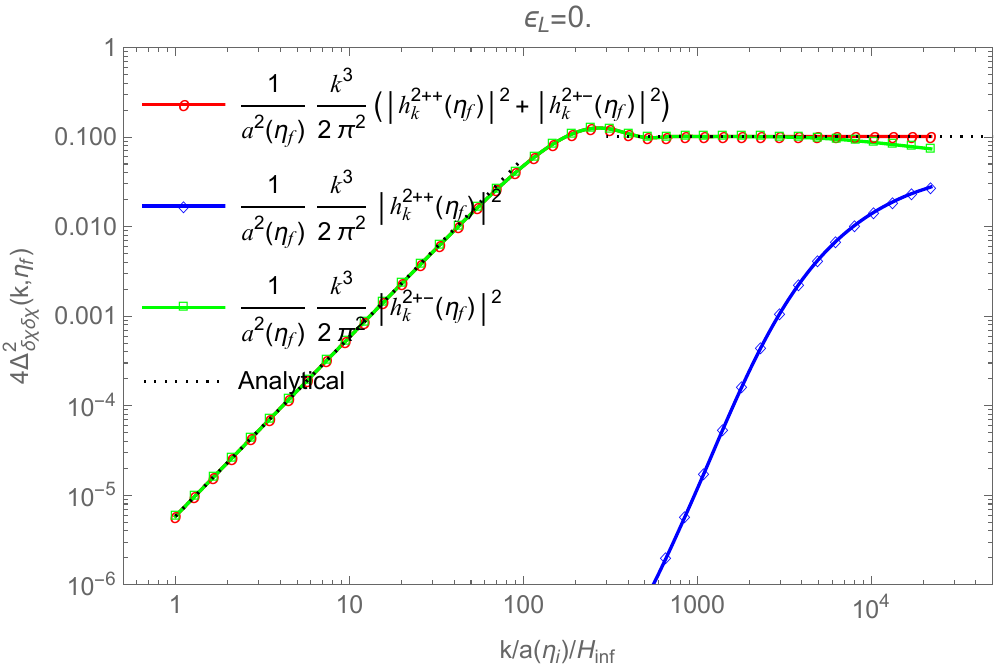}\medskip{}
\par\end{centering}
\begin{centering}
\includegraphics[scale=0.55]{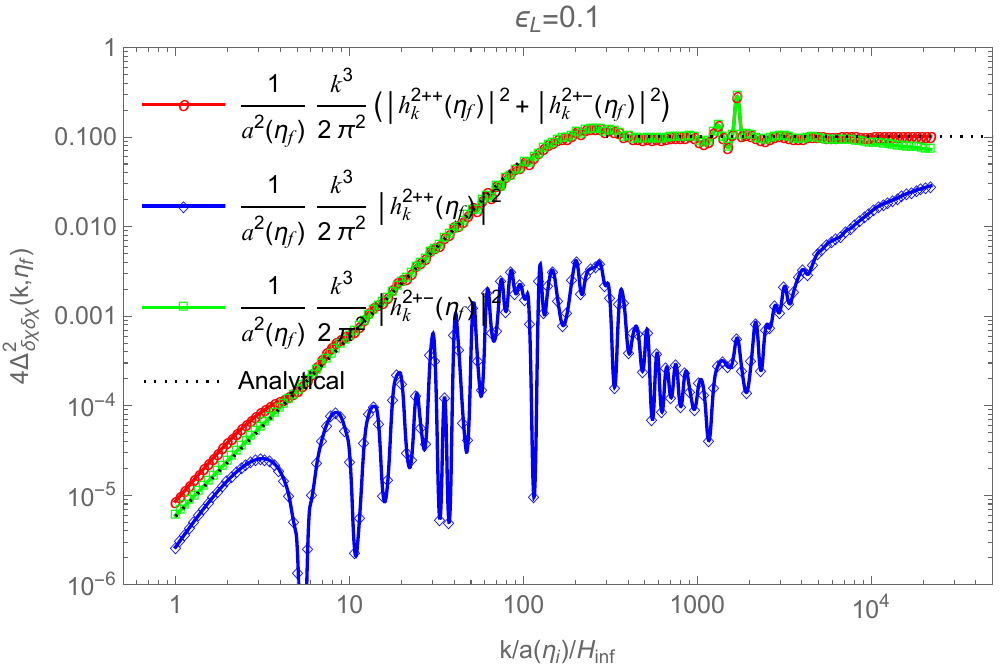}
\par\end{centering}
\caption{\label{fig:isocurv_ex1and2}Plots showing the late-time conserved
superhorizon isocurvature power spectra (normalized with $\left(f_{{\rm PQ}}\theta_{i}\right)^{2}/\omega_{a}^{2}$
in Eq.~(\ref{eq:isospec})) of the axial field $\delta\chi$ for
the two examples presented in Secs.~\ref{subsec:Adiabatic-time-evolution-example}
and \ref{subsec:Quasi-adiabatic-time-evolution-e}. The plots are
generated by numerically evolving the radial and axial mode fluctuations
from $\eta_{i}$ to $\eta_{f}\rightarrow0$ and evaluating the final
isocurvature spectrum using Eq.~(\ref{eq:isospec}). The plots on
the top (bottom) rows correspond to the parameter $\epsilon_{L}$ set to
$0$ ($0.1)$ while keeping $\kappa=0$. The power spectra have a spectral index $n_{I}\approx3$
for modes $k<k_{{\rm tr}}$ where $k_{{\rm tr}}/a_{i}/H\approx\Gamma_{i}/f_{{\rm PQ}}$. In each plot we show the final power spectrum (red curve, circular
markers) and individual contributions from the $\omega_{++}$ (blue
curve, diamond markers) and $\omega_{+-}$ (green curve, square
markers) frequency modes. The spectrum is dominated by the Goldstone
mode. Using Eqs.~(\ref{eq:analytic-blue}) and (\ref{eq:analytic-massless})
we plot the analytic spectrum (black dotted curve) in the $k<k_{{\rm tr}}$ and  $k>3k_{{\rm tr}}$ regions of the spectra respectively. The small-amplitude oscillations seen in the $\epsilon_{L}=0.1$ scenario are explained in the main text.}
\end{figure}
For the isocurvature spectrum normalized in this way, the amplitude
of blue-tilted region only depends upon the ratio $\Gamma_{0}(\eta_{i})/f_{{\rm PQ}}$.

 In Fig.~\ref{fig:isocurv_ex1and2} we illustrate the isocurvature
spectra for the two examples discussed in Secs.~\ref{subsec:Adiabatic-time-evolution-example}
and \ref{subsec:Quasi-adiabatic-time-evolution-e}. These plots highlight
that the isocurvature power spectrum has a blue index $n_{I}\approx3$
for modes $k<k_{{\rm tr}}$ where $k_{{\rm tr}}/a_{i}/H_{{\rm inf}}\approx\Gamma_{i}/f_{{\rm PQ}}$.
In each plot we show the contributions from the $\omega_{++}$ and
$\omega_{+-}$ frequency modes, with the spectrum being predominantly
influenced by the lighter $\left(\omega_{+-}\right)$ mode due to
the mode normalization. For comparison we also include our analytic
spectrum in the blue-tilted $k<k_{{\rm tr}}$ and massless-plateau
$k>3k_{{\rm tr}}$ regions of the spectrum. We lack an analytic prediction
for the intermediate (bumpy) region. Due to the sub-dominant deviations
from the conformal background solution in the $\epsilon_{L}=0.1$
scenario, we observe tiny oscillations in the spectrum that can be
attributed to the oscillation of the background radial field around
the conformal background. Below we explore the impact of a non-zero
$\epsilon_{L}$ on the isocurvature spectrum.

\subsection{$\epsilon_{L}$ dependence}

\begin{figure}[tpbh]
\begin{centering}
\includegraphics[scale=0.7]{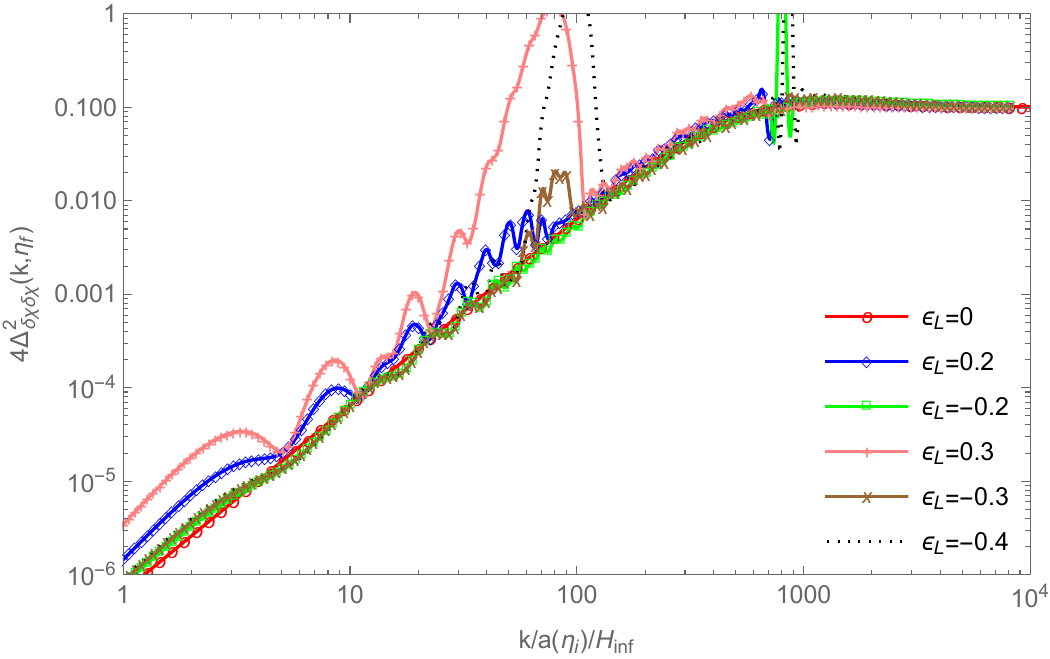}\medskip{}
\par\end{centering}
\begin{centering}
\includegraphics[scale=0.7]{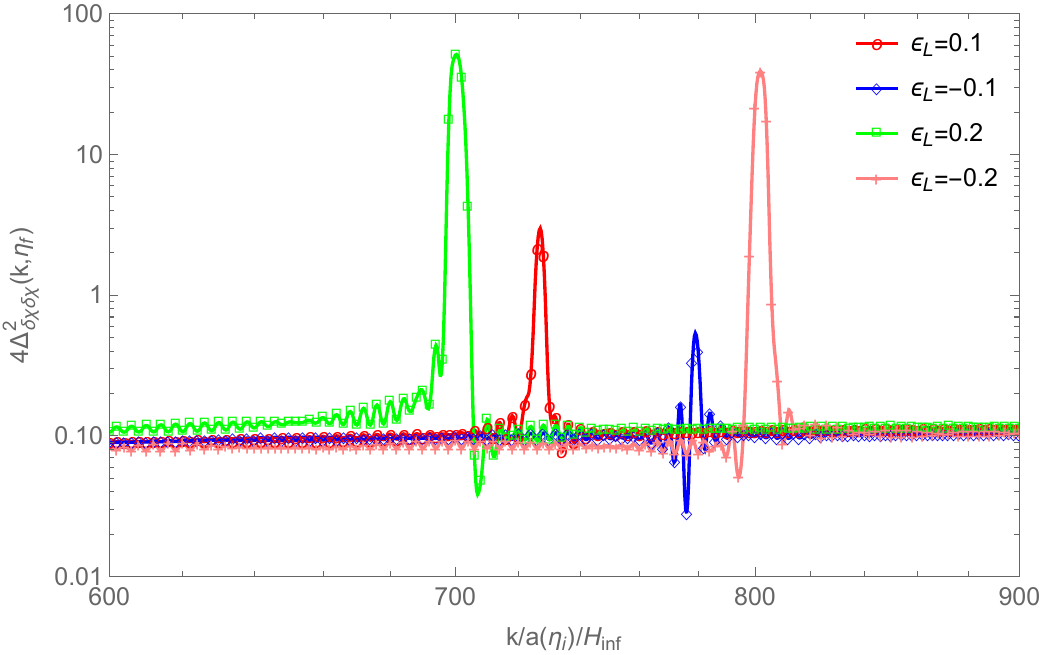}
\par\end{centering}
\caption{\label{fig:Plot_eL}Plots showing isocurvature power spectrum for
rotating complex scalar $\Phi$ for different values of $\epsilon_{L}$.
For generating these plots, we set $\lambda=10^{-4}$, $M=H_{{\rm inf}}$
and $\kappa=0$. The initial radial displacement at $\eta_{i}$ is
set at $300f_{{\rm PQ}}$ such that the $k$-range of the blue part
of the spectrum is approximately $k_{{\rm tr}}/k_{i}\sim10^{3}$.
The power spectrum has a blue index $n_{I}\approx3$. We note that
for $\epsilon_{L}\protect\neq0$, the spectrum exhibits oscillations
that grow rapidly with $\epsilon_{L}$. The plot in the top row shows
parametric enhancement of the spectrum within the $k^{2}$ region
for values of $\epsilon_{L}=+0.3$ and $\epsilon_{L}=-0.4$. In the
bottom row, we plot the isocurvature spectra on a much finer $k$-bin
to highlight sharp parameterically enhanced peaks at $k/H_{{\rm inf}}/a_{i}\approx0.9\left(2\sqrt{\lambda}Y_{c}\right)$
within the flat region.}
\end{figure}

In Fig.~\ref{fig:Plot_eL}, we plot several examples of isocurvature
power spectra for the rotating complex scalar $\Phi$ for different
values of $\epsilon_{L}$ highlighting the effect of a non-zero $\epsilon_{L}$
on the blue-tilted part of the spectrum. For all cases, the vacuum
boundary conditions for the fluctuations are set according to Eqs.~(\ref{eq:IC-1})
and (\ref{eq:IC-2}). As we will discuss below, the oscillations in
the spectrum arise due to the deviation of the background radial field
from a conformal background solution. There is also a contribution
from the residual non-adiabaticity in the Bunch-Davies-like vacuum
definition for the radial and axial fluctuations coming from our choice
of initial conditions.

In Sec.~\ref{sec:Explicit-quantization-in}, we showed that for a
time-independent conformal background solution, the Hamiltonian for
the coupled radial-axial field fluctuations can be diagonalized with
frequency solutions $\omega_{H}=\omega_{\pm+}$ and $\omega_{L}=\omega_{\pm-}$
as given in Eq.~(\ref{eq:freq}). The isocurvature power spectrum
for the IR modes is dominated by the lower frequency solution $\omega_{L}$
and is blue-tilted, $\Delta_{s}^{2}(k\lesssim k_{{\rm IR}})\propto k^{2}$.
In the $\epsilon_{L}\neq0$ scenario, the conformal symmetry is broken
in a time-dependent way through the choice of boundary condition leading
to the background radial field $\Gamma_{0}$ having $O\left(\epsilon_{L}\right)$
amplitude high-frequency oscillations for $|\epsilon_{L}|\ll1$. This
is described by the approximate analytic solution given in Eq.~(\ref{eq:full_radial_solution_Y0}).
Compared to the constant solution $Y_{c}$, the amplitude of the oscillations
can be defined by a new parameter $x$ as 
\begin{equation}
\left(\frac{Y_{i}-Y_{c}}{Y_{c}}\right)=x\sim O\left(\epsilon_{L}/3\right)\ll1.
\end{equation}

Under these conditions, the oscillations of the background radial
field induce coupling between the axial and radial field fluctuations
which is $O\left(x\right)$ magnitude and time-dependent. This leads
to a mixing between the normal mode-states $e_{L}$ and $e_{H}$ corresponding
to $\omega_{L}$ and $\omega_{H}$ respectively. Qualitatively, it
suggests that an initial excitation of the lighter frequency state
at $\eta_{i}$ will generate $O\left(x\right)$ excitations of the
remaining frequency solutions through the time-dependent mixing term.
Quantitatively, if the axial fluctuation is excited with the positive-frequency
lighter eigenstate, $e^{-i\omega_{L}\eta}$, then the time-dependent
mixing will generate a mixed state expressed approximately as
\begin{equation}
\delta\chi_{k}(\eta)\sim C_{k}\left[\left(1+O(x)\right)e^{-i\omega_{L}\eta}+O(x)e^{-i\omega_{H}\eta}+O(x)e^{+i\omega_{H}\eta}+O(x)e^{+i\omega_{L}\eta}\right]
\end{equation}
where the $C_{k}$ is an overall normalization of mode function $\chi_{k}$.
The isocurvature power spectrum during this phase ($\eta\ll\eta_{tr}$)
can be approximately given as
\begin{align}
\Delta_{\delta\chi\delta\chi}^{2}(k,\eta) & \sim k^{3}\delta\chi_{k}^{*}\delta\chi_{k}\\
 & \sim k^{3}\left|C_{k}\right|^{2}\left[1+2O(x)+2O(x)\cos\left(2\omega_{L}\left(\eta-\eta_{i}\right)\right)+\right.\nonumber \\
 & \left.2O(x)\cos\left(\left(\omega_{H}+\omega_{L}\right)\left(\eta-\eta_{i}\right)\right)+2O(x)\cos\left(\left(\omega_{H}-\omega_{L}\right)\left(\eta-\eta_{i}\right)\right)\right]+O(x^{2})\,.\label{eq:analytic-spectrum}
\end{align}
 Hence, we note that $|\epsilon_{L}|\ll1$ deviations from a perfect
conformal boundary condition ``weakly'' break time-independent conformal
phase and generate $O(x)$-amplitude oscillatory signals on the blue-tilted
part of the spectrum. These oscillations are approximately linear
in $k$. In terms of normalized momentum $-k\eta_{i}$, the $k$-space
frequencies for these oscillations in the blue-tilted region can be
read from the above expression as 
\begin{equation}
\omega_{j}\approx\left\{ \frac{2}{\sqrt{3}},\frac{1}{\sqrt{3}}\pm\frac{5k}{6\sqrt{6\lambda Y_{c}^{2}}}\right\} \left(1-\eta/\eta_{i}\right)
\end{equation}
and the isocurvature spectrum can be conveniently expressed as 
\begin{align}
\Delta_{\delta\chi\delta\chi}^{2}(k\ll\sqrt{\lambda}Y_{c},\eta\ll\eta_{tr}) & \sim k^{3}\left|C_{k}\right|^{2}\left[1+2O(x)+\sum_{j=1}^{3}2O(x)\cos\left(\omega_{j}\left(-k\eta_{i}\right)\right)\right].
\end{align}
For $\eta\rightarrow0$ and $k\ll\sqrt{\lambda}Y_{c}$, these frequencies
are simply multiples of $1/\sqrt{3}$. Thus, the time-dependent conformal
symmetry breaking boundary conditions imprint an oscillatory signal
that is a signature of the Goldstone mode's dispersion relation. Since
the $k$-space wavelength of these oscillations is $\lambda_{k}\approx2\pi\sqrt{3}/\eta_{i}\sim O(10/\eta_{i})$,
these may be measurable.

To facilitate matching/fitting with the numerical/observational data,
we transform the expression in Eq.~(\ref{eq:analytic-spectrum})
into a semi-analytic empirical form by introducing $\sim O(1)$ unknown
coefficients $c_{0,1,2,3}$:
\begin{align}
\Delta_{\delta\chi\delta\chi}^{2}(k,\eta) & \propto1+2x\left[c_{0}+c_{3}\cos\left(2\omega_{L}\left(\eta-\eta_{i}\right)\right)+\right.\nonumber \\
 & \left.c_{1}\cos\left(\left(\omega_{H}+\omega_{L}\right)\left(\eta-\eta_{i}\right)\right)+c_{2}\cos\left(\left(\omega_{H}-\omega_{L}\right)\left(\eta-\eta_{i}\right)\right)\right].\label{eq:semi-analytic-empirical-eqn}
\end{align}
If $c_{1}\approx c_{2}$, the above expression takes the form 
\begin{equation}
\Delta_{\delta\chi\delta\chi}^{2}(k,\eta)\propto1+2x\left[c_{0}+c_{3}\cos\left(2\omega_{L}\left(\eta-\eta_{i}\right)\right)+2c_{1}\cos\left(\omega_{H}\left(\eta-\eta_{i}\right)\right)\cos\left(\omega_{L}\left(\eta-\eta_{i}\right)\right)\right].
\end{equation}
To isolate the oscillations present in the data, we can normalize
it with the smoother (no-wiggle (nw)) spectrum. The resulting normalized
spectrum can then be fitted using the following empirical expression
\begin{align}
\frac{\Delta_{\delta\chi\delta\chi}^{2}(k)}{\Delta_{\delta\chi\delta\chi,{\rm nw}}^{2}(k)} & =1+2x\left[c_{0}+c_{3}\cos\left(2\omega_{L}\left(-\eta_{i}\right)\right)+c_{1}\cos\left(\left(\omega_{H}+\omega_{L}\right)\left(-\eta_{i}\right)\right)\right.\nonumber \\
 & \left.+c_{2}\cos\left(\left(\omega_{H}-\omega_{L}\right)\left(-\eta_{i}\right)\right)\right].\label{eq:empirical-fit}
\end{align}

\begin{figure}[H]
\begin{centering}
\includegraphics[scale=0.4]{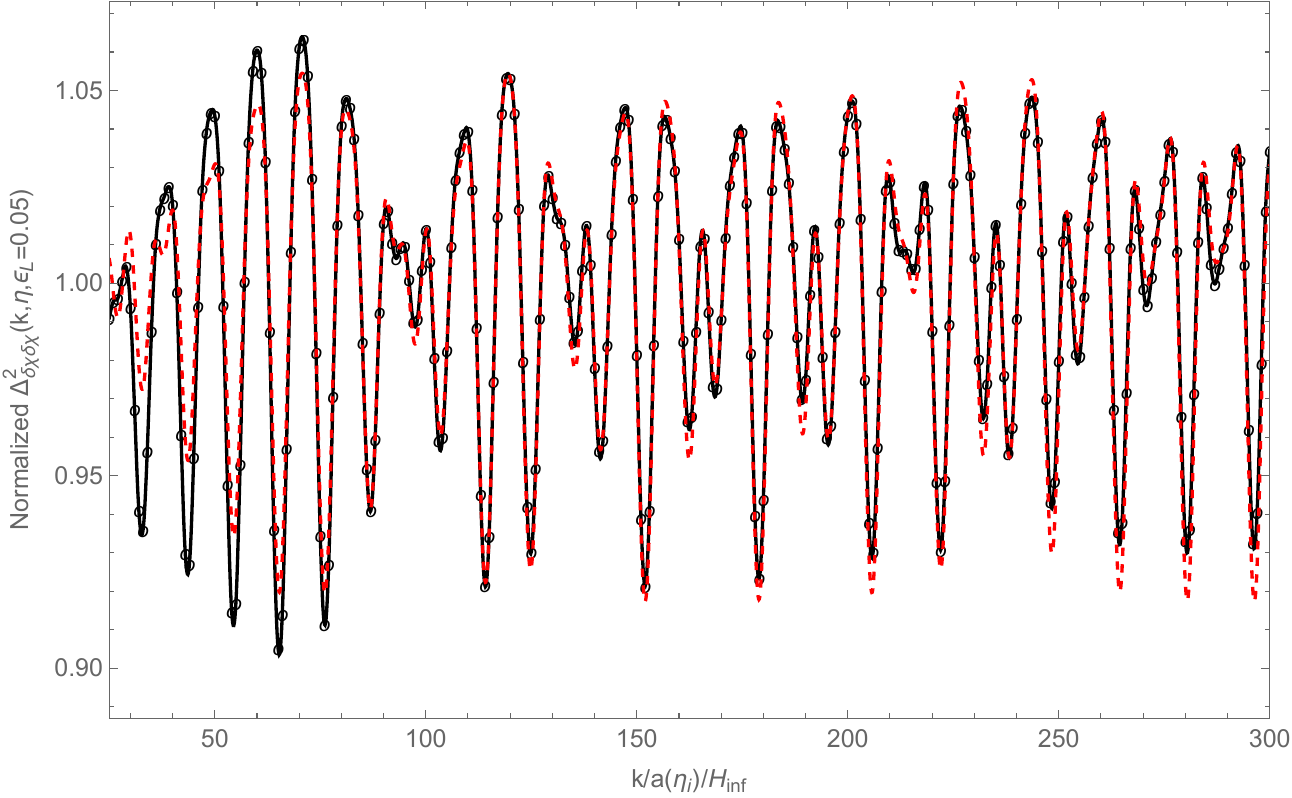}
\par\end{centering}
\vspace{0.1in}
\begin{centering}
\includegraphics[scale=0.4]{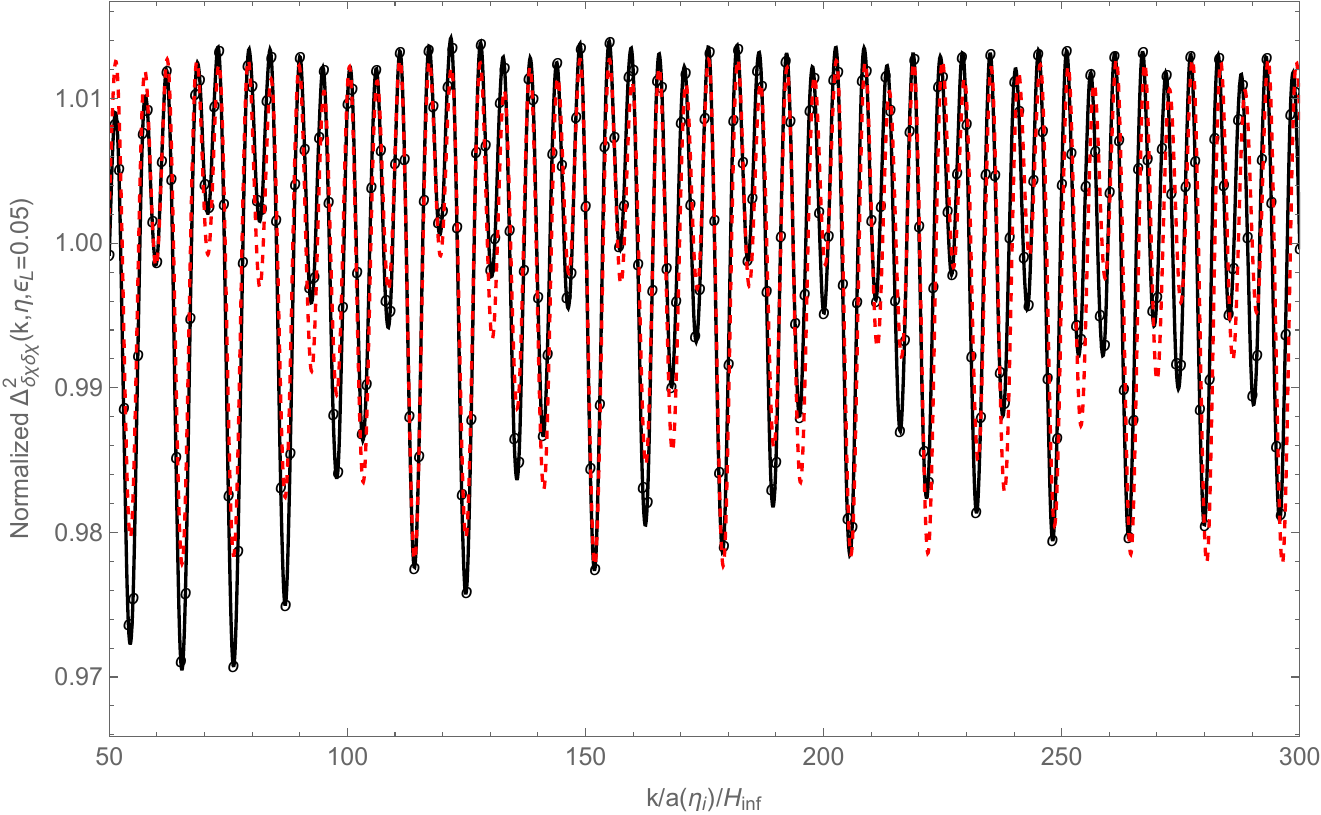}
\par\end{centering}
\centering{}\caption{\label{fig:small_amp_osc_spectrum}Plots showing the normalized isocurvature
power spectra for $\epsilon_{L}=0.05$. These plots are generated
for a fiducial set of model parameters with $\lambda=1$ and $\sqrt{2}M/H_{{\rm inf}}=10$
such that $f_{{\rm PQ}}=10H_{{\rm inf}}$. To highlight the oscillatory
signal in the power spectra for a nonzero $\epsilon_{L}$ parameter,
we have normalized the spectrum with the smooth, non-wiggly part of
the spectrum. The numerical data is plotted in black color with circular
markers and our semi-analytic empirical expression from Eq.~(\ref{eq:semi-analytic-empirical-eqn})
is depicted by the red-dashed curve. The top (bottom) plot shows the
isocurvature spectrum before (after) the radial field reaches the
$f_{PQ}$.}
\end{figure}

In Fig.~\ref{fig:small_amp_osc_spectrum}, we illustrate the normalized
isocurvature spectra for $\epsilon_{L}=0.05$. Through the figure,
we highlight the comparison between the numerical data and our semi-analytic
empirical expression in Eq.~(\ref{eq:empirical-fit}). For the axial
fluctuations initially excited with the positive-frequency lighter
eigenstate $e^{-i\omega_{L}\eta}$, the time-dependent $O(\epsilon_{L})$
oscillations will generate a mixed state with other frequencies, where
the mixing is controlled by the $O(\epsilon_{L}/3)$ parameter as
shown in Eq.~(\ref{eq:empirical-fit}). By fitting the numerical
data, we obtain the best fit values of the coefficients as $\{c_{0}=-0.0585,c_{1}=0.9544,c_{2}=1.0354,c_{3}=-0.3518\}$.
The normalized amplitude of the oscillation during this phase is $O(2x)\sim\epsilon_{L}\equiv0.05$.
After transition, the axial fluctuations corresponding to the heavier
frequency state, $\pm\omega_{H}$, decay by a factor $\propto\min\left[1,O(H/M)\right]$.
This is represented by the best fit values of the coefficients $\{c_{0}=-0.0029,c_{1}=0.1444,c_{2}=0.1684,c_{3}=-0.3311\}$
for the oscillations of the late-time spectrum as illustrated in the
bottom plot.

Let's now go back to Fig.~\ref{fig:Plot_eL} and discuss its features
for larger values of $\epsilon_{L}$. The spectrum shows parameteric
resonance enhancement of the mode amplitude for values of $\epsilon_{L}=+0.3$
and $\epsilon_{L}=-0.4$ in the blue-tilted region. To understand
the onset of the PR and its dependence on $\epsilon_{L}$, let us
consider the uncoupled mode equation for the scaled radial fluctuations
\begin{equation}
\partial_{\eta}^{2}\delta Y_{k}+\left(k^{2}-\left(2M^{2}+2H^{2}\right)a^{2}+3\lambda\left(Y_{c}+\Delta Y_{0}\right)^{2}-\frac{L^{2}}{\left(Y_{c}+\Delta Y_{0}\right)^{4}}\right)\delta Y_{k}=0.\label{eq:uncoupld_Y}
\end{equation}
In this simplified discussion, we focus solely on the effect of deformation
from the conformal background on the mass-squared term of the radial
mode, neglecting any coupling with the axial mode. By neglecting the
sub-dominant order Hubble mass terms and taking $\kappa=0$, we expand
up to linear order in $\Delta Y_{0}$. This yields the reduced EoM:
\begin{equation}
\partial_{\eta}^{2}\delta Y_{k}+\left(k^{2}+2\lambda Y_{c}^{2}+10\lambda Y_{c}^{2}\left(\frac{1-\left(1-\epsilon_{L}\right)^{1/3}}{\left(1-\epsilon_{L}\right)^{1/3}}\right)\cos\left(f\left(\eta-\eta_{i}\right)\right)\right)\delta Y_{k}=0
\end{equation}
where we recognize the mass-squared term for the radial fluctuations
as
\begin{align}
m_{\delta Y}^{2} & \approx k^{2}+2\lambda Y_{c}^{2}+10\lambda Y_{c}^{2}\left(\frac{1-\left(1-\epsilon_{L}\right)^{1/3}}{\left(1-\epsilon_{L}\right)^{1/3}}\right)\cos\left(f\left(\eta-\eta_{i}\right)\right).
\end{align}
In the above differential equation, we can identify the term $f_{N}=\sqrt{k^{2}+2\lambda Y_{c}^{2}}$
as the natural frequency of the oscillator and $f_{D}\approx f$ as
the frequency driving the parametric excitation. Through a variable
change $z=f\left(\eta-\eta_{i}\right)/2$, we reframe the above equation
in terms of a general Matheiu system:
\begin{equation}
\frac{d^{2}u}{dz^{2}}+\left(\alpha-2q\cos\left(2z\right)\right)u=0
\end{equation}
and find the corresponding Mathieu parameters as
\begin{align}
\alpha & =\frac{4\left(k^{2}+2\lambda Y_{c}^{2}\right)}{6\lambda Y_{c}^{2}},\label{eq:alpha-Mathieu}
\end{align}
\begin{equation}
q=-\frac{10\left(\frac{1-\left(1-\epsilon_{L}\right)^{1/3}}{\left(1-\epsilon_{L}\right)^{1/3}}\right)}{3}.\label{eq:q-epsilon}
\end{equation}
In terms of the original model parameters, we find that $\alpha$
depends only on one combination
\begin{equation}
\alpha=\alpha\left(\frac{k}{\left(1-\epsilon_{L}\right)^{1/3}\sqrt{\lambda}\Gamma_{i}}\right)
\end{equation}
while $q$ depends only on $\epsilon_{L}$. If $f^{2}\gg k^{2}$,
then
\begin{equation}
\alpha\approx\frac{4}{3}+O\left(k^{2}/f^{2}\right)\label{eq:a-value}
\end{equation}
causing $\alpha$ to be approximately independent of the parameters.
Thus, we find that the parameter $\alpha$ is approximately a constant
for modes that exit the horizon before axial field becomes massless,
while $|q|$ increases linearly with $\epsilon_{L}$.

\begin{figure}
\begin{centering}
\includegraphics[scale=0.6]{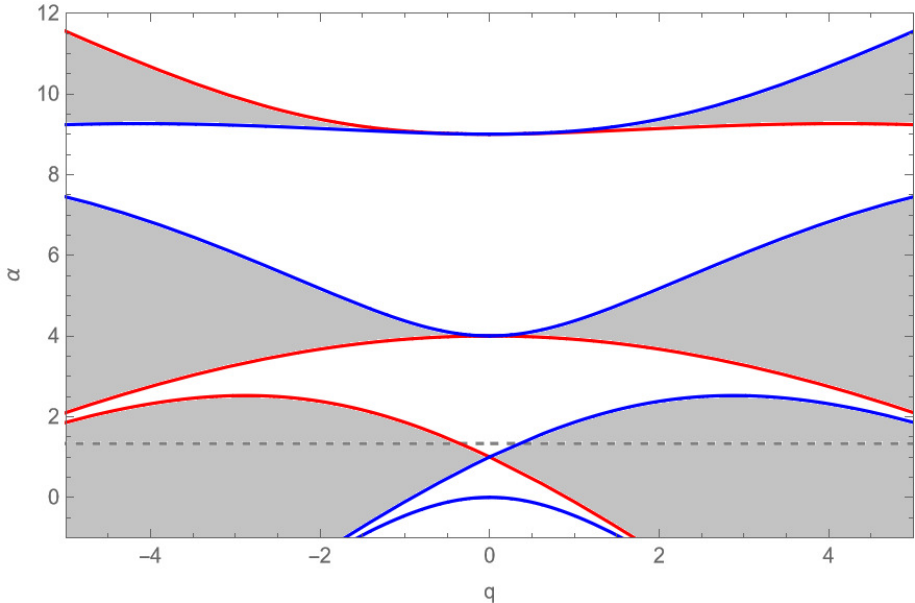}
\par\end{centering}
\caption{\label{fig:a-epsilon-chart}Plot showing Mathieu stability chart in
the $\alpha-q$ parameteric space for the first few stability bands.
The instability occurs within the shaded (unbounded) regions. For
a fixed value of $\alpha\approx4/3$ (gray dashed line), we find that
the system enters the first resonance band when the parameter $|q|\gtrsim0.35$.
The above figure is obtained by plotting even (blue) and odd (red)
Mathieu functions.}
\end{figure}
\begin{figure}
\begin{centering}
\includegraphics[scale=0.65]{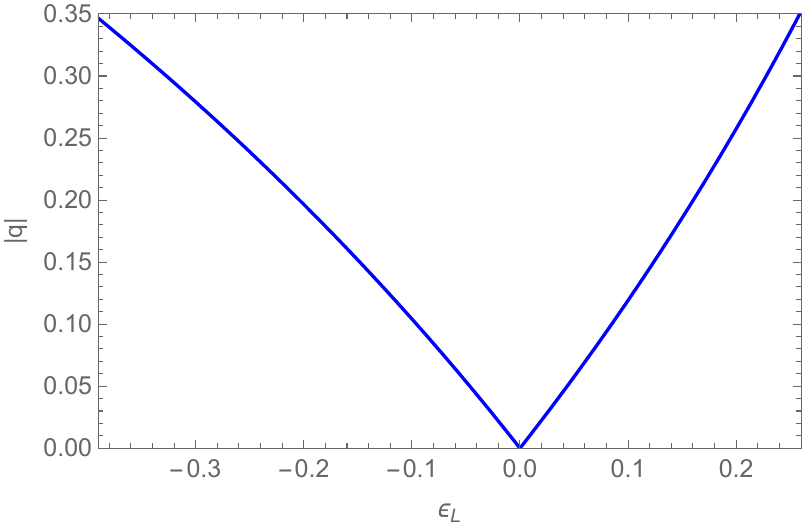}
\par\end{centering}
\caption{\label{fig:q-epsilon}Plot of Mathieu parameter $|q|$ as a function
of the rotational parameter $\epsilon_{L}$ using Eq.~(\ref{eq:q-epsilon}).
In terms of rotational parameter $\epsilon_{L}$, the oscillator becomes
unstable for $\epsilon_{L}\lesssim-0.37$ and $\epsilon_{L}\gtrsim+0.25$.
Also see Eq.~(\ref{eq:q-epsilon}).}
\end{figure}
An interesting behavior of a Mathieu oscillator is the excitation
via parameteric resonance for a range of parameters $\alpha$ and
$q$. In Fig.~\ref{fig:a-epsilon-chart} we plot a stability chart
of the Mathieu system highlighting regions/bands of stable and unstable
solutions in the $\alpha-q$ parametric space. In the plot we fix
$\alpha\approx4/3$ as derived in Eq.~(\ref{eq:a-value}). When the
oscillator system falls within an unstable resonance band it leads
to an almost exponential excitation of the amplitude. For the radial
mode fluctuations of our rotating complex field, Eq.~(\ref{eq:a-value})
suggests that the value of $\alpha$ is approximately a constant for
small values of $\epsilon_{L}$ and Eq.~(\ref{eq:q-epsilon}) indicates
that $q$ increases almost linearly with $\epsilon_{L}$ as shown
in Fig.~\ref{fig:q-epsilon}. For a fixed value of $\alpha\approx4/3$
(red dashed line), we find that the system enters the first resonance
band and becomes unstable when the parameter $|q|\gtrsim0.35$. From
Fig.~\ref{fig:q-epsilon}, we infer that the uncoupled radial mode
fluctuations $\delta Y_{k}$ become unstable for $\epsilon_{L}\lesssim-0.37$
and $\epsilon_{L}\gtrsim+0.25$ for modes $k^{2}\ll f^{2}$. The oscillator
amplitude is resonantly enhanced and results in a nearly exponential
amplification. This observation aligns with our findings in Fig\@.~\ref{fig:Plot_eL}.
 A similar analysis for the uncoupled axial field yields a much smaller
value of $\alpha\approx2k^{2}/\left(3\lambda Y_{c}^{2}\right)$ such
that the instability in the blue region occurs only for values of
$|q|$ close to unity.

Eq.~(\ref{eq:alpha-Mathieu}) also suggests that modes close to $k^{2}\approx4\lambda Y_{c}^{2}$
yield a value of $\alpha\approx4$, pushing the system towards the
next resonance band. From Fig.~\ref{fig:a-epsilon-chart}, we infer
that unlike the first, the second resonance band is significantly
narrow for small values of $|q|\sim|\epsilon_{L}|\ll1$. Consequently,
only finely tuned values of $k$ undergo PR, as depicted by the plot
in the bottom row of Fig.~\ref{fig:Plot_eL}, where we observe narrow
parameterically enhanced peaks for modes $k\approx O\left(2\sqrt{\lambda}Y_{c}\right)$.
Similarly, PR linked to the $n$th resonance band for $|q|\ll1$ would
manifest for correspondingly higher $k$ modes, with the width and
amplitude of the peaks decreasing with $n$.

The above discussion has a simple interpretation. In terms of the
natural and driving frequencies of a parameteric oscillator (defined
below Eq.~(\ref{eq:uncoupld_Y})), large exponential PR occurs when
\begin{equation}
f_{N}=n\frac{f_{D}}{2}\qquad n\in\{1,2,3,...\}
\end{equation}
where $n$ refers to the $n^{th}$ resonance/instability band. For
$q\ll1$, the bands have the usual width $\sim q^{n}$ and hence the
most important and broadest instability band is $n=1$ when $q\ll1$.
In the first band, resonance occurs close to $f_{N}=f_{D}/2$. Hence
resonance occurs when the mass of the oscillating radial field is
exactly twice the effective mass for the quantum modes $\delta Y_{k}$.

Due to the coupling between the radial and angular fluctuations, the
parametrically enhanced radial fluctuations can drive angular fluctuations
$\delta\chi$ to large amplitudes. This enhancement lasts as long
as the radial mode stays within the first resonance band, a duration
of about $O(1)$ Hubble time, after which the oscillatory mass behavior
ceases in Eq.~(\ref{eq:uncoupld_Y}). It's essential to note that
the above discussion on PR relies on the simplified ``uncoupled''
EoM for the radial mode $\delta Y_{k}$. However, the presence of
a strong derivative coupling with the axial mode can notably alter
the PR dynamics. Our numerical investigations across various Lagrangian
parameters and initial conditions indicate that PR generally does
not manifest within the blue region of the spectra for $|\epsilon_{L}|\lesssim0.1$.
A more comprehensive examination of PR's dynamics is reserved for
future studies.

\subsection{Maximum $k$-range\label{subsec:Max-k-range}}

As the background radial field approaches its stable vacuum, the effective
mass-squared term $m_{\delta\chi}^{2}$ for the axial fluctuations
becomes approximately massless. If $M^{2}a^{2}/\left(6\lambda Y_{c}^{2}\right)<1$
for $N_{\mathrm{blue}}$ number of e-folds, the mass term behaves
as in Eq.~(\ref{eq:mr2_eqn2}), during which time, we have a blue
spectrum. Hence, the range of scales across which the spectrum remains
strongly blue-tilted is approximately $\exp\left(N_{{\rm blue}}\right)$.
Starting from the condition $\lambda Y^{2}\gg\max\left(2M^{2},a''/a\right)$
for a blue spectral index, one can show that 
\begin{align}
\exp\left(N_{{\rm blue}}\right) & \approx\frac{\Gamma_{i}}{f_{{\rm PQ}}\sqrt{1+H^{2}/M^{2}}}.\label{eq:kc}
\end{align}

Using the spectator energy condition in Eq.~(\ref{eq:maxradial_bound}),
we can give an approximate upper bound on the maximum radial displacement
$\Gamma_{i}$ at $t_{i}$ for a rotating complex scalar $\Phi$ with
$|\epsilon_{L}|\ll1$:
\begin{equation}
\frac{3\lambda}{4}\Gamma_{\max}^{4}\approx r_{a}3M_{P}^{2}H^{2}
\end{equation}
or equivalently
\begin{align}
\frac{\Gamma_{\max}}{H} & \approx10^{3}\sqrt{0.2}\left(\frac{r_{a}}{0.01}\right)^{1/4}\left(\frac{1}{\lambda}\right)^{1/4}\sqrt{\frac{M_{P}/H}{10^{6}}}\label{eq:Rmax_bound}
\end{align}
where we have assumed a negligible radial velocity at $t_{i}$. The
parameter $r_{a}$ gives the ratio of spectator energy density to
that of inflaton's and must be much less than $1$. Also, Eq.~(\ref{eq:Rmax_bound})
states that the spectator energy bound is setting $\Gamma_{\mathrm{max}}\ll M_{P}$.
This is a significant departure from \citep{Chung:2021lfg,Chung:2017uzc,Chung:2016wvv,Kasuya:2009up}
in which the flat direction allowed the analog of the $\Gamma_{\mathrm{max}}$
field to reach $O(M_{P})$ while the axionic sector still remained
a spectator. Hence, even though the conformal limit liberated the
quartic model from the constraints associated with the fast roll,
the spectator condition has become more severe with the introduction
of the quartic coupling, limiting $\max\left(\Gamma_{i}\right)$.
Using Eq.~(\ref{eq:Rmax_bound}), the maximum range for the blue
part of the isocurvature spectrum for $M\sim O(H)$ is given by the
expression
\begin{align}
\max\left(k_{{\rm tr}}/k_{i}\right)\approx\exp\left(\max N_{{\rm blue}}\right) & \approx10^{3}\sqrt{0.2}\left(\frac{r_{a}}{0.01}\right)^{1/4}\left(\frac{1}{\lambda}\right)^{1/4}\sqrt{\frac{M_{P}/H}{10^{6}}}\frac{1}{f_{{\rm PQ}}/H}.\label{eq:max_krange}
\end{align}

\subsection{Spectral bump and $M$ dependence\label{subsec:M-dependnce}}

\begin{figure}
\begin{centering}
\includegraphics[scale=0.5]{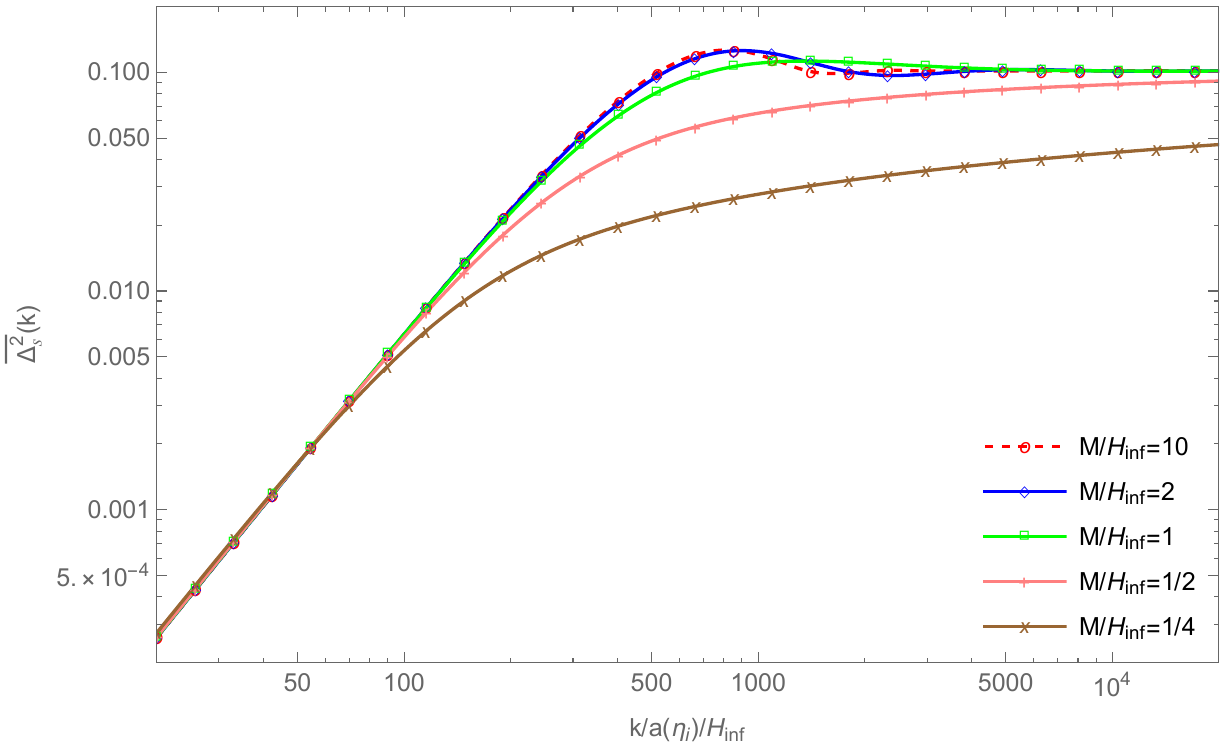}
\par\end{centering}
\caption{\label{fig:M_plot}Plot showing comparison of the normalized isocurvature
power spectra for different values of $M$ with $\lambda=10^{-4}$
and $\Gamma_{i}=300f_{{\rm PQ}}$. For these choice of $\lambda$
and $M$, the PQ scale $f_{{\rm PQ}}\approx100\,O(M)$. The transition
from a spectral index $n_{I}=3$ occurs at the transition scale $k_{{\rm tr}}/k_{i}\approx\Gamma_{i}/f_{{\rm PQ}}/\sqrt{1+H^{2}/M^{2}}$.
The plot also highlights the deviation of the isocurvature shapes
at the transition to the massless plateau for different values of
$M$. We observe the appearance of the spectral bump for $M\gtrsim3H/4$
as given in Eq.~(\ref{eq:mu_critical-1}).}
\end{figure}

In Fig.~\ref{fig:M_plot}, we show comparison between the isocurvature
power spectra for different values of $M$ while keeping $\lambda=10^{-4}$,
$\Gamma_{i}=300f_{{\rm PQ}}$ fixed and $\ensuremath{\epsilon_{L}},\kappa=0$.
From Eqs.~(\ref{eq:analytic-blue}) and (\ref{eq:analytic-massless}),
we note that the normalized isocurvature power spectra $\overline{\Delta_{s}^{2}}(k)$
is independent of $\lambda$ and hence we do not study variation of
$\lambda$ parameter. The plot highlights the deviation in the shape
of the power spectra for different values of $M$ as the spectrum
transitions from a blue region to a massless plateau. We observe the
appearance of a spectral bump (irrespective of $\lambda$) for values
of $M\gtrsim M_{c}$ where $M_{c}=3H/4$ is an approximate cutoff
derived in Appendix \ref{sec:mu_cutoff}. This cutoff is essentially
the usual dS oscillator equation having a critical $\mathrm{mass}/H=3/2$
but the mass at the asymptotic future minimum of the radial field
effective potential is $2M.$ As the value of $M$ rises above the
cutoff $M_{c}$, the asymptotic (late-time) behavior of the background
radial field transitions from an exponential to oscillatory similar
to the critical transition observed in damped oscillators. Thus, as
the radial field $\Gamma$ rolls down the potential and approaches
its stable vacuum $f_{{\rm PQ}}$, for values of $M\gtrsim M_{c}$
the radial field oscillates momentarily around the stable vacuum $f_{{\rm PQ}}$
before settling down. The oscillation of the radial field around $f_{{\rm PQ}}$
translates into oscillations of the mass-squared term $m_{\Gamma}^{2}\equiv m_{\delta\chi}^{2}$
around zero. These ``non-adiabatic'' oscillations give rise to a
bump in the power spectrum. However, due to the presence of the tachyonic
drag force from the non-zero angular velocity term, the amplitude
of the oscillations and the corresponding height of the spectral bump
become saturated for larger values of $M$. Numerically, we find that
for $M\gg H$, the amplitude of the bump is approximately a factor
of $1.3$ larger than the flat spectrum. On the other hand, when $M\lesssim M_{c}$,
the radial field settles to the vacuum exponentially slow ($\Gamma\rightarrow f_{{\rm PQ}}\left(1+\exp\left(-\frac{3}{4}\left(M/M_{c}\right)^{2}t\right)\right)$)
and hence the power spectrum gradually converges, without any bump,
to the massless plateau over a large range of modes $k$ as seen from
the plots in Fig.~\ref{fig:M_plot}.

Blue-tilted isocurvature power spectra with spectral bumps have been
discussed previously in \citep{Chung:2017uzc,Chung:2021lfg} for a
SUSY embedding of the axion model as presented in \citep{Kasuya:2009up}
which we will refer to as the KK model. In the KK model, a blue power
spectrum with $n_{I}=3$ occurs when the Lagrangian parameter is fixed
at $c_{+}=2$ (corresponding to the dynamical axion mass squared of
$c_{+}H^{2}$). Notably, a bump in the power spectrum at the transition
``always'' exists for the KK model, unlike the model discussed in
this paper, where the bump vanishes for $M\lesssim M_{c}$ despite
the blue spectral index remaining $n_{I}=3$. This is an important
distinguishing feature between the model discussed in this work and
the flat-direction models like the KK model. This distinction arises
from the proximity of the mass $\sqrt{c_{+}}H=\sqrt{2}H$ in the KK
model to the critical mass $3H/2$. Moreover, the KK model lacks an
additional drag force from a non-zero angular velocity term, unlike
the model discussed in this work. This drag force slows down the motion
of the radial field in our model towards the minimum of the potential,
resulting in a gradual transition of the spectrum to the massless
plateau. Consequently, the presence of a bump at the transition from
a $k^{2}$-spectrum is a generic feature in the KK model due to its
near-critical mass and absence of an additional drag force. Unlike
the model discussed in this work, the height of the spectral bump
in the overdamped KK model for $c_{+}=2$ can be larger than the flat
spectrum by at most a factor of $3$ where the height is governed
by the parameter $c_{-}$ (\citep{Chung:2021lfg}).

\begin{figure}
\begin{centering}
\includegraphics[scale=0.55]{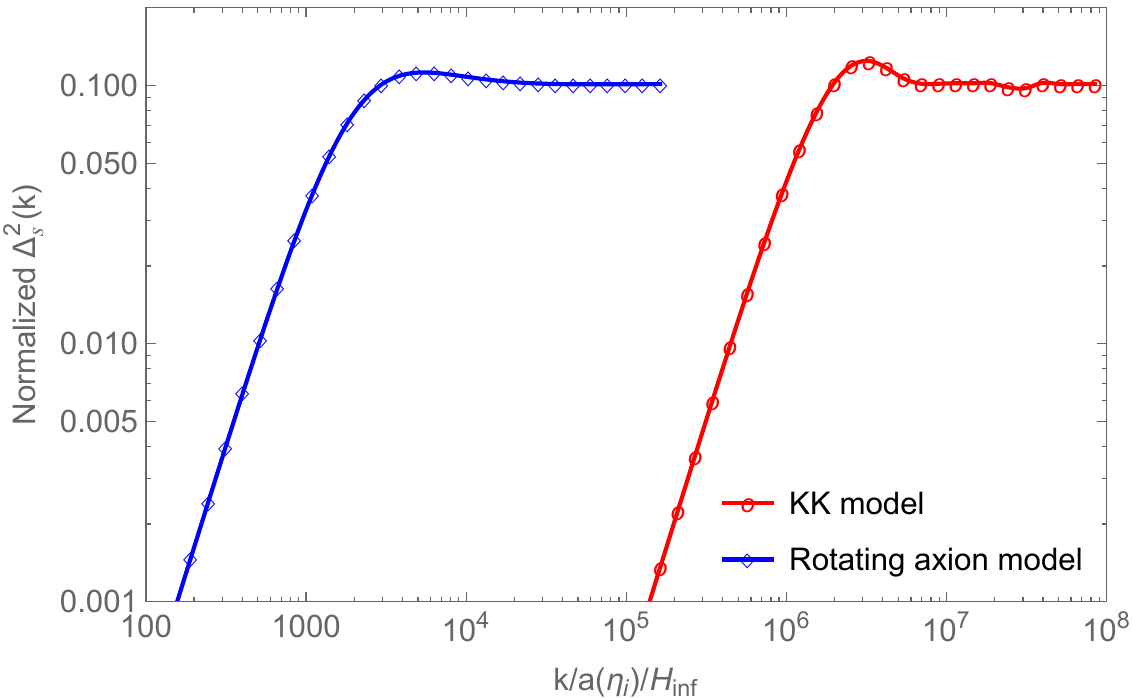}
\par\end{centering}
\caption{\label{fig:KK-comparison}We present examples of normalized isocurvature
power spectra for both the KK model and the rotating axion model.
As discussed in the main text, the KK model can exhibit a significantly
larger $\max\left(k_{{\rm tr}}/k_{i}\right)$ for the same values
of $f_{{\rm PQ}}$ and $H$. This is due to the comparatively larger
radial displacement allowed by the spectator condition in the KK model.
For these plots, we set $f_{{\rm PQ}}/H=100$ and $H/M_{p}=10^{-9}$.
The remaining Lagrangian parameters are set at $\{c_{+}=2,c_{-}=2\}$
for the KK model and $\{\lambda=2\times10^{-4},M/H=1\}$ for the current
model. In both models, the initial radial velocity is set to zero.
Without any adverse tuning of $\lambda$, a large $\max\left(k_{{\rm tr}}/k_{i}\right)$
in the KK model can serve as a distinguishing feature between the
two models.}
\end{figure}

Additionally, the maximum $k$-range for the blue part of the spectrum
in the KK model can be much larger than that achievable from the rotating
axion model. This difference arises because the potential of the KK
model is quadratically dominated, compared to the quartic potential
of the rotating axion model. In Fig.~\ref{fig:KK-comparison}, we
plot examples of normalized isocurvature power spectra for both the
KK model and the rotating axion model. We emphasize that the KK model
can exhibit a significantly larger $\max\left(k_{{\rm tr}}/k_{i}\right)$.
In the absence of any adverse tuning of $\lambda$, such a large $\max\left(k_{{\rm tr}}/k_{i}\right)$
can serve as another distinguishing feature between the two models.

\subsection{Bounds on the conformal axion model}

Since the bounds for the blue isocurvature spectrum is weak for $k_{{\rm tr}}/a_{\mathrm{today}}\gtrsim1$
Mpc$^{-1}$ \citep{Chluba:2013dna,Takeuchi2014,Dent:2012ne,Chung:2015pga,Chung:2015tha,Chluba:2016bvg,Chung:2017uzc,Planck:2018jri,Chabanier:2019eai,Lee:2021bmn,Kurmus:2022guy},
the plateau part of isocurvature spectrum $\Delta_{s}^{2}$ can be
much larger than $O(10^{-2})\Delta_{\zeta}^{2}$ (where $\Delta_{\zeta}^{2}$
is the adiabatic spectrum) if $k_{{\rm tr}}/k_{i}\gtrsim10^{4}$.
Nonetheless, there is a constraint on the isocurvature for this shape
of the spectrum as explored by \citep{Chung:2017uzc,Planck:2018jri}.
To this end, we will discuss how all of following conditions being
satisfied simultaneously within this conformal scenario \emph{applied
to QCD axions }is difficult, although relaxing any one constraint
gives a sizeable parameter region:
\begin{enumerate}
\item The axion is a QCD axion
\item $\max\left(k_{{\rm tr}}/k_{i}\right)\gg O(10^{3})$
\item $\lambda=O(1)$
\item All of DM being composed of axions.
\item Isocurvature not violating the current bounds.
\end{enumerate}
The second condition is something that is desired for the interest
of future observations and allows much larger signals than the current
bound of $\Delta_{s}^{2}/\Delta_{\zeta}^{2}\lesssim0.02$ associated
with the scale invariant CDM-photon isocurvature spectrum. The third
condition comes from naturalness/simplicity of axion models. On the
other hand, if the fourth condition is relaxed to CDM fraction being
$\omega_{a}=0.1$ (which is still quite sizeable and may even be detectable
depending on the size of the electromagnetic coupling \citep{Semertzidis:2021rxs}),
then an appreciable parameter region opens up where the isocurvature
primordial amplitude can be larger than the adiabatic amplitude. Of
course, for non-QCD axions, depending on the dark matter scenario,
the rest of the conditions can be satisfied.

\begin{figure}
\centering{}\includegraphics[scale=0.5]{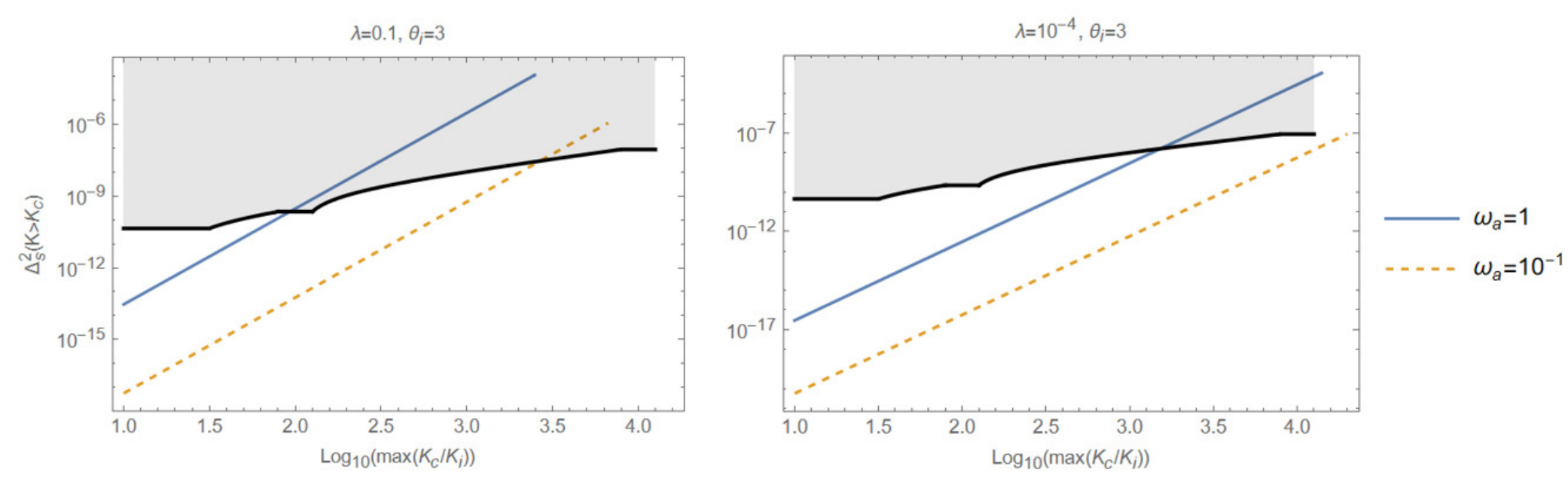}\caption{\label{fig:Left-figure-illustrates}Left figure illustrates a natural
coupling $\lambda=0.1$ scenario with spectator energy fraction taken
as $r_{a}=10^{-2}$. As $\max\left(k_{{\rm tr}}/k_{i}\right)\sim\sqrt{M_{P}H/\sqrt{\lambda}}/f_{{\rm PQ}}$
increases, the inflationary expansion rate $H$ has to become larger
with a fixed $\lambda$ and $f_{{\rm PQ}}$ (where the latter is fixed
by the axion fraction of CDM denoted as $\omega_{a}$), making the
isocurvature amplitude rise. The isocurvature bound in the shaded
region is an approximate extrapolation based on \citep{Chung:2017uzc}
assuming that the data induced bounds for that work applies to the
current scenario because of the similarity in the spectral shape.
Only the small segment near $k_{{\rm tr}}/k_{i}\approx10^{2}$ and
$10^{4}$ can be read off easily from \citep{Chung:2017uzc}, and
the rest of the black curve above $k_{{\rm tr}}/k_{i}=10^{1.5}$ represents
a smooth interpolation. The horizontal solid black curve below $k_{{\rm tr}}/k_{i}=10^{1.5}$
represent the phenomenological \citep{Chung:2017uzc,Planck:2018jri}
$\Delta_{s}^{2}/\Delta_{\zeta}^{2}\approx2\times10^{-2}$ CDM-photon
isocurvature bound applicable to \emph{scale invariant spectrum}.
The allowed region in the $\left(\Delta_{s}^{2}(k>k_{{\rm tr}}),\max\left(k_{{\rm tr}}/k_{i}\right)\right)$
plane is the left of the solid and the dashed diagonal lines (and
below the shaded region), and the exact location for the model prediction
depends on the $\Gamma$ initial conditions. The reason why the solid
and dashed diagonal lines cut off before reaching the top of the plot
is because we impose $H/f_{{\rm PQ}}<0.1$ bound such that there is
no symmetry restoration during inflation. The right figure is similar
to the left figure, except $\lambda$ has been decreased to a more
tuned value of $10^{-4}$. The main lesson from these plots is that
$\lambda\sim O(1)$ coupling is incompatible with large break $k_{{\rm tr}}/k_{i}$
(such as $k_{{\rm tr}}/k_{i}\sim10^{4}$ which phenomenologically
allows larger isocurvature signal) if all of the dark matter is made
of axion dark matter.}
\end{figure}

Shown in Fig.~\ref{fig:Left-figure-illustrates} is an illustration
of the predictions from the present scenario assuming that the axions
are QCD axions. The break in the blue spectrum given by Eq.~(\ref{eq:max_krange})
contains the following parametric dependences:
\begin{equation}
\max\left(\frac{k_{{\rm tr}}}{k_{i}}\right)\sim\frac{\sqrt{M_{P}H}}{\lambda^{1/4}f_{{\rm PQ}}}.
\end{equation}
The prediction for the break spectral value depends on the initial
value $\Gamma(t_{i})$ and $k_{{\rm tr}}/k_{i}$ can be maximally
as large as what is shown in the solid and dashed diagonal lines.
Hence, the conformal scenario of $\Gamma(t)\propto1/a$ lives to the
left of these diagonal lines.

To derive the diagonal curves in Fig.~\ref{fig:Left-figure-illustrates},
note that fixing $\omega_{a}$ and $\theta_{i}$ essentially fix $f_{{\rm PQ}}$
since
\begin{align}
\omega_{a} & =\frac{\Omega_{a}h^{2}}{0.12}\\
 & \approx2\left(\theta_{i}^{2}+\left(\frac{H}{2\pi f_{{\rm PQ}}}\right)^{2}\right)\left[\ln\left(\frac{e}{1-\theta_{i}^{2}/\pi^{2}}\right)\right]^{7/6}\left(\frac{f_{{\rm PQ}}}{10^{12}{\rm GeV}}\right)^{7/6}\label{eq:omega}
\end{align}
according to \citep{Visinelli:2009zm}.\footnote{Here the $\ln$ factor approximately taking into account the anharmonic
effects of axion oscillations has been included to obtain the $O(10)$
enhancement that exists for $\theta_{i}=3$.}\footnote{Some models in the literature explore scenarios where $f_{{\rm PQ}}$
during inflation differs from the late-time $f_{a}$ for the axions,
potentially relaxing isocurvature constraints. For further details,
see \citep{Kearney:2016vqw} and the references therein.} Combining this fact with our knowledge that the plateau part of the
spectrum ($k\gg k_{{\rm tr}}$ part) is given by
\begin{equation}
\Delta_{\delta a}^{2}\left(k\gg k_{{\rm tr}}\right)\approx4\left(\frac{H}{2\pi f_{{\rm PQ}}\theta_{i}}\right)^{2},\label{eq:axion-isocurv}
\end{equation}
we see that the isocurvature amplitude in the plateau depends just
on $H$ with $f_{{\rm PQ}}$ fixed. Since we now want the axion sector
to be a spectator to inflation and $H$ controls the energy density
during inflation, a larger $H$ is needed if the initial $\Gamma(t_{i})$
displacement that we desire for a larger
\begin{equation}
\max\left(\frac{k_{{\rm tr}}}{k_{i}}\right)\approx\exp\left(H\Delta t\right)\approx\frac{\mathrm{max}\left[\Gamma(t_{i})\right]}{f_{{\rm PQ}}}\label{eq:gammamaxdef}
\end{equation}
carries a larger energy, where the $\exp(H\Delta t)$ comes from the
conformal scaling behavior of $\Gamma(t)$.

The spectator condition with the initial energy in $\Gamma(t_{i})$
being $r_{a}$ fraction of the total sets the maximum $\Gamma(t_{i})$
value to
\begin{equation}
\Gamma_{i}^{\mathrm{max}}(k_{{\rm tr}}/k_{i})\approx\left(\frac{4M_{P}^{2}H^{2}r_{a}}{\lambda}\right)^{1/4}\label{eq:Hasfuncofgammax}
\end{equation}
where we are assuming that the kinetic energy is negligible initially.
In practice, we take $r_{a}\lesssim10^{-2}$ since the slow roll parameters
are of $O(10^{-2})$. Hence, we find that $H$ is a function of $k_{{\rm tr}}/k_{i}$
in Eq.~(\ref{eq:axion-isocurv}). More explicitly, Eqs.~(\ref{eq:gammamaxdef})
and (\ref{eq:Hasfuncofgammax}) give
\begin{equation}
\frac{H^{2}}{f_{{\rm PQ}}^{2}}=\frac{\lambda f_{{\rm PQ}}^{2}}{4M_{P}^{2}r_{a}}\left[\max\left(\frac{k_{{\rm tr}}}{k_{i}}\right)\right]^{4}
\end{equation}
or when inserted into Eq.~(\ref{eq:axion-isocurv})
\begin{equation}
\Delta_{s}^{2}\left(k>k_{{\rm tr}}\right)\approx10^{-5}\left(\frac{3\lambda/r_{a}}{10^{-2}}\right)\left(\frac{\theta_{i}}{3}\right)^{-2}\left(\frac{z(\theta_{i})}{38}\right)^{-12/7}\left(\frac{\max\left(k_{{\rm tr}}/k_{i}\right)}{10^{4}}\right)^{4}\omega_{a}^{26/7}\label{eq:isobound}
\end{equation}
where
\begin{equation}
z(\theta_{i})\equiv\theta_{i}^{2}\left[\ln\left(\frac{e}{1-\theta_{i}^{2}/\pi^{2}}\right)\right]^{7/6}
\end{equation}
and we have neglected the $H$ dependence in Eq.~(\ref{eq:omega}):
i.e.~the formula applies to $\left\{ \theta_{i}\gg0.02,H/f_{{\rm PQ}}\lesssim0.1\right\} $.
Note that $\lambda\sim O(1)$ is in tension with $\max\left(k_{{\rm tr}}/k_{i}\right)\sim10^{4}$
with the axions being all of the CDM since the isocurvature at the
break would then be already five orders of magnitude larger than the
adiabatic perturbations. Making $\theta_{i}$ smaller does not help
to alleviate the isocurvature constraints while maintaining a large
$k_{{\rm tr}}/k_{i}$ and $\omega_{a}$.

Note that the bound of
\begin{equation}
f_{{\rm PQ}}\gtrsim10^{9}\mathrm{GeV}\label{eq:Fabound}
\end{equation}
coming from white dwarf cooling time merely sets a lower bound on
$\omega_{a}$ for a given $\theta_{i}$ according to Eq.~(\ref{eq:omega}):
\begin{equation}
\omega_{a}\gtrsim6\times10^{-4}\left(\theta_{i}^{2}+\left(\frac{H}{2\pi f_{{\rm PQ}}}\right)^{2}\right)\left[\ln\left(\frac{e}{1-\theta_{i}^{2}/\pi^{2}}\right)\right]^{7/6}\left(\frac{f_{{\rm PQ}}^{\min}}{10^{9}{\rm GeV}}\right)^{7/6}.
\end{equation}
Hence, with $\theta_{i}=3$ and neglecting the $H/(2\pi f_{{\rm PQ}})$
term, we find $\omega_{a}\approx0.02$ which rules out $10^{-3}$
as a possibility to plot in Fig.~\ref{fig:Left-figure-illustrates}.
With smaller $\theta_{i}$, a smaller $\omega_{a}$ is certainly consistent
with Eq.~(\ref{eq:Fabound}), but that leads to a more stringent
constraint associated with the existing isocurvature bounds because
of Eq.~(\ref{eq:isobound}).

\section{Conclusion}

In this paper, we have shown that modifying the initial conditions
of a generic $U(1)$ symmetric quartic potential complex scalar model
can lead to a novel axion isocurvature scenario in which a transition
takes place from a time-independent conformal phase to the time-dependent
conformal phase, the latter being the usual equilibrium axion scenario.
Such time-independent spontaneously broken conformal phase initial
condition is controlled by a large classical background phase angular
momentum $\partial_{\eta}\theta_{0}(\eta_{i})\gg Ma(\eta_{i})\gtrsim Ha(\eta_{i})$
and a large radial field displacement $\Gamma_{0}(\eta_{i})\sim\partial_{\eta}\theta_{0}(\eta_{i})/\sqrt{\lambda}$.
With such initial conditions for the background, the quantum perturbations
remarkably enter a nontrivial time-independent spontaneously symmetry-broken
conformal phase characterized by a long wavelength spectral index
of $n_{I}-1=2$ and a Goldstone dispersion with a sound speed of $1/\sqrt{3}$.
Interestingly, the cross-correlation function $\left\langle \partial_{\eta}\delta\Gamma\partial_{\eta}\delta\Sigma\right\rangle $
during this time-independent conformal phase between the radial and
the axion fields does not vanish even though $\left\langle \delta\Gamma\delta\Sigma\right\rangle =0$
to leading order in perturbation theory.

After the $\Gamma_{0}$ reaches the usual spontaneous PQ symmetry
breaking minimum, the theory enters the usual time-dependent conformal
phase characterized by the time-dependent effective mass term $a''/a=2/\eta^{2}$.
For $k$ values corresponding to this time region, denoted as $k>k_{\mathrm{tr}}$,
the isocurvature spectrum is the well-known $n_{I}-1=0$ flat plateau,
and the Goldstone dispersion has a sound speed of $1$. One nontrivial
phenomenological result established in this paper is that the spectral
transition from $n_{I}=3$ to $n_{I}=1$ for realistic parameter ranges
can be sudden such that there is no large bump connecting these two
regions. This means that this quartic potential model can behave qualitatively
differently from the overdamped supersymmetric (SUSY) scenarios of
\citep{Kasuya:2009up} where there is a bump \citep{Chung:2016wvv}.
Furthermore, if the $k$ range over which the blue spectral index
sets in is sufficiently large, then the present model becomes more
fine tuned compared to the flat direction models. In the sense of
making parameters less tuned, the SUSY models in this context can
be considered analogous to the low-energy SUSY models solving the
Higgs mass hierarchy problem.

With two-parameter initial condition perturbations away from those
generating the time-independent spontaneously broken conformal phase,
we have shown that the smooth $n_{I}=3$ to $n_{I}=1$ spectra transition
scenarios are stable with $O(0.1)$ deformations of $\sqrt{\lambda}a(\eta_{i})\Gamma_{0}(\eta_{i})/\partial_{\eta}\theta_{0}(\eta_{i})$
and $\partial_{\eta}[a\Gamma_{0}]/\left(\sqrt{6\lambda}a^{2}(\eta_{i})\Gamma_{0}^{2}(\eta_{i})\right)$.
On the other hand, small deviations of the spectral amplitudes linear
in these deformation ratios eventually gain a nonlinear dependence
as these deviations grow beyond magnitudes of around $0.3$. Afterwards
parametric resonances strongly set in and destroy the original qualitative
shape of the spectra. The small oscillatory features apparent in small
deformation cases are well-fit by a simple formula characteristic
of the $1/\sqrt{3}$ sound speed of the conformal phase.

We have also explored the parametric region for which this scenario
is phenomenologically interesting. Requiring the \emph{simultaneous}
satisfaction of constraint of the axion being a QCD axion, maximum
$n_{I}=3$ blue spectral interval $[k_{i},k_{\mathrm{tr}}]$ satisfying
$k_{{\rm tr}}/k_{i}\gg O(10^{3})$, quartic coupling of order unity,
all of the DM being composed of axions, and isocurvature not violating
the current bounds, no viable parameter region exists. On the other
hand, with the relaxation of these constraints, there is a phenomenologically
viable parametric region as shown in Fig.~\ref{fig:Left-figure-illustrates}.
Because the energy density rises steeply compared to the flat direction
scenarios as the radial field is displaced, the spectator condition
imposes a significant constraint that makes this scenario sensitive
to the quartic coupling.

There are many natural future directions to explore. Given the natural
similarities between this model and the SUSY flat direction model
of \citep{Kasuya:2009up}, it would be interesting to see whether
non-Gaussianities can break the degeneracy. Indeed, there is a peculiar
feature of the time-independent conformal spectra which kinetically
cross correlates the radial mode and the axial mode, and this kinetic
mixing does not exist in the SUSY flat direction model. Hence, we
would expect the non-Gaussianities to be different between the two
models even if the isocurvature spectra are similar. Another interesting
direction is in exploring the observability of the oscillatory features
in the power spectra. As noted above, in the quasi-conformal model,
there are oscillatory features in the isocurvature spectra for small
deviations away from time-independent conformality and since those
oscillations encode the $1/\sqrt{3}$ sound speed information, it
would be interesting to see if observations can measure this sound
speed. Of course, work even remains to be done in assessing the observability
of the oscillatory features in the underdamped SUSY models \citep{Chung:2021lfg}
as noted in \citep{Chung:2023syw}.

\appendix

\section{\label{sec:Conformal-limit-for}Conformal limit for the background}

In this section, we describe how a large $\Gamma/M$ and $\Gamma/H$
limit together with a certain classical boundary condition corresponds
to a spontaneously broken approximate conformal limit of the field
theory of $\Gamma$ and $\Sigma$ during which $\Gamma a=\mbox{nonzero constant}+\delta(\Gamma a)$
where $a$ is the scale factor corresponding to the metric
\begin{equation}
ds^{2}=a^{2}(\eta)\left[-d\eta^{2}+|d\vec{x}|^{2}\right].
\end{equation}
We begin by deriving the effective action from a general $U(1)$ symmetric
renormalizable theory that spontaneously breaks an approximate conformal
symmetry with a large phase angular momentum. We then use the conformal
symmetry parameterization to generate an automorphism of the correlation
functions. This allows one to derive a differential equation for the
correlation functions whose general solution is given. We will then
use the spontaneously broken $U(1)$ coset representation to derive
$|\vec{x}-\vec{y}|^{2}$ for the $\delta X$ correlators and use the
absence of this symmetry for $\delta Y$ correlators to argue for
the $|\vec{x}-\vec{y}|^{-3}\left(\partial_{\eta}\theta_{0}\right)^{-1}$
dependence.

Start with a general renormalizable $U(1)$ invariant action action
given by Eq.~(\ref{eq:action1}):
\begin{align}
S & =\int d\eta d^{3}x\left\{ -\frac{1}{2}\eta^{\mu\nu}\partial_{\mu}Y\partial_{\nu}Y-\frac{1}{2}\eta^{\mu\nu}Y^{2}\partial_{\mu}\theta\partial_{\nu}\theta-\left(-\frac{1}{2}\frac{a''}{a}Y^{2}-M^{2}a^{2}Y^{2}+\frac{\lambda}{4}Y^{4}\right)\right\} \label{eq:conformal}
\end{align}
where $Y\equiv a\Gamma$. Note that this theory is almost invariant
under the following constant $u$ scaling conformal (dilatation) transform:
\begin{equation}
a\rightarrow au^{-1}
\end{equation}
were it not for the $M^{2}a^{2}Y^{2}$ term. Look for $Y=$constant
solutions to the equation of motion for $Y(x)=Y_{0}(\eta)$:
\begin{equation}
\frac{1}{\eta^{00}}Y_{0}^{-3}L^{2}-\left(\frac{a''}{a}+2M^{2}a^{2}\right)Y_{0}+\lambda Y_{0}^{3}=0.\label{eq:EOMconst}
\end{equation}
where we used the $U(1)$ generated conservation law to set
\begin{equation}
-\eta^{00}Y_{0}^{2}\partial_{\eta}\theta=L\label{eq:constantthetatimederiv}
\end{equation}
with $L$ being a constant. Because metric scaling will be involved
later, here we have chosen to keep $\eta^{00}$ explicit coming from
the conserved quantity being proportional to the $U(1)$ charge density
$j^{0}$ and not its associated 1-form $j_{0}$. Eq.~(\ref{eq:EOMconst})
has an approximately time-independent solution
\begin{equation}
Y_{0}=\frac{L^{1/3}}{\left(-\eta^{00}\lambda\right)^{1/6}}\label{eq:Y0sol}
\end{equation}
 when
\begin{equation}
\lambda Y_{0}^{2}\gg\frac{a''}{a}+2M^{2}a^{2}.\label{eq:hierarchy0}
\end{equation}
 Substituting Eq.~(\ref{eq:constantthetatimederiv}) into (\ref{eq:Y0sol}),
we find
\begin{equation}
Y_{0}=Y_{c}\equiv\frac{\sqrt{-\eta^{00}}\left|\partial_{\eta}\theta_{0}\right|}{\sqrt{\lambda}}\label{eq:constant}
\end{equation}
which is a constant by the virtue of Eq.~(\ref{eq:Y0sol}). Since
$Y_{c}$ is a constant, we know from this equation that $\theta_{0}'(\eta)$
is a constant. In terms of $\Gamma$ and $\theta$ fields, these solutions
represent 
\begin{equation}
\Gamma\approx\Gamma_{0}(\eta)=\frac{\sqrt{-\eta^{00}}\left|\theta_{0}'(\eta)\right|}{a(\eta)\sqrt{\lambda}}\label{eq:gaminversea}
\end{equation}
\begin{equation}
\theta\approx\theta_{0}(\eta)=\theta_{0}(\eta_{i})+\eta\partial_{\eta}\theta_{0}(\eta_{i})\label{eq:thetahomog}
\end{equation}
indicating that this is a nontrivial approximate time-dependent background
in terms of the canonical real radial field.

Now, define 
\begin{equation}
X\equiv Y_{c}\theta.
\end{equation}
By neglecting $a''/a$ and $M^{2}a^{2}$ consistently with Eq.~(\ref{eq:hierarchy0}),
the action now turns out to be completely independent of the scale
factor $a$:
\begin{equation}
S[X,Y,\eta_{\mu\nu},Y_{c}]\approx\int d\eta d^{3}x\sqrt{\eta}\left(-\frac{1}{2}\eta^{\mu\nu}\left[\partial_{\mu}Y\right]\left[\partial_{\nu}Y\right]-\frac{1}{2}\eta^{\mu\nu}\left[\partial_{\mu}X\right]\left[\partial_{\nu}X\right]\left(\frac{Y}{Y_{c}}\right)^{2}-\frac{\lambda Y^{4}}{4}\right)\label{eq:actioneffective}
\end{equation}
showing explicitly that we have a conformal theory enjoying the symmetry
\begin{align}
S[Xu,Yu,\eta_{\mu\nu}u^{-2},Y_{c}u] & =S[X,Y,\eta_{\mu\nu},Y_{c}].\label{eq:symmetry1}
\end{align}
where $u$ is a constant.

Here, the arguments $\eta_{\mu\nu}$ and $Y_{c}$ of $S$ are viewed
as externally input parameters, and we transform them as we would
a spurion. Note that the conformal representation here is different
from the dilatation subgroup representation of diffeomorphism (see
e.g.~\citep{Ginsparg:1988ui}) especially because we are scaling
$Y_{c}$ which is a parameter, as in a spurion representation of the
conformal group in a free massive scalar theory. On the other hand,
rewriting Eq.~(\ref{eq:gaminversea}) as
\begin{equation}
Y_{c}=a(\eta)\Gamma_{0}(\eta)
\end{equation}
shows that the transform $\{\eta_{\mu\nu},X,Y,Y_{c}\}\rightarrow\{\eta_{\mu\nu}u^{-2},Xu,Yu,Y_{c}u\}$
comes from scaling the scale factor by a constant as $a\rightarrow au^{-1}$,
such that the symmetry of Eq.~(\ref{eq:symmetry1}) being an element
of the conformal group is evident.\footnote{If we had used $Y_{c}=\sqrt{-\eta^{00}}\left|\theta_{0}'(\eta)\right|/\sqrt{\lambda}$,
we would have still ended up with the conformal transform $a^{2}(\eta_{i})\eta_{00}=\eta_{00}\rightarrow a^{2}(\eta_{i})u^{-2}\eta_{00}=\eta_{00}u^{-2}$
giving $Y_{c}\rightarrow uY_{c}$.} We will see how these symmetries together with diffeomorphism will
give rise to constraints on the correlation functions of interest
below.

Expand the fields as 
\begin{equation}
X(x)=\theta_{0}(\eta)Y_{c}+\delta X(x)
\end{equation}
\begin{equation}
Y(x)=Y_{c}+\delta Y(x)
\end{equation}
where $\theta_{0}(\eta)$ is given by Eq.~(\ref{eq:thetahomog})
and look for the effective action governing the perturbations only.
The perturbation-only action is
\begin{align}
S_{2}[\delta X,\delta Y,\eta_{\mu\nu},Y_{c},\partial_{\eta}\theta_{0}] & =\int d\eta d^{3}x\sqrt{\eta}\left(\frac{-1}{2}\eta_{\mu\nu}\partial^{\mu}\delta Y\partial^{\nu}\delta Y-\frac{1}{2}\eta_{\mu\nu}\partial^{\mu}\delta X\partial^{\nu}\delta X\right.\\
 & \left.-2\delta Y\eta_{\mu\nu}\partial^{\mu}\delta X\partial^{\nu}\theta_{0}-\frac{1}{2}\left(\delta Y\right)^{2}\eta^{\mu\nu}\left(\partial_{\mu}\theta_{0}\partial_{\nu}\theta_{0}\right)-\left(\frac{3\lambda}{2}Y_{c}^{2}\right)\left(\delta Y\right)^{2}\right)\label{eq:S2sym}
\end{align}
which enjoys the conformal symmetry
\begin{equation}
S_{2}\left[\delta Xu,\delta Yu,\eta_{\mu\nu}u^{-2},Y_{c}u,\partial_{\eta}\theta_{0}\right]=S_{2}\left[\delta X,\delta Y,\eta_{\mu\nu},Y_{c},\partial_{\eta}\theta_{0}\right]\label{eq:spurionsymmetryhere}
\end{equation}
where the constant $\partial_{\eta}\theta_{0}$ does not transform.
However, as we will see below, $\partial_{\eta}\theta_{0}$ will transform
under diffeomorphism because of the time derivative.

Carry out a coordinate change (diffeomorphism) $d\underline{x}^{\mu}=udx^{\mu}$
leading to
\begin{align}
\underline{\eta_{\mu\nu}} & =u^{-2}\eta_{\mu\nu}
\end{align}
 and $\underline{\phi}(\underline{x})=\phi(x)=\phi(\underline{x}u^{-1})$
leading to the diffeomorphism invariant action transforming as 
\begin{align}
S_{2}[\delta X,\delta Y,\eta_{\mu\nu},Y_{c},\partial_{\eta}\theta_{0}] & =\int d^{4}\underline{x}\sqrt{\underline{\eta}}\mathcal{L}(\underline{\delta X}(\underline{x}),\underline{\delta Y}(\underline{x}),\underline{\eta_{\mu\nu}},\underline{Y_{c}},\partial_{\underline{\eta}}\underline{\theta_{0}})\\
 & =\int d^{4}\underline{x}u^{-4}\sqrt{\eta}\mathcal{L}\left(\delta X(\underline{x}u^{-1}),\delta Y(\underline{x}u^{-1}),u^{-2}\eta_{\mu\nu},Y_{c},u^{-1}\partial_{\eta}\theta(\eta_{i})\right)\label{eq:prevdiffeo-1}
\end{align}
Scaling variables, we find

\begin{align}
S_{2}[\delta Xu^{-1},\delta Yu^{-1},\eta_{\mu\nu}u^{2},Y_{c}u^{-1},\partial_{\eta}\theta_{0}] & =\int d^{4}x\sqrt{\eta}\mathcal{L}(u^{-1}\delta X(xu^{-1}),u^{-1}\delta Y(xu^{-1}),\eta_{\mu\nu},Y_{c}u^{-1},\partial_{\eta}\theta_{0}u^{-1}).
\end{align}
 Because of Eq.~(\ref{eq:spurionsymmetryhere}), this is equivalent
to 
\begin{equation}
S_{2}[\delta Xu^{-1},\delta Yu^{-1},\eta_{\mu\nu}u^{2},Y_{c}u^{-1},\partial_{\eta}\theta_{0}]=S_{2}[\delta X,\delta Y,\eta_{\mu\nu},Y_{c},\partial_{\eta}\theta_{0}]
\end{equation}
and thus
\begin{align}
S_{2}[\delta X,\delta Y,\eta_{\mu\nu},uY_{c},u\partial_{\eta}\theta_{0}] & =\int d^{4}x\sqrt{\eta}\mathcal{L}(u^{-1}\delta X(xu^{-1}),u^{-1}\delta Y(xu^{-1}),\eta_{\mu\nu},Y_{c},\partial_{\eta}\theta_{0})\\
 & =S_{2}[u^{-1}\delta X(xu^{-1}),u^{-1}\delta Y(xu^{-1}),\eta_{\mu\nu},Y_{c},\partial_{\eta}\theta_{0}]\label{eq:betterformagain}
\end{align}

Let's see the implication of this on the Feynman correlator which
will be equivalent to the in-in equal time correlator that we seek
at free field level:
\begin{equation}
\left\langle \delta X(\eta,\vec{x})\delta X(\eta,\vec{y})\right\rangle _{g,Y_{c},\partial_{\eta}\theta_{0}}=\frac{\int D\delta XD\delta Ye^{iS_{2}[\delta X,\delta Y,g,Y_{c},\partial_{\eta}\theta_{0}]}\delta X(\eta,\vec{x})\delta X(\eta,\vec{y})}{\int D\delta XD\delta Ye^{iS_{2}[\delta X,\delta Y,g,Y_{c},\partial_{\eta}\theta_{0}]}}
\end{equation}
Change variables
\begin{equation}
\int D\delta XD\delta Y=\int D\left[u^{-1}\delta\tilde{X}(zu^{-1})\right]D\left[u^{-1}\delta\tilde{Y}(zu^{-1})\right]
\end{equation}
to conclude
\begin{align}
\left\langle \delta X(\eta,\vec{x})\delta X(\eta,\vec{y})\right\rangle _{\eta_{\mu\nu},Y_{c},\partial_{\eta}\theta_{0}} & =\frac{\int D\delta XD\delta Ye^{iS_{2}[\delta X,\delta Y,\eta_{\mu\nu},uY_{c},u\partial_{\eta}\theta_{0}]}u^{-2}\delta X(\eta u^{-1},\vec{x}u^{-1})\delta X(\eta u^{-1},\vec{y}u^{-1})}{\int D\delta XD\delta Ye^{iS_{2}[\delta X,\delta Y,\eta_{\mu\nu},uY_{c},u\partial_{\eta}\theta_{0}]}}
\end{align}
or more explicitly
\begin{equation}
\left\langle \delta X(\eta,\vec{x})\delta X(\eta,\vec{y})\right\rangle _{\eta_{\mu\nu},Y_{c},\partial_{\eta}\theta_{0}}=u^{-2}\left\langle \delta X(\eta u^{-1},\vec{x}u^{-1})\delta X(\eta u^{-1},\vec{y}u^{-1})\right\rangle _{\eta_{\mu\nu},uY_{c},u\partial_{\eta}\theta_{0}}.\label{eq:correlator}
\end{equation}

To derive the differential equation governing the correlator by assuming
$SO(3)$ and spatial translation invariance, start by writing
\begin{equation}
\left\langle \delta X(\eta,\vec{x})\delta X(\eta,\vec{y})\right\rangle _{\eta_{\mu\nu},Y_{c},\partial_{\eta}\theta_{0}}=f(\eta,\left|\vec{x}-\vec{y}\right|,Y_{c},\partial_{\eta}\theta_{0})\label{eq:def1-1}
\end{equation}
where $f(\eta,w,s,P)$ is a function of variables $\{\eta,w,s,P\}$.
This and Eq.~(\ref{eq:correlator}) says 
\begin{equation}
u^{-2}f(\eta u^{-1},u^{-1}\left|\vec{z}\right|,uY_{c},u\partial_{\eta}\theta_{0})=f(\eta,\left|\vec{x}-\vec{y}\right|,Y_{c},\partial_{\eta}\theta_{0})\label{eq:mainresbeforederiv-1}
\end{equation}
Taking a derivative with respect to $u$ and setting $u=1$, we find
\begin{align}
-2f(\eta,z,Y_{c},\partial_{\eta}\theta_{0})-\eta\partial_{\eta}f(\eta,z,Y_{c},\partial_{\eta}\theta_{0})|_{w_{1}=\eta}-z\frac{\partial}{\partial z}f(\eta,z,Y_{c},\partial_{\eta}\theta_{0})\nonumber \\
+Y_{c}\partial_{Y_{c}}f(\eta,z,Y_{c},\partial_{\eta}\theta_{0})+\partial_{\eta}\theta_{0}\partial_{\left(\partial_{\eta}\theta_{0}\right)}f(\eta,z,Y_{c},\partial_{\eta}\theta_{0}) & =0
\end{align}
governing the correlator. A general solution to this equation is
\begin{equation}
f(\eta,z,Y_{c},\partial_{\eta}\theta_{0})=\frac{\exp\left(C_{f}\left(\ln\frac{z}{-\eta},\ln\left[\left(-\eta\right)Y_{c}\right],\ln\left[\left(\partial_{\eta}\theta_{0}(\eta_{i})\right)\left(-\eta\right)\right]\right)\right)}{\eta^{2}}\label{eq:sol-1}
\end{equation}
where $C_{f}(w_{1},w_{2})$ is a general function of two variables.
Now, $S_{2}$ time translation invariance implies$f$ being time-translation-invariant.
This results in the differential equation
\begin{equation}
-2+\sum_{i=1}^{3}\partial_{w_{i}}C_{f}(w_{1},w_{2},w_{3})=0
\end{equation}
whose general solution is
\begin{equation}
C_{f}(w_{1},w_{2},w_{3})=-2w_{1}+B_{f}(w_{1}+w_{2},w_{1}+w_{3})
\end{equation}
giving
\begin{align}
f(\eta,z,Y_{c},\partial_{\eta}\theta_{0}) & =\frac{\exp\left(B_{f}(\ln\left[zY_{c}\right],\ln\left[z\partial_{\eta}\theta_{0}(\eta_{i})\right]\right)}{z^{2}}.
\end{align}
Furthermore, we know for circular orbits
\begin{equation}
Y_{c}\lambda^{1/2}=\sqrt{-\eta^{00}}\left(\partial_{\eta}\theta_{0}\right)
\end{equation}
and thus combine $Y_{c}$ and $\partial_{\eta}\theta_{0}$ dependence
to conclude
\begin{equation}
\left\langle \delta X(\eta,\vec{x})\delta X(\eta,\vec{y})\right\rangle _{\eta_{\mu\nu},Y_{c},\partial_{\eta}\theta_{0}}=\frac{E_{f}^{X}\left(\ln\left[\left|\vec{x}-\vec{y}\right|\left(\partial_{\eta}\theta_{0}(\eta_{i})\right)\right]\right)}{\left|\vec{x}-\vec{y}\right|^{2}}\label{eq:delX}
\end{equation}
\begin{equation}
\left\langle \delta Y(\eta,\vec{x})\delta Y(\eta,\vec{y})\right\rangle _{\eta_{\mu\nu},Y_{c},\partial_{\eta}\theta_{0}}=\frac{E_{f}^{Y}\left(\ln\left[\left|\vec{x}-\vec{y}\right|\left(\partial_{\eta}\theta_{0}(\eta_{i})\right)\right]\right)}{\left|\vec{x}-\vec{y}\right|^{2}}\label{eq:delY}
\end{equation}
which are manifestly time-independent but contain arbitrary $|\vec{x}-\vec{y}|$
dependences unlike in the case of massless scalar fields in Minkowski
space.

Next, note the $U(1)$ induced shift symmetry $\delta X\rightarrow\delta X+C$
has an associated current
\begin{equation}
j^{\mu}=\eta^{\mu\nu}\left\{ \partial_{\nu}\delta X+2\delta Y\partial_{\nu}\theta_{0}\right\} .
\end{equation}
whose conservation is
\begin{equation}
-\partial_{0}^{2}\delta X-2\partial_{0}\delta Y\partial_{0}\theta_{0}+\partial_{i}^{2}\delta X=0\label{eq:currentconserve-1}
\end{equation}
which is notably linear. In normal mode-Fourier space, Eq.~(\ref{eq:currentconserve-1})
is
\begin{equation}
\omega^{2}\delta X_{\omega,k}+2i\omega\delta Y_{\omega,k}\partial_{0}\theta_{0}-k^{2}\delta X_{\omega,k}=0.
\end{equation}
In the large $k$ limit, we have the usual Goldstone condition $\omega^{2}=k^{2}$
for any nonvanishing $\delta X_{\omega,k}$. The fact that the two
modes have $\omega=\pm k$ dispersion possibilities can be viewed
as a consequence of approximately in tact CPT symmetry in that limit.
For small $k$, there is at least one massless mode that is independent
of $\delta X_{\omega,k}$ and $\delta Y_{\omega,k}$ as long as $\omega\delta X_{\omega,k}$
and $\delta Y_{\omega,k}$ do not diverge. Let's label that massless
mode frequency as $\omega_{0}$. Indeed, because of the $\partial_{\eta}\theta_{0}$
dependent mixing in Eq.~(\ref{eq:S2sym}), both $\delta X_{\omega_{0},0}$
and $\lim_{k\rightarrow0}\delta Y_{\omega_{0},k}/\omega_{0}$ do not
vanish.

According to Eq.~(\ref{eq:S2sym}), we see that the Lagrangian has
a mode contribution
\begin{align}
-2\delta Y\eta^{\mu\nu}\partial_{\mu}\delta X\partial_{\nu}\theta_{0}-\frac{1}{2}\left(\delta Y\right)^{2}\eta^{\mu\nu}\left(\partial_{\mu}\theta_{0}\partial_{\nu}\theta_{0}\right)-\left(\frac{3\lambda}{2}Y_{c}^{2}\right)\left(\delta Y\right)^{2}\\
\sim\left(\begin{array}{cc}
\delta X_{\omega_{0},k}^{*} & \delta Y_{\omega_{0},k}^{*}\end{array}\right)\left(\begin{array}{cc}
0 & -i\omega_{0,k}\\
i\omega_{0,k} & -2\left(\partial_{\eta}\theta_{0}(\eta_{i})\right)^{2}
\end{array}\right)\left(\begin{array}{c}
\delta X_{\omega_{0},k}\\
\delta Y_{\omega_{0},k}
\end{array}\right)\label{eq:massmat}
\end{align}
which in the $\partial_{\eta}\theta_{0}(\eta_{i})/\omega_{0,k}\rightarrow\infty$
limit has $\delta X$ decoupling from $\delta Y$ and $\left(\partial_{\eta}\theta_{0}(\eta_{i})\right)\left|\vec{x}-\vec{y}\right|\rightarrow\infty$
not changing the approximate correlation function for $\left\langle \delta X\delta X\right\rangle $.
In that approximation, the factor $E_{f}^{X}\left(\ln\left[\left|\vec{x}-\vec{y}\right|\left(\partial_{\eta}\theta_{0}(\eta_{i})\right)\right]\right)$
in Eq.~(\ref{eq:delX}) has an expansion for large $|\vec{x}-\vec{y}|$
as
\begin{equation}
E_{f}^{X}\left(\ln\left[\left|\vec{x}-\vec{y}\right|\left(\partial_{\eta}\theta_{0}(\eta_{i})\right)\right]\right)=c_{1}+W_{X}\left(\frac{1}{\left|\vec{x}-\vec{y}\right|\left(\partial_{\eta}\theta_{0}(\eta_{i})\right)}\right)
\end{equation}
where $c_{1}$ is independent of $|\vec{x}-\vec{y}|$ and $W_{X}(w)$
is a function where $W_{X}(0)=0$. This and Eq.~(\ref{eq:delX})
implies the correlator in Fourier space behaving as $k^{2}$ corresponding
to a spectral index of $n_{I}=3$ matching Eq.~(\ref{eq:conformalresult}).
Hence, unlike the ordinary massless Minkowski field, one has to use
$U(1)$ Goldstone dynamical information contained in Eq.~(\ref{eq:massmat})
to fix the $|\vec{x}-\vec{y}|^{-2}$ scaling for the $\left\langle \delta X\delta X\right\rangle $
correlator.

With this same $\partial_{\eta}\theta_{0}(\eta_{i})/\omega_{0,k}\rightarrow\infty$
approximation, we see from Eq.~(\ref{eq:massmat}) that a decoupled
$\delta Y$ becomes infinitely heavy as $\left(\partial_{\eta}\theta_{0}(\eta_{i})\right)\rightarrow\infty$.
Since this means that $\left\langle \delta Y(\eta,\vec{x})\delta Y(\eta,\vec{y})\right\rangle $
should vanish with the same limit, we expect
\begin{equation}
E_{f}^{Y}\left(\ln\left[\left|\vec{x}-\vec{y}\right|\left(\partial_{\eta}\theta_{0}(\eta_{i})\right)\right]\right)\propto\frac{1}{\left|\vec{x}-\vec{y}\right|\left(\partial_{\eta}\theta_{0}(\eta_{i})\right)}+O\left(\frac{1}{\left|\vec{x}-\vec{y}\right|^{2}\left(\partial_{\eta}\theta_{0}(\eta_{i})\right)^{2}}\right)
\end{equation}
if the expansion is analytic in inverse powers of $\left|\vec{x}-\vec{y}\right|\left(\partial_{\eta}\theta_{0}(\eta_{i})\right)$.
Explicit computations justify the analyticity.

Now, the sound speed of $\delta X$ cannot quite be read off from
this expression since $X=Y_{c}(\theta(\eta_{i})+\eta\partial_{\eta}\theta(\eta_{i}))+\delta X$
generates a mixing between $\delta Y$ and $\partial_{\eta}\delta X$.
This mixing generated change in the sound speed, which is the most
theoretically interesting aspect of the system studied in this paper,
and other aspects of this system are addressed in the main body of
the text when we quantize the theory.

\section{WKB approximation for oscillating potentials\label{sec:WKB-approximation-for}}

In this section, we explain Eq.~(\ref{eq:mainapprox}) which is a
generalization of the WKB ansatz applicable for dispersion relationships
with a fast time-oscillation component.

Consider the following differential equation
\begin{equation}
\partial_{\eta}^{2}y+m^{2}(\eta)y=0.\label{eq:diff_eqn1}
\end{equation}
The WKB method allows us to approximate the solution to the above
differential equation as
\begin{equation}
y(\eta)\approx\sum_{\pm}\frac{c_{\pm}}{\sqrt{m(\eta)}}\exp\left(\pm i\int^{\eta}d\eta m(\eta)\right)
\end{equation}
given 
\begin{equation}
\partial_{\eta}m(\eta)\ll m^{2}\left(\eta\right)
\end{equation}
and hence the mass-squared function must be slowly varying.

Let us now consider the situation where the mass-squared term is
\begin{equation}
m^{2}(\eta)=K^{2}+A\cos\left(f\eta\right)\label{eq:masssq}
\end{equation}
where $K,A$ are constants and we are interested in cases with $K^{2}<A$
and $f\gg1$ such that the mass-squared function is characterized
by high frequency large amplitude oscillations. It is easy to note
that the WKB approximation as given above is inappropriate and diverges
at the zeros of $m^{2}$. Although one may use matching solutions
at the zero crossings, such an approach is unwieldy for a fast oscillating
potential and doesn't capture the long-time characteristic behavior
of the system.

Another well-known approach begins with noting that Eq.~(\ref{eq:diff_eqn1})
with the mass-squared function given in Eq.~(\ref{eq:masssq}) satisfies
the Mathieu differential equation 
\begin{equation}
\partial_{x}^{2}y+\left(a-2q\cos\left(2x\right)\right)y=0
\end{equation}
where 
\begin{align}
2x & =f\eta\\
q & =\frac{-A/2}{f^{2}/4}\\
a & =\frac{K^{2}}{f^{2}/4}
\end{align}
with the generalized solution 
\begin{equation}
y=\sum_{\pm}c_{\pm}{\rm Me^{\pm}\left(r,q,x\right)}\label{eq:mathieusol}
\end{equation}
\begin{equation}
{\rm Me^{\pm}=}{\rm ce}\left(r,q,x\right)\pm i\,{\rm se}\left(r,q,x\right)
\end{equation}
for $ce$, $se$ as the generalized angular Mathieu functions. In
the limiting case $|q|\ll1$, the Mathieu function has the following
series expansion in powers of $q$
\begin{equation}
{\rm Me^{\pm}\left(r,q,x\right)}\approx\exp\left(\pm irx\right)+\frac{q}{4}\left(\frac{\exp\left(\pm i\left(r-2\right)x\right)}{r-1}+\frac{\exp\left(\pm i\left(r+2\right)x\right)}{r+1}\right)+O(q^{2})\quad\mbox{for \ensuremath{r\neq1}}
\end{equation}
where
\begin{equation}
r\approx\sqrt{a+\frac{1}{2\left(1-a\right)}q^{2}}.
\end{equation}
Using this, we can identify
\begin{equation}
rx\approx\eta\sqrt{K^{2}+\frac{A^{2}}{2\left(f^{2}-4K^{2}\right)}}
\end{equation}
in the limit 
\begin{equation}
\frac{2A}{f^{2}}\ll1,\mbox{ and }\frac{A}{f}<K\,,
\end{equation}
and the exact Mathieu solution given in Eq.~(\ref{eq:mathieusol})
can be approximated up to first order in $q$ as 
\begin{equation}
y\approx\sum_{\pm}c_{\pm}\left(\exp\left(\pm iK\eta\right)+\frac{A}{2f^{2}}\left(\frac{\exp\left(\pm i\left(K-f\right)\eta\right)}{\frac{2K}{f}-1}+\frac{\exp\left(\pm i\left(K+f\right)\eta\right)}{\frac{2K}{f}+1}\right)\right).
\end{equation}
Hence, in the limit $2A/f^{2}\ll1$, we observe that the solution
to the oscillatory mass-squared function in Eq. (\ref{eq:masssq})
is a superposition of states with the dominant state having a frequency
$\sim K$. Therefore, up to an accuracy of $r_{a}$, the WKB approximate
solution to the differential Eq\@.~(\ref{eq:diff_eqn1}) can be
given as
\begin{equation}
y_{{\rm WKB}}\approx\sum_{\pm}\frac{c_{\pm}}{\sqrt{K}}\exp\left(\pm i\int^{\eta}d\eta K\right)+O(r_{a})
\end{equation}
where 
\begin{equation}
r_{a}\approx\frac{A}{2f^{2}}\times\max\left(\frac{1}{\frac{2K}{f}-1},\frac{1}{\frac{2K}{f}+1}\right).
\end{equation}
Note that as long as $r_{a}\ll1$, the dominant frequency of the WKB
approximation is given by the slow-varying mass parameter such that
the WKB method no longer suffers from any oscillatory divergences.
The above results motivate us to draw following important conclusions.

Given a differential equation of the form
\begin{equation}
\partial_{\eta}^{2}y+\left(K^{2}(\eta)+A(\eta)\cos\left(f\eta\right)\right)y=0
\end{equation}
where $K(\eta)$ and $A(\eta)$ are slow-varying non-oscillatory functions,
the solution to the above differential equation can be approximate
through the following WKB ansatz
\begin{equation}
y_{{\rm WKB}}\approx\sum_{\pm}\frac{c_{\pm}}{\sqrt{K(\eta)}}\exp\left(\pm i\int^{\eta}d\eta K(\eta)\right)+O\left(\frac{A(\eta)}{2f^{2}}\times\max\left[\frac{1}{\frac{2K(\eta)}{f}-1},\frac{1}{\frac{2K(\eta)}{f}+1}\right]\right)\label{eq:mainapprox}
\end{equation}
if 
\begin{equation}
2A(\eta)/f^{2}\ll1,\mbox{ and }A(\eta)/f<K
\end{equation}
where 
\begin{equation}
\partial_{\eta}K(\eta)\ll K^{2}(\eta).
\end{equation}
For $f\gg K$, then the solution exhibits a large hierarchy in states
such that the system can be described as a superposition of IR and
UV states 
\begin{equation}
y\approx y_{{\rm IR}}+y_{{\rm UV}}
\end{equation}
with the corresponding frequencies as
\begin{align}
{\rm freq}_{{\rm IR}} & \approx K\\
{\rm freq}_{{\rm UV}} & \approx f
\end{align}
and the amplitudes
\begin{equation}
\left|\frac{y_{{\rm UV}}}{y_{{\rm IR}}}\right|_{\eta}\approx\frac{A(\eta)}{2f^{2}}\ll1.
\end{equation}

\section{Late time behavior and $M$ cutoff\label{sec:mu_cutoff}}

In this Appendix we discuss the asymptotic (late-time) behavior of
the background radial field as it settles to its stable vacuum and
explore the associated mass parameter $M$-dependence. We will see
that as the radial field settles to its vacuum, the system can be
characterized as an oscillator with various damping. The damping characteristic
is determined by $M/M_{c}$ where the critical value $M_{c}$ is what
we construct below

To study the late-time behavior of the radial field as it settles
to $f_{{\rm PQ}},$we consider displacements of the radial field around
$f_{{\rm PQ}}$parameterized as
\begin{equation}
\Gamma_{0}(\eta)=f_{{\rm PQ}}\left(1+r(\eta)\right)
\end{equation}
and substitute into the EoM for the radial field $\Gamma_{0}$ given
in Eq.~(\ref{eq:Gamma0-EoM}) to obtain an EoM for the function $r(\eta)$:
\begin{equation}
\partial_{\eta}^{2}r+2\frac{\partial_{\eta}a}{a}\partial_{\eta}r+2M^{2}a^{2}\left(-1+\left(1+r\right)^{2}-\frac{1}{2M^{2}a^{6}}\left(\frac{L/f_{{\rm PQ}}^{2}}{\left(1+r\right)^{2}}\right)^{2}\right)\left(1+r\right)=0.
\end{equation}
In the limit where $\Gamma_{0}\rightarrow f_{{\rm PQ}}$, the angular
velocity term can be neglected since it decays as $La^{-2}$ and so
we reduce the above expression to 
\begin{equation}
\partial_{\eta}^{2}r+2\frac{\partial_{\eta}a}{a}\partial_{\eta}r+2M^{2}a^{2}\left(-1+\left(1+r\right)^{2}\right)\left(1+r\right)\approx0.
\end{equation}
For small displacements $r\ll1$,
\begin{equation}
\partial_{\eta}^{2}r+2\frac{\partial_{\eta}a}{a}\partial_{\eta}r+4M^{2}a^{2}r\approx0.
\end{equation}
The above differential equation has the solution 
\begin{equation}
\lim_{\Gamma_{0}\rightarrow f_{{\rm PQ}}}r(\eta)\approx c_{1}\eta^{\frac{3}{2}\left(1-\left(M/H\right)\sqrt{\frac{1}{\left(M/H\right)^{2}}-\frac{16}{9}}\right)}+c_{2}\eta^{\frac{3}{2}\left(1+\left(M/H\right)\sqrt{\frac{1}{\left(M/H\right)^{2}}-\frac{16}{9}}\right)}\label{eq:Rsol}
\end{equation}
From the above asymptotic solution, we infer that
\begin{align}
M_{c}= & \frac{3}{4}H\label{eq:mu_critical-1}
\end{align}
is a critical value that characterizes the damped oscillations of
the radial field around the vacuum. For $M>M_{c}$, the argument of
the exponential in Eq.~(\ref{eq:Rsol}) obtains an imaginary part
and the radial field $\Gamma_{0}$ behaves as an underdamped oscillator.
For values of $M\ll M_{c}$, the radial field is overdamped and settles
to the vacuum as
\begin{equation}
r(\eta)\rightarrow c_{1}\eta^{\frac{3}{4}\left(\frac{M}{M_{c}}\right)^{2}}.
\end{equation}

In the underdamped case which occurs when $M>M_{c}$, the radial field
oscillates around the minimum at $f_{{\rm PQ}}$ until the oscillations
decay away. These oscillations are characterized by the expression:
\begin{align}
r(\eta) & \approx\eta^{\frac{3}{2}}\left(c_{1}\eta^{-i\nu}+c_{2}\eta^{+i\nu}\right)\\
 & \approx\eta^{\frac{3}{2}}\left(c_{1}\cos\left(\nu\ln|\eta|\right)+c_{2}\sin\left(\nu\ln|\eta|\right)\right)
\end{align}
where 
\begin{equation}
\nu=\frac{3/2}{M_{c}}\sqrt{M^{2}-M_{c}^{2}}
\end{equation}
is real.

\section{Details of quantization\label{sec:Details-of-quantization}}

Here we present the details of the quantization of the radial-axion
system in the presence of large phase angular momentum background
classical solution. The nontriviality will be coming from the nonvanishing
of the cross-commutator $[\partial_{\eta}\delta X,\partial_{\eta}Y]\propto\partial_{\eta}\theta_{0}$
even though $[\delta X,\delta Y]=0$. This quantization is what allows
us to compute the sound speed and the vacuum structure rigorously.

In terms of the linear order field fluctuations $\delta\phi^{n}=\left(\delta\Gamma,\delta\chi\right)$
where $\delta\chi=\Gamma_{0}\delta\theta$, we derive the EoM from
the action in Eq.~(\ref{eq:action}) using the Euler-Lagrange equation
\begin{equation}
\partial_{\mu}\frac{\partial\mathcal{L}_{2}}{\partial\left(\partial_{\mu}\phi^{n}\right)}=\frac{\partial\mathcal{L}_{2}}{\partial\phi^{n}}
\end{equation}
where $\mathcal{L}_{2}$ is the component of the Lagrangian which
is quadratic order in linear perturbations. The EoM for $\delta\Gamma$
and $\delta\chi$ are obtained from the above expression as
\begin{equation}
\partial_{\eta}^{2}\delta\Gamma-a^{-2}\partial_{i}^{2}\delta\Gamma+2\frac{\partial_{\eta}a}{a}\partial_{\eta}\delta\Gamma-2\partial_{\eta}\theta_{0}\partial_{\eta}\delta\chi+\left(-2M^{2}a^{2}+3\lambda\Gamma_{0}^{2}a^{2}-\left(\partial_{\eta}\theta_{0}\right)^{2}\right)\delta\Gamma+2\partial_{\eta}\theta_{0}\frac{\partial_{\eta}\Gamma_{0}}{\Gamma_{0}}\delta\chi=0,\label{eq:radial}
\end{equation}
\begin{equation}
\partial_{\eta}^{2}\delta\chi-a^{-2}\partial_{i}^{2}\delta\chi+2\frac{\partial_{\eta}a}{a}\partial_{\eta}\delta\chi+2\partial_{\eta}\theta_{0}\partial_{\eta}\delta\Gamma+\left(-2M^{2}a^{2}+\lambda\Gamma_{0}^{2}a^{2}-\left(\partial_{\eta}\theta_{0}\right)^{2}\right)\delta\chi-2\partial_{\eta}\theta_{0}\frac{\partial_{\eta}\Gamma_{0}}{\Gamma_{0}}\delta\Gamma=0.\label{eq:axial}
\end{equation}
Eqs.~(\ref{eq:radial}) and (\ref{eq:axial}) form a system of coupled
ODEs and can be expressed compactly as
\begin{align}
\partial_{\eta}^{2}\delta\phi^{n}-\partial_{i}^{2}\delta\phi^{n}+2\frac{\partial_{\eta}a}{a}\partial_{\eta}\delta\phi^{n}+\kappa^{nm}\partial_{\eta}\phi^{m}+\left(\mathcal{M}^{2}\right)^{nm}\delta\phi^{m} & =0\label{eq:EoM-dphin}
\end{align}
where
\begin{equation}
\kappa^{nm}=\left(\begin{array}{cc}
0 & -2\partial_{\eta}\theta_{0}\\
2\partial_{\eta}\theta_{0} & 0
\end{array}\right),
\end{equation}
and
\begin{equation}
\left(\mathcal{M}^{2}\right)^{nm}=\left(\begin{array}{cc}
a^{2}\left(-2M^{2}+3\lambda\Gamma_{0}^{2}\right)-\left(\partial_{\eta}\theta_{0}\right)^{2} & 2\partial_{\eta}\theta_{0}\frac{\partial_{\eta}\Gamma_{0}}{\Gamma_{0}}\\
-2\partial_{\eta}\theta_{0}\frac{\partial_{\eta}\Gamma_{0}}{\Gamma_{0}} & a^{2}\left(-2M^{2}+\lambda\Gamma_{0}^{2}\right)-\left(\partial_{\eta}\theta_{0}\right)^{2}
\end{array}\right).
\end{equation}
Note that the linear perturbations $\delta\phi^{n}$ are kinetically
coupled through the coefficient $\partial_{\eta}\theta_{0}$. We will
refer to all scenarios where $\partial_{\eta}\theta_{0}\gg H^{2}$
as strongly coupled. By defining new field variables $\delta\psi=a\delta\phi^{n}=\left(a\delta\Gamma,a\delta\chi\right)\equiv\left(\delta Y,\delta X\right)$,
we can rewrite the above system of equations as
\begin{align}
\partial_{\eta}^{2}\delta\psi^{n}-\partial_{i}^{2}\delta\psi^{n}+\kappa^{nm}\partial_{\eta}\psi^{m}+\left(\mathcal{M}^{2}\right)^{nm}\delta\psi^{m} & =0\label{eq:EoM-dYdX}
\end{align}
where we modify $\left(\mathcal{M}^{2}\right)^{nm}$ as
\begin{align}
\left(\mathcal{M}^{2}\right)^{nm} & =\left(\begin{array}{cc}
-2M^{2}a^{2}+3\lambda Y_{0}^{2}-\left(\partial_{\eta}\theta_{0}\right)^{2}-\frac{\partial_{\eta}^{2}a}{a} & 2\partial_{\eta}\theta_{0}\left(\frac{\partial_{\eta}Y_{0}}{Y_{0}}\right)\\
-2\partial_{\eta}\theta_{0}\left(\frac{\partial_{\eta}Y_{0}}{Y_{0}}\right) & -2M^{2}a^{2}+\lambda Y_{0}^{2}-\left(\partial_{\eta}\theta_{0}\right)^{2}-\frac{\partial_{\eta}^{2}a}{a}
\end{array}\right)
\end{align}
and $Y_{0}=a\Gamma_{0}$.

We will quantize this system of coupled fields $\delta\psi^{n}\equiv\left(\delta Y,\delta X\right)^{n}$
using the commutator relations defined in Eq.~(\ref{eq:picommutator}).
From the Lagrangian, we find the canonical momenta as
\begin{equation}
\pi^{1}=\frac{\partial\mathcal{L}_{2}}{\partial\left(\partial_{\eta}\delta Y\right)}=-\partial^{0}\delta Y=\partial_{\eta}\delta Y,\label{eq:pi-1}
\end{equation}
and
\begin{align}
\pi^{2} & =\frac{\partial\mathcal{L}_{2}}{\partial\left(\partial_{\eta}\delta X\right)}=\partial_{\eta}\delta X+2\delta Y\partial_{\eta}\theta_{0}.\label{eq:pi-2}
\end{align}
Hence, we arrive at the following commutator expressions for the fields
$\delta\psi^{n}$ and their time-derivatives $\partial_{\eta}\delta\psi^{n}$:
\begin{align}
\left[\delta\psi^{n},\delta\psi^{m}\right] & =0,\nonumber \\
\left[\delta\psi^{n},\partial_{\eta}\delta\psi^{m}\right] & =i\delta^{nm}\delta^{(3)}\left(\vec{x}-\vec{y}\right)\nonumber \\
\left[\partial_{\eta}\delta\psi^{n},\partial_{\eta}\delta\psi^{m}\right] & =i\delta^{(3)}\left(\vec{x}-\vec{y}\right)\left[\begin{array}{cc}
0 & 2\partial_{\eta}\theta_{0}\\
-2\partial_{\eta}\theta_{0} & 0
\end{array}\right]\label{eq:commutator-2}
\end{align}
which is remarkable since $[\delta X,\delta Y]=0$ while $[\partial_{\eta}\delta X,\partial_{\eta}\delta Y]\neq0$.
As expected in the decoupling limit when $\partial_{\eta}\theta_{0}\rightarrow0$,
the kinetic cross commutators vanish.

Next we write the most general solution for $\delta\psi^{n}$ in terms
of time-independent non-Hermitian ladder operators $a_{k}^{nr}$ and
mode function $h_{k}^{nr}$ as
\begin{equation}
\delta\psi^{n}\left(\eta,\vec{x}\right)=\int\frac{d^{3}k}{(2\pi)^{3/2}}\delta\psi^{n}(\eta,\vec{k})e^{i\vec{k}\cdot\vec{x}}=\sum_{r}\int\frac{d^{3}k}{(2\pi)^{3/2}}\left(a_{\vec{k}}^{r}h_{k}^{nr}(\eta)+a_{-\vec{k}}^{r\dagger}h_{k}^{nr*}(\eta)\right)e^{i\vec{k}\cdot\vec{x}}\label{eq:psi}
\end{equation}
where $n$ is the flavor index and $r$ counts the number of distinct
frequency solutions. The time-derivative of the field is 
\begin{equation}
\partial_{\eta}\delta\psi^{n}=\sum_{r}\int\frac{d^{3}k}{(2\pi)^{3/2}}\left(a_{\vec{k}}^{r}\partial_{\eta}h_{k}^{nr}(\eta)+a_{-\vec{k}}^{r\dagger}\partial_{\eta}h_{k}^{nr*}(\eta)\right)e^{i\vec{k}\cdot\vec{x}}.
\end{equation}
The ladder operators $a_{\vec{k}}^{r}$ satisfy relation
\begin{equation}
[a_{\vec{k}}^{r},a_{\vec{p}}^{s\dagger}]=F^{rs}\delta^{(3)}(\vec{k}-\vec{p})
\end{equation}
where the coefficients $F^{rs}$ must be determined by solving for
mode function $h_{k}^{nr}$ and using canonical commutator relation
given in Eq.~(\ref{eq:commutator-2}). Substituting our general solution
from Eq.~(\ref{eq:psi}) into Eq.~(\ref{eq:EoM-dYdX}) we obtain
\begin{align}
\left[\delta^{nj}\partial_{\eta}^{2}+\kappa^{nj}\partial_{\eta}+\left(\mathcal{W}^{2}\right)^{nj}\right]h_{k}^{jm}(\eta) & =0\label{eq:orig_eom}
\end{align}
where
\begin{equation}
\kappa^{nj}=\left(\begin{array}{cc}
0 & -2\partial_{\eta}\theta_{0}\\
2\partial_{\eta}\theta_{0} & 0
\end{array}\right)
\end{equation}
\begin{align}
\left(\mathcal{W}^{2}\right)^{nj} & =\left(\begin{array}{cc}
k^{2}+\left(-2M^{2}a^{2}+3\lambda Y_{0}^{2}\right)-\left(\partial_{\eta}\theta_{0}\right)^{2}-\frac{\partial_{\eta}^{2}a}{a} & 2\partial_{\eta}\theta_{0}\left(\frac{\partial_{\eta}Y_{0}}{Y_{0}}\right)\\
-2\partial_{\eta}\theta_{0}\left(\frac{\partial_{\eta}Y_{0}}{Y_{0}}\right) & k^{2}+\left(-2M^{2}a^{2}+\lambda Y_{0}^{2}\right)-\left(\partial_{\eta}\theta_{0}\right)^{2}-\frac{\partial_{\eta}^{2}a}{a}
\end{array}\right).
\end{align}

\subsection{Normal modes}

We will now solve the system of equations given in Eq.~(\ref{eq:orig_eom})
during the conformal regime. Hence, we propose the following ansatz
\begin{align}
\left(\begin{array}{c}
h_{k}^{1r}(\eta)\\
h_{k}^{2r}(\eta)
\end{array}\right) & =\sum_{j}c_{j}\delta_{j}^{r}\left(\begin{array}{c}
A_{j}^{1}\\
A_{j}^{2}
\end{array}\right)e^{-i\omega_{j}\eta}\label{eq:ansatz}
\end{align}
for time-independent mode vectors $\left(A_{j}^{1},A_{j}^{2}\right)$.
Substituting this ansatz into Eq.~(\ref{eq:orig_eom}) we obtain
\begin{align}
\sum_{r}\left[\delta^{nj}\partial_{\eta}^{2}+\kappa^{nj}\partial_{\eta}+\left(\mathcal{W}^{2}\right)^{nj}\right]\left(\begin{array}{c}
A_{r}^{1}\\
A_{r}^{2}
\end{array}\right)e^{-i\omega_{r}\eta} & =0,
\end{align}
which is rewritten as
\begin{equation}
\left(\begin{array}{cc}
b+2\lambda Y_{0}^{2} & \frac{2L}{Y_{0}^{2}}\left(i\omega_{r}+\frac{\partial_{\eta}Y_{0}}{Y_{0}}\right)\\
\frac{2L}{Y_{0}^{2}}\left(-i\omega_{r}-\frac{\partial_{\eta}Y_{0}}{Y_{0}}\right) & b
\end{array}\right)\left(\begin{array}{c}
A_{r}^{1}\\
A_{r}^{2}
\end{array}\right)=0\label{eq:big_system}
\end{equation}
where we have replaced $\partial_{\eta}\theta_{0}$ with the conserved
angular momentum $L$ from Eq.~(\ref{eq:Ldef}) and $b=-\omega_{r}^{2}+k^{2}+\left(-2M^{2}a^{2}+\lambda Y_{0}^{2}\right)-\left(\frac{L}{Y_{0}^{2}}\right)^{2}-\frac{\partial_{\eta}^{2}a}{a}$.

Since we solve Eq.~(\ref{eq:orig_eom}) at an early time $\eta_{i}\rightarrow-\infty$
and in the conformal limit $Y_{0}\approx Y_{c}=\left(L^{2}/\lambda\right)^{1/6}$
and $\partial_{\eta}\theta_{0}=L/Y_{0}^{2}$, we can neglect any small
amplitude oscillations of the background radial field. In the conformal
limit $\lambda Y_{0}^{2}\gg M^{2}a^{2}\sim H^{2}a^{2}\sim\partial_{\eta}^{2}a/a$,
and hence we arrive at the reduced expression for Eq.~(\ref{eq:big_system})
given as
\begin{equation}
\left(\begin{array}{cc}
-\omega_{r}^{2}+k^{2}+2\lambda Y_{c}^{2} & i\frac{2L}{Y_{c}^{2}}\omega_{r}\\
-i\frac{2L}{Y_{c}^{2}}\omega_{r} & -\omega_{r}^{2}+k^{2}
\end{array}\right)\left(\begin{array}{c}
A_{r}^{1}\\
A_{r}^{2}
\end{array}\right)=0.\label{eq:reduced_matrix}
\end{equation}

We note that in the conformal limit, our system is defined by time-independent
coefficients. The distinct ``real'' frequency solutions obtained
by solving\footnote{Obtained by equating the determinant of the coefficient matrix in
Eq.~(\ref{eq:reduced_matrix}) to zero and solving for $\omega$} Eq.~(\ref{eq:reduced_matrix}) are
\begin{equation}
\omega_{r}=\{\omega_{--},\omega_{+-},\omega_{-+},\omega_{++}\}
\end{equation}
where
\begin{equation}
\omega_{s_{1}s_{2}}\equiv s_{1}\sqrt{k^{2}+3\lambda Y_{c}^{2}+s_{2}Y_{c}\sqrt{\lambda\left(4k^{2}+9\lambda Y_{c}^{2}\right)}},\label{eq:freq-1}
\end{equation}
and $s_{1,2}\in[+,-]$. In the IR limit 
\begin{equation}
1\ll k^{2}\ll\lambda Y_{c}^{2},
\end{equation}
the two distinct frequency squared
\begin{equation}
\omega_{\pm-}^{2}\approx\frac{k^{2}}{3}+O\left(\frac{k^{4}}{\lambda Y_{c}^{2}}\right),
\end{equation}
and
\begin{equation}
\omega_{\pm+}^{2}\approx6\lambda Y_{c}^{2}+\frac{5k^{2}}{3}+O\left(\frac{k^{4}}{\lambda Y_{c}^{2}}\right)
\end{equation}
correspond to low and high frequency solutions and are separated by
the large $O\left(\lambda Y_{c}^{2}/k^{2}\right)$ hierarchy. In the
UV limit,
\begin{equation}
\lim_{k\gg\lambda Y_{c}^{2}}\omega_{\pm\pm}^{2}\rightarrow k^{2}
\end{equation}
and the two frequency solutions become degenerate. 

We write the full mode function $h^{nr}$ as
\begin{align}
\left(\begin{array}{c}
h_{k}^{1r}(\eta)\\
h_{k}^{2r}(\eta)
\end{array}\right) & =c_{++}\delta_{++}^{r}\left(\begin{array}{c}
A_{++}^{1}\\
A_{++}^{2}
\end{array}\right)e^{-i\omega_{++}\eta}+\;c_{+-}\delta_{+-}^{r}\left(\begin{array}{c}
A_{+-}^{1}\\
A_{+-}^{2}
\end{array}\right)e^{-i\omega_{+-}\eta}\nonumber \\
 & +c_{-+}\delta_{-+}^{r}\left(\begin{array}{c}
A_{-+}^{1}\\
A_{-+}^{2}
\end{array}\right)e^{-i\omega_{-+}\eta}+\;c_{--}\delta_{--}^{r}\left(\begin{array}{c}
A_{--}^{1}\\
A_{--}^{2}
\end{array}\right)e^{-i\omega_{--}\eta}
\end{align}
where $r\in\left[++,+-,-+,--\right]$ counts distinct frequencies
given by Eq.~(\ref{eq:freq-1}). The normal vectors corresponding
to each frequency are given by the ratios
\begin{align}
\frac{A_{++}^{2}}{A_{++}^{1}} & =-\frac{A_{-+}^{2}}{A_{-+}^{1}}=i\left(\frac{-2\left(\frac{L}{Y_{c}^{2}}\right)\omega_{++}}{\frac{1}{2}\left(\omega_{++}^{2}-\omega_{+-}^{2}\right)+\left(\lambda Y_{c}^{2}\right)+2\left(\frac{L}{Y_{c}^{2}}\right)^{2}}\right),\label{eq:A++-1}\\
\frac{A_{+-}^{2}}{A_{+-}^{1}} & =-\frac{A_{--}^{2}}{A_{--}^{1}}=i\left(\frac{2\left(\frac{L}{Y_{c}^{2}}\right)\omega_{+-}}{\frac{1}{2}\left(\omega_{++}^{2}-\omega_{+-}^{2}\right)-\left(\lambda Y_{c}^{2}\right)-2\left(\frac{L}{Y_{c}^{2}}\right)^{2}}\right),\label{eq:A+--1}
\end{align}
which are purely ``imaginary''. Hence, we can rewrite the solution
for the mode function as
\begin{align}
\left(\begin{array}{c}
h_{k}^{1r}(\eta)\\
h_{k}^{2r}(\eta)
\end{array}\right) & =c_{++}\delta_{++}^{r}\left(\begin{array}{c}
1\\
\frac{A_{++}^{2}}{A_{++}^{1}}
\end{array}\right)e^{-i\omega_{++}\eta}+\;c_{+-}\delta_{+-}^{r}\left(\begin{array}{c}
1\\
\frac{A_{+-}^{2}}{A_{+-}^{1}}
\end{array}\right)e^{-i\omega_{+-}\eta}+\nonumber \\
 & +c_{-+}\delta_{-+}^{r}\left(\begin{array}{c}
1\\
\frac{A_{++}^{2}}{A_{++}^{1}}
\end{array}\right)^{*}e^{i\omega_{++}\eta}+\;c_{--}\delta_{--}^{r}\left(\begin{array}{c}
1\\
\frac{A_{+-}^{2}}{A_{+-}^{1}}
\end{array}\right)^{*}e^{i\omega_{+-}\eta}\label{eq:e1}
\end{align}
where the frequencies $\omega_{++}$ are constant, real and positive
and the ratios $A_{r}^{2}/A_{r}^{1}$ are purely imaginary. In the
decoupling limit where $L\rightarrow0$ the kinetic terms mixing $\delta\Gamma$
and $\delta\chi$ vanish and the decoupled solution becomes :
\begin{align*}
\lim_{L\rightarrow0}\left(\begin{array}{c}
h_{k}^{1r}(\eta)\\
h_{k}^{2r}(\eta)
\end{array}\right) & =c_{++}\delta_{++}^{r}\left(\begin{array}{c}
1\\
0
\end{array}\right)e^{-i\omega_{++}\eta}+c_{-+}\delta_{-+}^{r}\left(\begin{array}{c}
1\\
0
\end{array}\right)e^{-i\omega_{-+}\eta}\\
 & +c_{+-}\delta_{+-}^{r}\left(\begin{array}{c}
0\\
1
\end{array}\right)e^{-i\omega_{+-}\eta}+c_{--}\delta_{--}^{r}\left(\begin{array}{c}
0\\
1
\end{array}\right)e^{-i\omega_{--}\eta},\\
h_{k}^{1r}(\eta) & =c_{++}\delta_{++}^{r}e^{-i\omega_{++}\eta}+c_{-+}\delta_{-+}^{r}e^{-i\omega_{-+}\eta},\\
h_{k}^{2r}(\eta) & =c_{+-}\delta_{+-}^{r}e^{-i\omega_{+-}\eta}+c_{--}\delta_{--}^{r}e^{-i\omega_{--}\eta}.
\end{align*}
where the decoupled ``instantaneous'' frequencies\footnote{In the decoupled limit, the background solution for the radial field
is not conformal due to the quartic potential which will induce large
amplitude oscillations. Hence, we cannot assume that $Y_{0}\approx{\rm constant}$.
Therefore, we give instantaneous frequencies.} are
\begin{equation}
\lim_{L\rightarrow0}\omega_{s_{1}s_{2}}\equiv s_{1}\sqrt{k^{2}+2\lambda Y_{c}^{2}+s_{2}\left(\lambda Y_{c}^{2}\right)}
\end{equation}
for
\begin{align}
\omega_{\pm+} & =\pm\sqrt{k^{2}+3\lambda Y_{c}^{2}}\\
\omega_{\pm-} & =\pm\sqrt{k^{2}+\lambda Y_{c}^{2}}.
\end{align}
Substituting our solution for $h_{k}^{nr}$ from Eq.~(\ref{eq:e1})
into the expression for $\delta\psi^{n}$ we obtain
\begin{align}
\delta\psi^{n} & =\int\frac{d^{3}pe^{i\vec{p}\cdot\vec{x}}}{(2\pi)^{3/2}}\left[a_{\vec{p}}^{++}c_{++}V_{++}^{n}e^{-i\omega_{++}\eta}+a_{\vec{p}}^{+-}c_{+-}V_{+-}^{n}e^{-i\omega_{+-}\eta}\right.  \nonumber \\
 & \qquad \qquad  \qquad + \left.a_{-\vec{p}}^{++\dagger}c_{++}^{*}V_{++}^{n*}e^{i\omega_{++}\eta}+a_{-\vec{p}}^{+-\dagger}c_{+-}^{*}V_{+-}^{n*}e^{i\omega_{+-}\eta}\right]\label{eq:final_psi_solution}
\end{align}
where
\begin{equation}
V_{++}^{n}=\left(\begin{array}{c}
1\\
\mathcal{R}_{++}
\end{array}\right),\,\,V_{+-}^{n}=\left(\begin{array}{c}
1\\
\mathcal{R}_{+-}
\end{array}\right),\,\,V_{-+}^{n}=V_{++}^{*}=\left(\begin{array}{c}
1\\
-\mathcal{R}_{++}
\end{array}\right),\,\,V_{--}^{n}=V_{+-}^{*}=\left(\begin{array}{c}
1\\
-\mathcal{R}_{+-}
\end{array}\right)\label{eq:Vnr}
\end{equation}
and
\begin{equation}
\mathcal{R}_{++}=\frac{A_{++}^{2}}{A_{++}^{1}},\quad\mathcal{R}_{+-}=\frac{A_{+-}^{2}}{A_{+-}^{1}}.
\end{equation}

\subsection{Ladder algebra}

To evaluate the ladder algebra, we first express the ladder operators:
$a_{\vec{p}}^{++}$, $a_{\vec{p}}^{+-}$ and their conjugates in terms
of the fields $\delta\psi^{n}$ and its conjugate momenta $\pi^{n}$.
To this end, we define 
\begin{equation}
L^{n}[w,\vec{q}]=\frac{1}{(2\pi)^{3/2}}\int d\eta e^{iw\eta}\int d^{3}xe^{-i\vec{q}\cdot\vec{x}}\delta\psi^{n}
\end{equation}
and 
\begin{equation}
N^{n}[w,\vec{q}]\equiv\frac{1}{(2\pi)^{3/2}}\int d\eta e^{iw\eta}\int d^{3}xe^{-i\vec{q}\cdot\vec{x}}\partial_{\eta}\delta\psi^{n}.
\end{equation}
From the above definition and Eq.~(\ref{eq:final_psi_solution}),
we conclude that 
\begin{align}
J^{n}=\int_{-\infty}^{\infty}\frac{dw}{2\pi}L^{n}[w,\vec{q}] & =a_{\vec{q}}^{++}c_{++}V_{++}^{n}+a_{\vec{q}}^{+-}c_{+-}V_{+-}^{n}+a_{-\vec{q}}^{\dagger++}c_{++}^{*}V_{++}^{n*}+a_{-\vec{q}}^{\dagger+-}c_{+-}^{*}V_{+-}^{n*},\label{eq:Jn}\\
J^{n+2}=i\int_{-\infty}^{\infty}\frac{dw}{2\pi}N^{n}[w,\vec{q}] & =a_{\vec{q}}^{++}\omega_{++}c_{++}V_{++}^{n}+a_{\vec{q}}^{+-}\omega_{+-}c_{+-}V_{+-}^{n} \nonumber \\
& -a_{-\vec{q}}^{\dagger++}\omega_{++}c_{++}^{*}V_{++}^{n*}-a_{-\vec{q}}^{\dagger+-}\omega_{+-}c_{+-}^{*}V_{+-}^{n*},\label{eq:Jn+2}
\end{align}
where $n$ is the flavor index of our two-field coupled system and
runs from $1$ to $2$. From Eqs.~(\ref{eq:Jn}) and (\ref{eq:Jn+2})
we setup the following system of equations to solve for the ladder
operators
\begin{equation}
\left(\begin{array}{cccc}
c_{++}V_{++}^{1} & c_{+-}V_{+-}^{1} & c_{++}^{*}V_{++}^{1*} & c_{+-}^{*}V_{+-}^{1*}\\
c_{++}V_{++}^{2} & c_{+-}V_{+-}^{2} & c_{++}^{*}V_{++}^{2*} & c_{+-}^{*}V_{+-}^{2*}\\
\omega_{++}c_{++}V_{++}^{1} & \omega_{+-}c_{+-}V_{+-}^{1} & -\omega_{++}c_{++}^{*}V_{++}^{1*} & -\omega_{+-}c_{+-}^{*}V_{+-}^{1*}\\
\omega_{++}c_{++}V_{++}^{2} & \omega_{+-}c_{+-}V_{+-}^{2} & -\omega_{++}c_{++}^{*}V_{++}^{2*} & -\omega_{+-}c_{+-}^{*}V_{+-}^{2*}
\end{array}\right)\left(\begin{array}{c}
a_{\vec{q}}^{++}\\
a_{\vec{q}}^{+-}\\
a_{-\vec{q}}^{++\dagger}\\
a_{-\vec{q}}^{+-\dagger}
\end{array}\right)=\left(\begin{array}{c}
J^{1}\\
J^{2}\\
J^{3}\\
J^{4}
\end{array}\right).\label{eq:matrix-system_n=00003D1}
\end{equation}
Solving the above equation yields
\begin{align}
a_{\vec{q}}^{++} & =-\frac{\mathcal{R}_{+-}\omega_{+-}J^{1}}{2c_{++}\left(\mathcal{R}_{++}\omega_{++}-\mathcal{R}_{+-}\omega_{+-}\right)}+\frac{\omega_{+-}J^{2}}{2c_{++}\left(\mathcal{R}_{++}\omega_{+-}-\mathcal{R}_{+-}\omega_{++}\right)}\nonumber \\
 & +\frac{J^{4}}{2c_{++}\left(\mathcal{R}_{++}\omega_{++}-\mathcal{R}_{+-}\omega_{+-}\right)}-\frac{\mathcal{R}_{+-}J^{3}}{2c_{++}\left(\mathcal{R}_{++}\omega_{+-}-\mathcal{R}_{+-}\omega_{++}\right)}\nonumber \\
a_{-\vec{q}}^{++\dagger} & =-\frac{\mathcal{R}_{+-}\omega_{+-}J^{1}}{2c_{++}^{*}\left(\mathcal{R}_{++}\omega_{++}-\mathcal{R}_{+-}\omega_{+-}\right)}+\frac{\mathcal{R}_{+-}J^{3}}{2c_{++}^{*}\left(\mathcal{R}_{++}\omega_{+-}-\mathcal{R}_{+-}\omega_{++}\right)}\nonumber \\
 & +\frac{J^{4}}{2c_{++}^{*}\left(\mathcal{R}_{++}\omega_{++}-\mathcal{R}_{+-}\omega_{+-}\right)}-\frac{\omega_{+-}J^{2}}{2c_{++}^{*}\left(\mathcal{R}_{++}\omega_{+-}-\mathcal{R}_{+-}\omega_{++}\right)}\nonumber \\
a_{\vec{q}}^{+-} & =-\frac{\mathcal{R}_{+-}\omega_{+-}J^{1}}{2c_{+-}\left(\mathcal{R}_{+-}\omega_{+-}-\mathcal{R}_{++}\omega_{++}\right)}+\frac{\mathcal{R}_{++}\omega_{+-}J^{2}}{2\mathcal{R}_{+-}c_{+-}\left(\mathcal{R}_{+-}\omega_{++}-\mathcal{R}_{++}\omega_{+-}\right)}\nonumber \\
 & +\frac{J^{4}}{2c_{+-}\left(\mathcal{R}_{+-}\omega_{+-}-\mathcal{R}_{++}\omega_{++}\right)}-\frac{\mathcal{R}_{++}J^{3}}{2c_{+-}\left(\mathcal{R}_{+-}\omega_{++}-\mathcal{R}_{++}\omega_{+-}\right)}+\frac{J^{2}}{2\mathcal{R}_{+-}c_{+-}}+\frac{J^{1}}{2c_{+-}}\nonumber \\
a_{-\vec{q}}^{+-\dagger} & =-\frac{\mathcal{R}_{+-}\omega_{+-}J^{1}}{2c_{+-}^{*}\left(\mathcal{R}_{+-}\omega_{+-}-\mathcal{R}_{++}\omega_{++}\right)}+\frac{\mathcal{R}_{++}J^{3}}{2c_{+-}^{*}\left(\mathcal{R}_{+-}\omega_{++}-\mathcal{R}_{++}\omega_{+-}\right)}\nonumber \\
 & +\frac{J^{4}}{2c_{+-}^{*}\left(\mathcal{R}_{+-}\omega_{+-}-\mathcal{R}_{++}\omega_{++}\right)}-\frac{\mathcal{R}_{++}\omega_{+-}J^{2}}{2\mathcal{R}_{+-}c_{+-}^{*}\left(\mathcal{R}_{+-}\omega_{++}-\mathcal{R}_{++}\omega_{+-}\right)}-\frac{J^{2}}{2\mathcal{R}_{+-}c_{+-}^{*}}+\frac{J^{1}}{2c_{+-}^{*}}\label{eq:a-set}
\end{align}
where we use Eq.~(\ref{eq:Vnr}) to set
\begin{equation}
V_{++}^{1}=V_{+-}^{1}=1\quad\mbox{and,\ensuremath{\quad}}V_{++}^{2}=\mathcal{R}_{++},\quad V_{+-}^{2}=\mathcal{R}_{+-}.
\end{equation}
We can summarize these results as 
\begin{align}
a_{\vec{p}}^{r} & =\sum_{n=1}^{4}U_{n}^{r}J^{n},\\
a_{-\vec{p}}^{r\dagger} & =\sum_{n=1}^{4}W_{n}^{r}J^{n},
\end{align}
where $r=(++,+-)$. The $J^{n}$ and $J^{n+2}$ operators appearing
on the RHS of Eq.~(\ref{eq:matrix-system_n=00003D1}) can be evaluated
as
\begin{align}
J^{n} & \equiv\int_{-\infty}^{\infty}\frac{dw}{2\pi}\left\{ L^{n}[w,\vec{q}]\right\} \\
 & =\int_{-\infty}^{\infty}\frac{dw}{2\pi}\left\{ \frac{1}{(2\pi)^{3/2}}\int d\eta e^{iw\eta}\int d^{3}xe^{-i\vec{q}\cdot\vec{x}}\delta\psi^{n}\right\} \\
 & =\frac{1}{(2\pi)^{3/2}}\int d^{3}xe^{-i\vec{q}\cdot\vec{x}}\delta\psi^{n}(0,\vec{x})
\end{align}
and,
\begin{align}
J^{n+2} & \equiv i\int_{-\infty}^{\infty}\frac{dw}{2\pi}\left\{ N^{n}[w,\vec{q}]\right\} \\
 & =i\int_{-\infty}^{\infty}\frac{dw}{2\pi}\left\{ \frac{1}{(2\pi)^{3/2}}\int d\eta e^{iw\eta}\int d^{3}xe^{-i\vec{q}\cdot\vec{x}}\partial_{\eta}\delta\psi^{n}\right\} \\
 & =i\frac{1}{(2\pi)^{3/2}}\int d^{3}xe^{-i\vec{q}\cdot\vec{x}}\partial_{\eta}\delta\psi^{n}(0,\vec{x})
\end{align}
where we used $\int_{-\infty}^{\infty}\frac{dw}{2\pi}\int d\eta e^{iw\eta}f(\eta)=\int d\eta f(\eta)\int_{-\infty}^{\infty}\frac{dw}{2\pi}e^{iw\eta}=\int d\eta f(\eta)\delta(\eta)=f(0)$.
It follows then that the commutators of $J^{n}$ and $J^{n+2}$ operators
can be obtained from the commutator relations in Eq.~(\ref{eq:commutator-2}).
Hence, 
\begin{align}
\left[J^{1},J^{2}\right] & =0\nonumber \\
\left[J^{1},J^{3}\right] & =-\delta^{(3)}\left(\vec{q}+\vec{q}'\right)\nonumber \\
\left[J^{1},J^{4}\right] & =0.\label{eq:J-commutator}\\
\left[J^{2},J^{3}\right] & =0.\nonumber \\
\left[J^{2},J^{4}\right] & =-\delta^{(3)}\left(\vec{q}+\vec{q}'\right)\nonumber \\
\left[J^{3},J^{4}\right] & =-i2\partial_{\eta}\theta_{0}\delta^{(3)}\left(\vec{q}+\vec{q}'\right)\nonumber 
\end{align}

\subsection{Ladder commutators}

Using the solution for the ladder operators $a_{\vec{p}}^{++},a_{\vec{p}}^{+-}$
and their conjugates as given in Eq.~(\ref{eq:a-set}), and the
commutator algebra of $J$ operators given in Eq.~(\ref{eq:J-commutator}),
we can evaluate the ladder commutators. There are 6 unique combinations
of the commutators for $+\pm$ frequencies that we present below.
We work out the first commutator in detail and leave the remaining
for the readers as an exercise:

\begin{align}
\left[a_{\vec{q}}^{++},a_{-\vec{q}'}^{++\dagger}\right] & =\left[U_{1}^{++}J^{1},W_{2}^{++}J^{2}+W_{3}^{++}J^{3}+W_{4}^{++}J^{4}\right]\nonumber 
  +\left[U_{2}^{++}J^{2},W_{1}^{++}J^{1}+W_{3}^{++}J^{3}+W_{4}^{++}J^{4}\right]\nonumber \\
 & +\left[U_{3}^{++}J^{3},W_{1}^{++}J^{1}+W_{2}^{++}J^{2}+W_{4}^{++}J^{4}\right]\nonumber 
  +\left[U_{4}^{++}J^{4},W_{1}^{++}J^{1}+W_{2}^{++}J^{2}+W_{3}^{++}J^{3}\right]\\
 & =\left(U_{1}^{++}W_{3}^{++}-U_{3}^{++}W_{1}^{++}\right)\left(-1\right)\nonumber \\
 & +\left(U_{2}^{++}W_{4}^{++}-U_{4}^{++}W_{2}^{++}\right)\left(-1\right)+\left(U_{3}^{++}W_{4}^{++}-U_{4}^{++}W_{3}^{++}\right)\left(-i2\partial_{\eta}\theta_{0}\right) \nonumber \\
 & =\frac{\left(\left(\mathcal{R}_{+-}\right)^{2}-1\right)\omega_{+-}+2i\partial_{\eta}\theta_{0}\mathcal{R}_{+-}}{2c_{++}c_{++}^{*}\left(\mathcal{R}_{+-}\omega_{+-}-\mathcal{R}_{++}\omega_{++}\right)\left(\mathcal{R}_{+-}\omega_{++}-\mathcal{R}_{++}\omega_{+-}\right)}\delta^{(3)}\left(\vec{q}+\vec{q}'\right).
\end{align}
\begin{align*}
\left[a_{\vec{q}}^{+-},a_{-\vec{q}'}^{+-\dagger}\right] & =\frac{\left(\left(\mathcal{R}_{++}\right)^{2}-1\right)\omega_{++}+2i\partial_{\eta}\theta_{0}\mathcal{R}_{++}}{2c_{+-}c_{+-}^{*}\left(\mathcal{R}_{+-}\omega_{+-}-\mathcal{R}_{++}\omega_{++}\right)\left(\mathcal{R}_{+-}\omega_{++}-\mathcal{R}_{++}\omega_{+-}\right)}\delta^{(3)}\left(\vec{q}+\vec{q}'\right) \\
\left[a_{\vec{q}}^{++},a_{\vec{q}'}^{+-}\right] & =\delta^{(3)}\left(\vec{q}+\vec{q}'\right)\frac{\left(\mathcal{R}_{++}\mathcal{R}_{+-}+1\right)\omega_{+-}-\left(\mathcal{R}_{++}\mathcal{R}_{+-}+1\right)\omega_{++}+2i\partial_{\eta}\theta_{0}\left(\mathcal{R}_{++}-\mathcal{R}_{+-}\right)}{4c_{+-}c_{++}\left(\mathcal{R}_{+-}\omega_{+-}-\mathcal{R}_{++}\omega_{++}\right)\left(\mathcal{R}_{+-}\omega_{++}-\mathcal{R}_{++}\omega_{+-}\right)}\nonumber\\
 & =0 \\
\left[a_{\vec{q}}^{1++},a_{-\vec{q}}^{2++\dagger}\right] & =\delta^{(3)}\left(\vec{q}+\vec{q}'\right)\frac{\left(1-\mathcal{R}_{++}\mathcal{R}_{+-}\right)\omega_{+-}+\left(1-\mathcal{R}_{++}\mathcal{R}_{+-}\right)\omega_{++}-2i\partial_{\eta}\theta_{0}\left(\mathcal{R}_{++}+\mathcal{R}_{+-}\right)}{4c_{++}c_{+-}^{*}\left(\mathcal{R}_{+-}\omega_{+-}-\mathcal{R}_{++}\omega_{++}\right)\left(\mathcal{R}_{+-}\omega_{++}-\mathcal{R}_{++}\omega_{+-}\right)}\nonumber \\ 
 & =0 \\
\left[a_{\vec{q}}^{+-},a_{-\vec{q}'}^{++\dagger}\right] &=0 \\
\left[a_{-\vec{q}}^{++\dagger},a_{-\vec{q}'}^{+-\dagger}\right] &=0\;.
\end{align*}
Hence for the operator set defined as
\begin{equation}
u_{\vec{q}}^{n}=\left(a_{\vec{q}}^{++},a_{-\vec{q}}^{++\dagger},a_{\vec{q}}^{+-},a_{-\vec{q}}^{+-\dagger}\right)^{n}
\end{equation}
the commutators are
\begin{equation}
\left[u_{\vec{q}}^{n},u_{\vec{q}'}^{m}\right]=\left[\begin{array}{cccc}
0 & 1 & 0 & 0\\
-1 & 0 & 0 & 0\\
0 & 0 & 0 & 1\\
0 & 0 & -1 & 0
\end{array}\right]^{nm}\delta^{(3)}\left(\vec{q}+\vec{q}'\right)
\end{equation}
where we set the coefficients $c_{+\pm}$ through the expressions
\begin{align}
c_{++}c_{++}^{*} & =-\frac{\left(1-\left(\mathcal{R}_{+-}\right)^{2}\right)\omega_{+-}-2i\partial_{\eta}\theta_{0}\mathcal{R}_{+-}}{2\left(\mathcal{R}_{+-}\omega_{+-}-\mathcal{R}_{++}\omega_{++}\right)\left(\mathcal{R}_{+-}\omega_{++}-\mathcal{R}_{++}\omega_{+-}\right)},\label{eq:cpp and cpm-1}\\
c_{+-}c_{+-}^{*} & =-\frac{\left(1-\left(\mathcal{R}_{++}\right)^{2}\right)\omega_{++}-2i\partial_{\eta}\theta_{0}\mathcal{R}_{++}}{2\left(\mathcal{R}_{+-}\omega_{+-}-\mathcal{R}_{++}\omega_{++}\right)\left(\mathcal{R}_{+-}\omega_{++}-\mathcal{R}_{++}\omega_{+-}\right)}.
\end{align}
Clearly, the ladder operators associated with distinct frequencies
commute with each other. Hence, we can define a common vacuum state
$\left|0\right\rangle $ which is annihilated simultaneously by both
$a_{\vec{q}}^{++}$ and $a_{\vec{q}}^{+-}$ for all $\vec{q}$:
\begin{align}
a_{\vec{q}}^{++}\left|0\right\rangle  & =0,\\
a_{\vec{q}}^{+-}\left|0\right\rangle  & =0.
\end{align}
Note that the normal mode vectors $V_{+\pm}^{n}$ associated with
the corresponding ladder operators $a^{+\pm}$ are not orthogonal.

\subsection{Hamiltonian\label{sec:Hamiltonian}}

We conclude our discussion on the quantization of the coupled system
by evaluating its Hamiltonian. We show that in the conformal limit,
the normal mode solution given in Eq.~(\ref{eq:e1}) diagonalizes
the Hamiltonian such that the vacuum state $\left|0\right\rangle $
is the state of minimum energy.

Hence, let us consider the Hamiltonian density defined through the
expression
\begin{equation}
\mathcal{H}_{2}=\sum_{n}\pi^{n}\partial_{\eta}\psi^{n}-\mathcal{L}_{2}
\end{equation}
and the Hamiltonian given by
\begin{equation}
H(\eta)=\int d^{3}x\mathcal{H}(\eta,\vec{x}).
\end{equation}
In the time-independent conformal regime when $Y_{0}=Y_{c}$ and $\partial_{\eta}\theta_{0}$
are constants, the Hamiltonian simplifies to 
\begin{align}
\mathcal{H} & =\frac{1}{2}\left(\partial_{\eta}\delta Y\right)^{2}+\frac{1}{2}\left(\partial_{\eta}\delta X\right)^{2}+\frac{1}{2}\left(\partial_{i}\delta\Gamma\right)^{2}+\frac{1}{2}\left(\partial_{i}\delta\chi\right)^{2}\nonumber \\
 & -\frac{1}{2}\left(\delta Y\right)^{2}\left(\partial_{\eta}\theta_{0}\right)^{2}+\left(\frac{3\lambda}{2}Y_{c}^{2}\right)\left(\delta Y\right)^{2}.\label{eq:reduced_Hamiltonian}
\end{align}
We introduce Fourier notation
\begin{align}
\delta\psi_{\vec{k}}^{n}(\eta) & \equiv\int d^{3}xe^{-i\vec{k}\cdot\vec{x}}\delta\psi^{n}(\eta,\vec{x}),\\
\delta\psi^{n}(\eta,\vec{x}) & \equiv\int\frac{d^{3}k}{\left(2\pi\right)^{3}}e^{i\vec{k}\cdot\vec{x}}\delta\psi_{\vec{k}}^{n}(\eta),\\
\partial_{\eta}\delta\psi^{n}(\eta,\vec{x}) & \equiv\int\frac{d^{3}k}{\left(2\pi\right)^{3}}e^{i\vec{k}\cdot\vec{x}}\partial_{\eta}\delta\psi^{n}(\eta),\\
\partial_{i}\psi^{n}(\eta,\vec{x}) & \equiv\int\frac{d^{3}k}{\left(2\pi\right)^{3}}i\vec{k}e^{i\vec{k}\cdot\vec{x}}\delta\psi_{\vec{k}}^{n}(\eta).
\end{align}
Taking the Fourier transform of the fields we write the Hamiltonian
as
\begin{align*}
H & =\int d^{3}x\mathcal{H}\\
 & =\int d^{3}x\int\frac{d^{3}k}{\left(2\pi\right)^{3}}e^{i\vec{k}\cdot\vec{x}}\int\frac{d^{3}q}{\left(2\pi\right)^{3}}e^{i\vec{q}\cdot\vec{x}}\left(A^{mn}\partial_{\eta}\delta\psi_{\vec{k}}^{n}\partial_{\eta}\delta\psi_{\vec{q}}^{m}-\left(\vec{k}\cdot\vec{q}\right)B^{mn}\delta\psi_{\vec{k}}^{n}\delta\psi_{\vec{q}}^{m}+C^{mn}\delta\psi_{\vec{k}}^{n}\delta\psi_{\vec{q}}^{m}\right)\\
 & =\int\frac{d^{3}k}{\left(2\pi\right)^{3}}\int\frac{d^{3}q}{\left(2\pi\right)^{3}}\int d^{3}xe^{i\left(\vec{k}+\vec{q}\right)\cdot\vec{x}}\left(A^{mn}\partial_{\eta}\delta\psi_{\vec{k}}^{n}\partial_{\eta}\delta\psi_{\vec{q}}^{m}-\left(\vec{k}\cdot\vec{q}\right)B^{mn}\delta\psi_{\vec{k}}^{n}\delta\psi_{\vec{q}}^{m}+C^{mn}\delta\psi_{\vec{k}}^{n}\delta\psi_{\vec{q}}^{m}\right)\\
 & =\int\frac{d^{3}k}{\left(2\pi\right)^{3}}\left(A^{mn}\partial_{\eta}\delta\psi_{\vec{k}}^{n}\partial_{\eta}\delta\psi_{-\vec{k}}^{m}+B^{mn}\delta\psi_{\vec{k}}^{n}\delta\psi_{-\vec{k}}^{m}\right)
\end{align*}
where
\begin{align}
A^{mn} & =\frac{1}{2}\left[\begin{array}{cc}
1 & 0\\
0 & 1
\end{array}\right],\\
B^{mn} & =\frac{1}{2}\left[\begin{array}{cc}
\left(k^{2}+3\lambda Y_{c}^{2}-\left(\partial_{\eta}\theta_{0}\right)^{2}\right) & 0\\
0 & k^{2}
\end{array}\right].
\end{align}
Hence, in the conformal limit the Hamiltonian is diagonal in terms
of the fields and its time-derivatives.

Using our general solution for the fields
\begin{equation}
\delta\psi^{n}=\int\frac{d^{3}k}{(2\pi)^{3/2}}\sum_{r}\left(a_{\vec{k}}^{r}h_{k}^{nr}(\eta)+a_{-\vec{k}}^{r\dagger}h_{k}^{nr*}\right)e^{i\vec{k}\cdot\vec{x}}
\end{equation}
it is possible to write
\begin{equation}
\left(A^{mn}\partial_{\eta}\delta\psi_{\vec{k}}^{n}\partial_{\eta}\delta\psi_{-\vec{k}}^{m}+B^{mn}\delta\psi_{\vec{k}}^{n}\delta\psi_{-\vec{k}}^{m}\right)=\sum_{r,s}\left[T_{1}^{rs}a_{\vec{k}}^{r}a_{-\vec{k}}^{s}+T_{2}^{rs}a_{-\vec{k}}^{r\dagger}a_{-\vec{k}}^{s}+T_{3}^{rs}a_{\vec{k}}^{r}a_{\vec{k}}^{s\dagger}+T_{4}^{rs}a_{-\vec{k}}^{r\dagger}a_{\vec{k}}^{s\dagger}\right]
\end{equation}
where the indices $r,s\in\left(++,+-\right)$. Below we evaluate $T_{1}^{rs}$:
\begin{align}
T_{1}^{rs} & =A^{mn}\left(\partial_{\eta}h_{k}^{nr}\right)\left(\partial_{\eta}h_{-k}^{ms}\right)+a^{-2}B^{mn}h_{k}^{nr}h_{-k}^{ms}\nonumber \\
 & =\frac{1}{2}\partial_{\eta}h_{k}^{1r}\partial_{\eta}h_{-k}^{1s}+\frac{1}{2}\left(k^{2}+3\lambda Y_{0}^{2}-\left(\partial_{\eta}\theta_{0}\right)^{2}\right)h_{k}^{1r}h_{-k}^{1s}+\frac{1}{2}\partial_{\eta}h_{k}^{2r}\partial_{\eta}h_{-k}^{2s}+\frac{1}{2}\left(k^{2}\right)h_{k}^{2r}h_{-k}^{2s}.\label{eq:Trs}
\end{align}
Using the mode solution
\begin{equation}
h_{k}^{nr}=c_{r}V_{r}^{n}e^{-i\omega_{r}\eta}\qquad\partial_{\eta}h_{k}^{nr}=-i\omega_{r}c_{r}V_{r}^{n}e^{-i\omega_{r}\eta},
\end{equation}
it follows that 
\begin{align*}
\frac{T_{1}^{rs}}{e^{-i\left(\omega_{r}+\omega_{s}\right)\eta}c_{r}c_{s}} & =\left[-\omega_{r}\omega_{s}+k^{2}+3\lambda Y_{0}^{2}-\left(\partial_{\eta}\theta_{0}\right)^{2}\right]V_{r}^{1}V_{s}^{1}+\left[-\omega_{r}\omega_{s}+k^{2}\right]V_{r}^{2}V_{s}^{2}\\
 & =\left(\omega_{r}-\omega_{s}\right)\left(\omega_{r}V_{r}^{1}V_{s}^{1}-\omega_{s}V_{r}^{2}V_{s}^{2}-i2\partial_{\eta}\theta_{0}V_{s}^{1}V_{r}^{2}\right).
\end{align*}
In the above equation, the expression in the second bracket goes to
when $r\neq s$. Hence, we conclude that $T_{1}^{rs}=0$ for all combinations
of $r,s$. Similar calculations show that $T_{4}^{rs}=0$.

Next, we evaluate $T_{2}^{rs}$:
\begin{align*}
T_{2}^{rs} & =A^{mn}\left(\partial_{\eta}h_{k}^{nr*}\right)\left(\partial_{\eta}h_{k}^{ms}\right)+B^{mn}h_{k}^{nr*}h_{k}^{ms}\\
 & =\frac{1}{2}\left[\partial_{\eta}h_{k}^{1r*}\partial_{\eta}h_{k}^{1s}+\left(k^{2}+3\lambda Y_{c}^{2}-\left(\partial_{\eta}\theta_{0}\right)^{2}\right)h_{k}^{1r*}h_{k}^{1s}\right]\\
 & +\frac{1}{2}\left[\partial_{\eta}h_{k}^{2r*}\partial_{\eta}h_{k}^{2s}+\left(k^{2}\right)h_{k}^{2r*}h_{k}^{2s}\right]\\
\frac{T_{2}^{rs}}{e^{i\left(\omega_{r}-\omega_{s}\right)\eta}c_{r}^{*}c_{s}} & =\left[\omega_{r}\omega_{s}+k^{2}+3\lambda Y_{c}^{2}-\left(\partial_{\eta}\theta_{0}\right)^{2}\right]V_{r}^{1*}V_{s}^{1}+\left[\omega_{r}\omega_{s}+k^{2}\right]V_{r}^{2*}V_{s}^{2}\\
 & =\left(\omega_{r}+\omega_{s}\right)\left(\omega_{r}V_{r}^{1}V_{s}^{1}+\omega_{s}V_{r}^{2*}V_{s}^{2}+i2\partial_{\eta}\theta_{0}V_{s}^{1}V_{r}^{2*}\right)
\end{align*}
which is non zero only when $r=s$. Similarly
\begin{align*}
\frac{T_{3}^{rs}}{e^{-i\left(\omega_{r}-\omega_{s}\right)\eta}c_{r}c_{s}^{*}} & =\left[\omega_{r}\omega_{s}+k^{2}+3\lambda Y_{c}^{2}-\left(\partial_{\eta}\theta_{0}\right)^{2}\right]V_{r}^{1}V_{s}^{1*}+\left[\omega_{r}\omega_{s}+k^{2}\right]V_{r}^{2}V_{s}^{2*}\\
 & =\left(\omega_{r}+\omega_{s}\right)\left(\omega_{r}V_{r}^{1}V_{s}^{1*}+\omega_{s}V_{r}^{2}V_{s}^{2*}-i2\partial_{\eta}\theta_{0}V_{s}^{1*}V_{r}^{2}\right)
\end{align*}
vanishes when $r\neq s$. Hence we find that for the amplitudes $V_{r}^{n}(k,\eta_{i})$,
the coefficients $T_{1,4}^{rs}$ vanish for all combinations of $r,s$.
Meanwhile, we find that $T_{2,3}^{rs}$ are non-zero only when $r=s$.
Thus, the normal frequency solutions corresponding to $\omega_{++}$
and $\omega_{+-}$ diagonalize our Hamiltonian, which we write as
\begin{align*}
H & =\int\frac{d^{3}k}{\left(2\pi\right)^{3}}\left(A^{mn}\partial_{\eta}\delta\psi_{\vec{k}}^{n}\partial_{\eta}\delta\psi_{-\vec{k}}^{m}+B^{mn}\delta\psi_{\vec{k}}^{n}\delta\psi_{-\vec{k}}^{m}\right)\\
 & =\sum_{r=++,+-}\int\frac{d^{3}k}{\left(2\pi\right)^{3}}\left[T_{2}^{rs}a_{-\vec{k}}^{r\dagger}a_{-\vec{k}}^{r}+T_{3}^{rs}a_{\vec{k}}^{r}a_{\vec{k}}^{r\dagger}\right]\\
 & =\sum_{r=++,+-}\int\frac{d^{3}k}{\left(2\pi\right)^{3}}c_{r}^{*}c_{r}\omega_{r}\left(\omega_{r}\left(1-\mathcal{R}_{r}\mathcal{R}_{r}\right)-i2\partial_{\eta}\theta_{0}\mathcal{R}_{r}\right)\left(2a_{-\vec{k}}^{r\dagger}a_{\vec{k}}^{r}+\left[a_{\vec{k}}^{r},a_{-\vec{k}}^{r\dagger}\right]\right).
\end{align*}
Using the expression derived for the coefficients $c_{+\pm}$ from
Eq.~(\ref{eq:cpp and cpm-1}) we find that
\begin{equation}
c_{r}^{*}c_{r}\left(\omega_{r}\left(1-\mathcal{R}_{r}\mathcal{R}_{r}\right)-i2\partial_{\eta}\theta_{0}\mathcal{R}_{r}\right)=\frac{1}{2},
\end{equation}
which allows us to write the final form of the Hamiltonian as
\begin{align*}
\lim_{\eta\rightarrow\eta_{i}}H & =\sum_{r=++,+-}\int\frac{d^{3}k}{\left(2\pi\right)^{3}}\frac{\omega_{r}}{2}\left(2a_{-\vec{k}}^{r\dagger}a_{\vec{k}}^{r}+\left[a_{\vec{k}}^{r},a_{-\vec{k}}^{r\dagger}\right]\right)\\
 & =\sum_{r=++,+-}\int\frac{d^{3}k}{\left(2\pi\right)^{3}}\omega_{r}\left(a_{-\vec{k}}^{r\dagger}a_{\vec{k}}^{r}+\frac{1}{2}\delta^{(3)}(0)\right).
\end{align*}
The vacuum state $\left|0\right\rangle $ when applied to the above
Hamiltonian results in the lowest energy state with ground state energy
$E_{0}=\frac{1}{2}\hbar\left(\omega_{++}+\omega_{+-}\right)$. Also,
we see that the one particle state $\left|r\right\rangle =a_{\vec{k}}^{r\dagger}\left|0\right\rangle $
is an eigenstate of the Hamiltonian with the energy eigenvalue $E_{0}+\hbar\omega_{r}$.
We note that any other choice for the mode amplitudes $V_{r}^{n}(k,\eta_{i})$
other than that given in Eqs.~(\ref{eq:A++-1}) and (\ref{eq:A+--1})
will lead to a higher energy for the vacuum state $\left|0\right\rangle $
and as such it would not be the correct ground state of our theory.

\section{Correlation function\label{sec:Correlation-function}}

Through the quantization presented in Appendix \ref{sec:Details-of-quantization},
we showed that our coupled system of mode functions $h_{k}(\eta)$
has two sets of normal frequencies $\omega_{++}$ and $\omega_{+-}$
with which they can be excited. Each frequency $\omega_{r}(k)$ corresponds
to an independent quantum oscillator solution. In this Appendix we
will evaluate the non-zero variance of these zero-point quantum fluctuations.
Hence we consider the following expression for the two-correlation
$\xi_{nm}=\left\langle \delta\psi^{n}\left(0,\eta\right)\delta\psi^{m}\left(0,\eta\right)\right\rangle $
and evaluate it as
\begin{align*}
\xi_{nm} & =\left\langle 0\left|\delta\psi^{n}\left(0,\eta\right)\delta\psi^{m}\left(0,\eta\right)\right|0\right\rangle \\
 & =\int\frac{d^{3}k}{(2\pi)^{3/2}}\int\frac{d^{3}p}{(2\pi)^{3/2}}\left\langle 0\left|\left(a_{\vec{k}}^{++}h_{k}^{n++}+a_{\vec{k}}^{+-}h_{k}^{n+-}+h.c.\right)\right.\right.\\
 & \times\left.\left.\left(a_{\vec{p}}^{++}h_{p}^{m++}+a_{\vec{p}}^{+-}h_{p}^{m+-}+h.c.\right)\right|0\right\rangle \\
 & =\int\frac{d^{3}k}{(2\pi)^{3/2}}\int\frac{d^{3}p}{(2\pi)^{3/2}}\left\langle 0\left|a_{\vec{k}}^{++}a_{-\vec{p}}^{++\dagger}h_{k}^{n++}h_{p}^{m++*}+a_{\vec{k}}^{+-}a_{\vec{-p}}^{+-\dagger}h_{k}^{n+-}h_{p}^{m+-*}\right|0\right\rangle \\
 & =\int\frac{d^{3}k}{(2\pi)^{3/2}}\int\frac{d^{3}p}{(2\pi)^{3/2}}h_{k}^{n++}h_{p}^{m++*}\left\langle 0\left|\left[a_{\vec{k}}^{++},a_{-\vec{p}}^{++\dagger}\right]\right|0\right\rangle +h_{k}^{n+-}h_{p}^{m+-*}\left\langle 0\left|\left[a_{\vec{k}}^{+-},a_{\vec{-p}}^{+-\dagger}\right]\right|0\right\rangle \\
 & =\int\frac{d^{3}k}{(2\pi)^{3/2}}\int\frac{d^{3}p}{(2\pi)^{3/2}}\left(h_{k}^{n++}h_{p}^{m++*}+h_{k}^{n+-}h_{p}^{m+-*}\right)\delta^{(3)}(\vec{k}-\vec{p})\\
 & =\int d\ln k\frac{k^{3}}{2\pi^{2}}\left(h_{k}^{n++}(\eta)h_{k}^{m++*}(\eta)+h_{k}^{n+-}(\eta)h_{k}^{m+-*}(\eta)\right)\\
 & =\int d\ln k\Delta_{\delta\psi^{n}\delta\psi^{m}}^{2}(k,\eta)
\end{align*}
where we define 
\begin{equation}
\Delta_{\delta\psi^{n}\delta\psi^{m}}^{2}(k,\eta)=\frac{k^{3}}{2\pi^{2}}\left(h_{k}^{n++}(\eta)h_{k}^{m++*}(\eta)+h_{k}^{n+-}(\eta)h_{k}^{m+-*}(\eta)\right)
\end{equation}
such that 
\begin{equation}
\Delta_{\delta\phi^{n}\delta\phi^{m}}^{2}(k,\eta)=\frac{1}{a^{2}(\eta)}\frac{k^{3}}{2\pi^{2}}\left(h_{k}^{n++}(\eta)h_{k}^{m++*}(\eta)+h_{k}^{n+-}(\eta)h_{k}^{m+-*}(\eta)\right).\label{eq:DELTAsq_phi_nm}
\end{equation}
During conformal regime $\eta<\eta_{{\rm tr}}$, the power spectra
of the individual fields are given as
\begin{align}
\Delta_{\delta\Gamma\delta\Gamma}^{2}(\eta<\eta_{{\rm tr}}) & =\frac{1}{a^{2}(\eta)}\frac{k^{3}}{2\pi^{2}}\left(\left|h_{k}^{1++}(\eta)\right|^{2}+\left|h_{k}^{1+-}(\eta)\right|^{2}\right)\\
 & =\frac{1}{a^{2}(\eta)}\frac{k^{3}}{2\pi^{2}}\left(\left|c_{++}\right|^{2}+\left|c_{+-}\right|^{2}\right)\\
\lim_{k\ll\partial_{\eta}\theta_{0}}\Delta_{\delta\Gamma\delta\Gamma}^{2}(\eta<\eta_{{\rm tr}}) & \approx\frac{1}{a^{2}(\eta)}\frac{k^{2}}{2\pi^{2}}\left(\frac{k/\partial_{\eta}\theta_{0}}{3^{1/2}2^{3/2}}\right)
\end{align}
and 
\begin{align}
\Delta_{\delta\chi\delta\chi}^{2}(\eta<\eta_{{\rm tr}}) & =\frac{1}{a^{2}(\eta)}\frac{k^{3}}{2\pi^{2}}\left(\left|h_{k}^{2++}(\eta)\right|^{2}+\left|h_{k}^{2+-}(\eta)\right|^{2}\right)\label{eq:dim_spec_axial}\\
 & =\frac{1}{a^{2}(\eta)}\frac{k^{3}}{2\pi^{2}}\left(\left|c_{++}\mathcal{R}_{++}\left(\eta\right)\right|^{2}+\left|c_{+-}\mathcal{R}_{+-}\left(\eta\right)\right|^{2}\right)\\
\lim_{k\ll\partial_{\eta}\theta_{0}}\Delta_{\delta\chi\delta\chi}^{2}(\eta<\eta_{{\rm tr}}) & \approx\frac{1}{a^{2}(\eta)}\frac{k^{2}}{2\pi^{2}}\left(\frac{1}{3^{1/2}2}\right).
\end{align}
Similarly one can show that the dimensionless cross-correlation (covariance)
vanishes,
\begin{align}
\Delta_{\delta\Gamma\delta\chi}^{2}(\eta<\eta_{{\rm tr}}) & =\frac{1}{a^{2}(\eta)}\frac{k^{3}}{2\pi^{2}}\left(\left|c_{++}\right|^{2}\mathcal{R}_{++}^{*}\left(\eta\right)+\left|c_{+-}\right|^{2}\mathcal{R}_{+-}^{*}\left(\eta\right)\right)\nonumber \\
 & =0.
\end{align}
The correlation between the time-derivatives of the fields $\Xi=\left\langle \partial_{\eta}\delta\psi^{n}\left(0,\eta\right)\partial_{\eta}\delta\psi^{m}\left(0,\eta\right)\right\rangle $
can be derived as 
\begin{align*}
\Xi & =\left\langle 0\left|\partial_{\eta}\delta\psi^{n}\left(0,\eta\right)\partial_{\eta}\delta\psi^{m}\left(0,\eta\right)\right|0\right\rangle \\
 & =\int\frac{d^{3}k}{(2\pi)^{3/2}}\int\frac{d^{3}p}{(2\pi)^{3/2}}\left\langle 0\left|\partial_{\eta}\left(a_{\vec{k}}^{++}h_{k}^{n++}+a_{\vec{k}}^{+-}h_{k}^{n+-}+h.c.\right)\right.\right.\\
 & \times\left.\left.\partial_{\eta}\left(a_{\vec{p}}^{++}h_{p}^{m++}+a_{\vec{p}}^{+-}h_{p}^{m+-}+h.c.\right)\right|0\right\rangle \\
 & =\int\frac{d^{3}k}{(2\pi)^{3/2}}\int\frac{d^{3}p}{(2\pi)^{3/2}}\left\langle 0\left|a_{\vec{k}}^{++}a_{-\vec{p}}^{++\dagger}\partial_{\eta}h_{k}^{n++}\partial_{\eta}h_{p}^{m++*}+a_{\vec{k}}^{+-}a_{\vec{-p}}^{+-\dagger}\partial_{\eta}h_{k}^{n+-}\partial_{\eta}h_{p}^{m+-*}\right|0\right\rangle \\
 & =\int\frac{d^{3}k}{(2\pi)^{3/2}}\int\frac{d^{3}p}{(2\pi)^{3/2}}\partial_{\eta}h_{k}^{n++}\partial_{\eta}h_{p}^{m++*}\left\langle 0\left|\left[a_{\vec{k}}^{++},a_{-\vec{p}}^{++\dagger}\right]\right|0\right\rangle +\partial_{\eta}h_{k}^{n+-}\partial_{\eta}h_{p}^{m+-*}\left\langle 0\left|\left[a_{\vec{k}}^{+-},a_{\vec{-p}}^{+-\dagger}\right]\right|0\right\rangle \\
 & =\int\frac{d^{3}k}{(2\pi)^{3/2}}\int\frac{d^{3}p}{(2\pi)^{3/2}}\left(\partial_{\eta}h_{k}^{n++}\partial_{\eta}h_{p}^{m++*}+\partial_{\eta}h_{k}^{n+-}\partial_{\eta}h_{p}^{m+-*}\right)\delta^{(3)}(\vec{k}-\vec{p})\\
 & =\int d\ln k\frac{k^{3}}{2\pi^{2}}\left(\partial_{\eta}h_{k}^{n++}(\eta)\partial_{\eta}h_{k}^{m++*}(\eta)+\partial_{\eta}h_{k}^{n+-}(\eta)\partial_{\eta}h_{k}^{m+-*}(\eta)\right).
\end{align*}
Using the above expression we find the cross-correlation 
\begin{align}
\left\langle \partial_{\eta}\delta Y\left(0,\eta\right)\partial_{\eta}\delta X\left(0,\eta\right)\right\rangle _{\eta<\eta_{{\rm tr}}} & =\int d\ln k\frac{k^{3}}{2\pi^{2}}\left(\omega_{++}^{2}\left|c_{++}\right|^{2}\mathcal{R}_{++}^{*}+\omega_{+-}^{2}\left|c_{+-}\right|^{2}\mathcal{R}_{+-}^{*}\right)\\
 & =\int d\ln k\frac{k^{3}}{2\pi^{2}}\left(i\partial_{\eta}\theta_{0}\right).
\end{align}
As expected, the two point correlation of the field velocities is
a non-vanishing observable at early-time. Likewise,
\begin{align}
\lim_{k\ll\partial_{\eta}\theta_{0}}\left\langle \partial_{\eta}\delta Y\left(0,\eta\right)\partial_{\eta}\delta Y\left(0,\eta\right)\right\rangle _{\eta<\eta_{{\rm tr}}} & \approx\int d\ln k\sqrt{\frac{3}{2}}\left(\frac{k^{3}}{2\pi^{2}}\partial_{\eta}\theta_{0}\right),\\
\lim_{k\ll\partial_{\eta}\theta_{0}}\left\langle \partial_{\eta}\delta X\left(0,\eta\right)\partial_{\eta}\delta X\left(0,\eta\right)\right\rangle _{\eta<\eta_{{\rm tr}}} & \approx\int d\ln k\sqrt{\frac{2}{3}}\left(\frac{k^{3}}{2\pi^{2}}\partial_{\eta}\theta_{0}\right).
\end{align}

\section{\label{sec:Relationship-between-radial}Relationship between radial
and angular modes}

Suppose we parameterize a $U(1)$ sigma model with the symmetry $\Phi\rightarrow e^{i\alpha}\Phi$
as
\begin{equation}
\Phi=\frac{1}{\sqrt{2}}(\Gamma_{0}+\delta\Gamma)e^{i\left(\theta_{0}+\frac{\delta\chi}{\Gamma_{0}}\right)}.
\end{equation}
There is a shift symmetry 
\begin{equation}
\delta\chi\rightarrow\delta\chi+\alpha\Gamma_{0}\label{eq:shift}
\end{equation}
where $\alpha$ is a constant. If $\delta\chi$ obeys a linear equation
of motion
\begin{equation}
\mathcal{O}\delta\chi=\beta\label{eq:EOM1}
\end{equation}
where $\mathcal{O}$ and $\beta$ are independent of $\delta\chi$
but can depend on $\Gamma_{0}$, then Eq.~(\ref{eq:shift}) implies
\begin{equation}
\mathcal{O}\delta\chi+\alpha\mathcal{O}\Gamma_{0}=\beta.
\end{equation}
Using Eq.~(\ref{eq:EOM1}), we conclude
\begin{equation}
\mathcal{O}\Gamma_{0}=0\label{eq:homogimplication}
\end{equation}
which can be a nonlinear equation.

This means that if $\beta$ is negligible, then $\delta\chi$ and
$\Gamma_{0}$ obey the same equation. In our model, the equation of
motion for $\delta\chi$ (see Eq.~(\ref{eq:radial})) makes
\begin{equation}
\mathcal{O}=\partial_{\eta}^{2}-a^{-2}\partial_{i}^{2}+2\frac{\partial_{\eta}a}{a}\partial_{\eta}+\left(-2M^{2}a^{2}+\lambda\Gamma_{0}^{2}a^{2}-\left(\partial_{\eta}\theta_{0}\right)^{2}\right)
\end{equation}
in Eq.~(\ref{eq:homogimplication}) which upon expansion gives
\begin{equation}
\partial_{\eta}^{2}\Gamma_{0}+2\frac{\partial_{\eta}a}{a}\partial_{\eta}\Gamma_{0}+\left(-2M^{2}a^{2}+\lambda\Gamma_{0}^{2}a^{2}-\left(\partial_{\eta}\theta_{0}\right)^{2}\right)\Gamma_{0}=0
\end{equation}
matching Eq.~(\ref{eq:radial_eom}). The $\beta$ in this system
is
\begin{equation}
\beta=-2\partial_{\eta}\theta_{0}\Gamma_{0}\partial_{\eta}\frac{\delta\Gamma}{\Gamma_{0}}.
\end{equation}
Because of the mismatch of $\beta$ between $\Gamma_{0}$ and $\delta\chi$,
we cannot conclude that $\delta\chi/\Gamma_{0}$ is constant from
this argument alone.

We will now show that the conservation equation from $U(1)$ symmetry
together with a mild assumption about the lack of resonance allows
one to conclude that $\delta\chi(\eta)/\Gamma_{0}(\eta)$ is approximately
frozen during and after the transition. Start with the linear perturbation
equation for the the $U(1)$ current conservation
\begin{equation}
\partial_{\eta}\delta q+\frac{\left|\vec{k}\right|^{2}\left(a\Gamma_{0}\right)\delta\chi}{Q^{(0)}}=0\label{eq:conservationwithksquared}
\end{equation}
where we have defined 
\begin{equation}
\delta q\equiv\frac{1}{\left(\partial_{0}\theta_{0}\right)}\frac{1}{a}\frac{\partial}{\partial\eta}\left(\frac{\delta\chi}{a\Gamma_{0}}\right)+2\frac{\delta\Gamma}{a\Gamma_{0}}\label{eq:defofconservedquantity}
\end{equation}
and gone to Fourier space. Let $\eta_{i}$ be the first time in the
time-independent conformal era when the $k^{2}$ term can be neglected
in Eq.~(\ref{eq:conservationwithksquared}). We can conclude that
$\delta q$ which is set during the time-independent conformal era
to be
\begin{align}
\delta q & \approx\frac{3c_{+-}}{a(\eta_{i})\Gamma_{0}(\eta_{i})}\approx\frac{3\delta\Gamma(\eta_{i},\vec{k})}{a(\eta_{i})\Gamma_{0}(\eta_{i})}
\end{align}
is conserved while our quantity of interest $\delta\chi/\Gamma_{0}$
is related to this constant through
\begin{equation}
\frac{\partial}{\partial\eta}\left(\frac{\delta\chi}{a\Gamma_{0}}\right)+2\frac{\delta\Gamma}{a\Gamma_{0}}a\left(\partial_{0}\theta_{0}\right)=\delta qa\left(\partial_{0}\theta_{0}\right)
\end{equation}
coming from Eq.~(\ref{eq:defofconservedquantity}). Integrating,
we find
\begin{align}
\frac{\delta\chi(\eta_{f})}{a(\eta_{f})\Gamma_{0}(\eta_{f})}-\frac{\delta\chi(\eta_{tr})}{a(\eta_{i})\Gamma_{0}(\eta_{tr})} & =\int^{\eta_{f}}d\eta'\left\{ \delta q\partial_{\eta'}\theta_{0}(\eta')-2\frac{\delta\Gamma(\eta',\vec{k})}{\Gamma_{0}(\eta')}\frac{1}{a(\eta')}\partial_{\eta'}\theta_{0}\right\} \\
 & \approx\int_{\eta_{tr}}^{\eta_{f}}d\eta'\left\{ \frac{3\delta\Gamma(\eta_{tr},\vec{k})}{a(\eta_{tr})\Gamma_{0}(\eta_{tr})}\partial_{\eta'}\theta_{0}(\eta')-2\frac{\delta\Gamma(\eta',\vec{k})}{\Gamma_{0}(\eta')}\frac{1}{a(\eta')}\partial_{\eta'}\theta_{0}\right\} 
\end{align}
where $\eta_{tr}$ is the time at which $\Gamma_{0}(\eta)a(\eta)$
starts to change in time (i.e. deviate from the conformal behavior).
Because the lighter energy mode $|\omega_{+-}\rangle$ becomes purely
the $\delta\chi$ after the $\Gamma_{0}$ settles to the minimum of
the potential, we know $\delta\Gamma(\eta',\vec{k})\rightarrow0$
asymptotically. Hence, for $t'>t_{tr}$ we shall assume 
\begin{equation}
\delta\Gamma(\eta',\vec{k})\lesssim\delta\Gamma(\eta_{tr},\vec{k}).
\end{equation}
Additionally we consider a smooth non-resonant adiabatic transition
of the background radial field such that 
\begin{equation}
\Gamma_{0}(\eta')a(\eta')\gtrsim\Gamma_{0}(\eta_{tr})a(\eta_{tr})\label{eq:gammabound}
\end{equation}
because the equation of motion near the transition time can be solved
to obtain
\begin{equation}
Y_{0}(\eta)\approx Y_{c}\left(1+\frac{\left(2M^{2}/H^{2}+2\right)}{f^{2}\eta^{2}}-\frac{\left(2M^{2}/H^{2}+2\right)}{f^{2}\eta_{i}^{2}}\cos\left(f\left(\eta-\eta_{i}\right)\right)\right)
\end{equation}
which shows that the coefficient of $\cos(f(\eta-\eta_{i}))$ is suppressed.
Hence, compared to the first term, the second term in the integral
falls off rapidly for $t>t_{tr}$. Thus, we simplify the integral
as
\begin{align}
\frac{\delta\chi(\eta_{f})}{a(\eta_{f})\Gamma_{0}(\eta_{f})}-\frac{\delta\chi(\eta_{tr})}{a(\eta_{i})\Gamma_{0}(\eta_{tr})} & \lesssim\left|\frac{3\delta\Gamma(\eta_{tr},\vec{k})}{a(\eta_{tr})\Gamma_{0}(\eta_{tr})}\right|\int_{\eta_{tr}}^{\eta_{f}}d\eta'\partial_{\eta'}\theta_{0}(\eta').
\end{align}
Using the conservation equation, this becomes
\begin{align}
\frac{\delta\chi(\eta_{f})}{a(\eta_{f})\Gamma_{0}(\eta_{f})}-\frac{\delta\chi(\eta_{tr})}{a(\eta_{i})\Gamma_{0}(\eta_{tr})} & \lesssim\left|\frac{3\delta\Gamma(\eta_{tr},\vec{k})}{a(\eta_{tr})\Gamma_{0}(\eta_{tr})}\right|\frac{Q^{(0)}H^{2}}{\Gamma_{0}^{2}(\eta_{tr})}\left(-\eta_{tr}^{3}\right)
\end{align}
where we used Eq.~(\ref{eq:gammabound}). Since we can solve during
the time-independent conformal era
\begin{equation}
\left|\frac{\delta\Gamma(\eta_{tr},\vec{k})}{a(\eta_{tr})\Gamma_{0}(\eta_{tr})}\right|\approx\left(\frac{k}{\partial_{\eta}\theta_{0}(\eta_{tr})}\right)\left|\frac{\delta\chi(\eta_{tr},\vec{k})}{a(\eta_{tr})\Gamma_{0}(\eta_{tr})}\right|
\end{equation}
we obtain the relation
\begin{align}
\frac{\delta\chi(\eta_{f})}{a(\eta_{f})\Gamma_{0}(\eta_{f})}-\frac{\delta\chi(\eta_{tr})}{a(\eta_{tr})\Gamma_{0}(\eta_{tr})} & \lesssim\left|\frac{\delta\chi(\eta_{tr},\vec{k})}{a(\eta_{tr})\Gamma_{0}(\eta_{tr})}\right|\left(\frac{k}{a_{tr}H}\right).
\end{align}
This indicates that in the long wavelength limit, the isocurvature
perturbation is conserved for modes outside of the horizon at the
transition time even in the presence of a large rotating background.

\bibliographystyle{JHEP2}
\phantomsection\addcontentsline{toc}{section}{\refname}\bibliography{axion_isocurvature_papers2,blue_isocurvature,blueiso_constraints,file,kasuya_citations_and_other_misc_papers2,misc,Inflation_general}

\end{document}